\documentclass[12pt,oneside]{book}
\usepackage[a4paper, top=3cm, bottom=3cm]{geometry}
\usepackage{fancyhdr}
\usepackage{cite}
\makeatletter
\def\cleardoublepage{\clearpage\if@twoside \ifodd\c@page\else%
    \hbox{}%
    \thispagestyle{empty}%              % Empty header styles
    \newpage%
    \if@twocolumn\hbox{}\newpage\fi\fi\fi}
\makeatother

\usepackage{fancyhdr}
\usepackage{tocloft}
\pdfoutput =1
\usepackage{graphics}
\usepackage{graphicx}
\DeclareGraphicsExtensions{.pdf}
\usepackage{float} %Include figure filesusepackage{graphicx} %Include figure files
\usepackage{amsmath}
\usepackage{amsfonts}   
\usepackage{amssymb}
\usepackage{subfloat}

\begin{document}
\date{}

%\pagestyle{empty}
%\pagenumbering{}
% Set book title
% Include Author name and Copyright holder name
%\author{Dibakar Roychowdhury}
%%%%%%%%%%%%%%%%%%%%%%%%%%%%%%%%%%%%%%%%%%%%%%%%%%%%%%%%%%%%%%%%%%%%
\begin{titlepage}
\begin{center}

%%%%%%%%%%%%%%%%%%%%%%%%%%%%%%%%%%%%%%%%%%%%%%%%%%%%%%%%%%%%%%%%%%%%%%%%%%%%%%%%%%%%%%%%%%%
 {\bf \uppercase{\huge I\Large NVESTIGATIONS \huge i\Large N \vspace{0.3cm}\\
\vspace{0.3cm}\huge H\Large IGHER \huge D\Large ERIVATIVE  \huge F\Large IELD \huge T\Large HEORIES\\
%\vspace{0.4 cm} \huge Q\Large uantum \huge T\Large unneling
}}
%(Corrected Copy)
%%%%%%%%%%%%%%%%%%%%%%%%%%%%%%%%%%%%%%%%%%%%%%%%%%%%%%%%%%%%%%%%%%%%%%%%%%%%%%%%%%%%%%%%%%
%\Large\textsc{Field theory aspects of  Cosmology \\and \\Black Holes}
%%%%%%%%%%%%%%%%%%%%%%%%%%%%%%%%%%%%%%%%%%%%%%%%%%%%%%%%%%%%%%%%%%%%%%%%%%%%%%%%%%%%%%%5
\vfill

\normalsize
{\Large Thesis Submitted For The Degree Of}\\[2.2ex]
\textbf{\Large Doctor of Philosophy (Science)}\\[2ex]
In\\[2ex] 
{\Large \textbf{Physics (Theoretical)} }\\[2ex]
By\\[2ex]
\textbf{\Large Biswajit Paul}

\vfill

\vfill

\textbf{{\Large \textbf{\large University of Calcutta, India} }}\\[2ex]
{\large 2014}
%{\large \mbox{Satyendra Nath Bose National Centre for Basic Sciences}}\\
%{\large JD Block, Sector 3, Salt Lake, Kolkata 700098, India}
%{Department of Theoretical Sciences}

\end{center}
\end{titlepage}

\newpage
%\chapter*{}
\thispagestyle{empty}
\vspace*{6.5 cm}
%%%%%%%%%%%%%%%%%%%%%%%%%%%%%%%%%%%%%%%%%%%%%%%%%%%%%%%%%%%%%
%\begin{flushright}
%{\Large \bf To\hspace*{2 cm}\\my\hspace*{1.5 cm}\\mother}
%\end{flushright}
%%%%%%%%%%%%%%%%%%%%%%%%%%%%%%%%%%%%%%%%%%%%%%%%%%%%%%%%%%%%
%\begin{figure}[th] 
{\centering
%\includegraphics[width=0.5\textwidth]{nep1.jpg}
%\end{figure}
%%%%%%%%%%%%%%%%%%%%%%%%%%%%%%%%%%%%%%%%%%%%%%%%%%%%%%%%%%%%%%
%\Large \bf \hspace*{8.2 cm}To\\ \hspace*{9.6 cm}my\\ \hspace*{10.5 cm}mother\\
\hspace*{8.6 cm} $\mathcal{TO}$\\ \hspace*{10 cm}$\mathcal{MY \ PARENTS}$}

%%%%%%%%%%%%%%%%%%%%%%%%%%%%%%%%%%%%%%%%%%%%%%%%%%%%%%%%%%%%%

\newpage

\chapter*{Acknowledgments}

This thesis is the outcome of the cumulative efforts that has been paid for the last four and half years during my stay at S. N. Bose National Centre for Basic Sciences, Kolkata. Council of Scientific and Industrial Research, India has provided me the financial support.

First of all, my sincere thanks and gratitude to my guide Prof. Rabin Banerjee,  who throughout the term of my PhD.  was actively involved with me in research and monitored my progress.

I thank and acknowledge my joint supervisor Prof. Pradip Mukherjee for the help he provided me throughout my PhD. career. I came to know many calculation skills from him.

I  express my sincere thanks to Prof. Amitabha Lahiri (SNBNCBS, kolkata) for the lectures on differential geometry. I also  thank Prof. Claus Kiefer (University of Koln, Cologne), Prof. Andrei V. Smilga (University of Nantes, Nantes), Prof. Hanno Sahlmann (University of Erlangen, Nuremberg), Dr. Golam Hossain (IISER Kolkata, Kolkata), Prof. Sayan Kar (IIT Kharagpur, Kharagpur), Dr. Sukanta Panda (IISER Bhopal) for organising short visits and discussions.

I am specially thankful to my schoolteacher  Mr. Premananda Paul as I am indebted to him for his  inspiration and help in many ways during  my school days. 

I take the opportunity to convey my sincere gratitude to  Suman Saha, my childhood friend for his constant encouragement and assistance. 
 
 I thank my cousin  Binoy Paul for being all the time with me.
  
I am grateful to Dr. Debraj Roy, who  always was helpful for me on physics and nonphysics matters,  whenever needed. 

I am specially thankful to Dr. Dibakar Roychowdhury, my group mate and cubicle mate for his  support throughout  and spending very valuable moments with me.

I am also thankful to my  group members   Dr. Bibhas Ranjan Majhi, Dr. Sujoy Kumar Modak, Dr. Sudhaker Upadhyay, Dr. Sunandan Gangopadhyay,   Arindam Lala, Shirsendu Dey, Arpan K. Mitra and Arpita Mitra for  discussion on physics and gossiping.

 I would like to thank my fellow friends  Arun Laxmanan, Rajiv Chauhan, Animesh Patra, Arghya Das, Abhijit Chakraborty, Debmalya Mukhopadhyay, Paulami Chakraborty, Shubhashish Chakraborty, Anshuman De, all mates from SNB football team  with whom I spent my times at S N Bose Centre involving nonphysics activities.
 %I also thank my keyboard(synthesizer) teacher Mr. B. N. Shur and my taekwondo teachers Master Pradipta kumar Roy (Hall of Fame) and Master Ruma Roy Chowdhury for all the 
 
  I owe my entire life including this thesis to my parents and younger brothers for all the love, affections and support they provided me.

Finally, it's to her, Debjani Paul, for making the struggle smoother and constantly  motivating me.

%%%%%%%%%%%%%%%%%%%%%%%%%%%%%%%%%%%%%%%%%%%%%%%%%%%%%%%%%%%%%

\chapter*{List of publications}
\thispagestyle{empty}

%%%%%%%%%%%%%%%%%%%%%%%%%%%%%%%%%%%%%%%%%%%%%%%%%%%%%%%%%%%%%%

%\begin{minipage}{13cm}
 %\raggedright
%\vskip -0.8cm

\begin{enumerate}
\item \textit{``Gauge symmetry and W-algebra in higher derivative systems"}\\
{Rabin Banerjee, Pradip Mukherjee, \textbf{Biswajit Paul}}\\
Published in \textbf{Journal of High Energy Physics \textbf{08} (2011) 085.}\\
e-Print: arXiv:1012.2969.\\

\item \textit{``Gauge invariances of higher derivative Maxwell-Chern-Simons field theory: A new Hamiltonian approach"}\\
{Pradip Mukherjee, \textbf{Biswajit Paul}}\\
Published in \textbf{Phys. Rev. D \textbf{85 }(2012) 045028.}\\
e-Print: arXiv:1111.0153.\\

\item \textit{``Gauge symmetry and Virasoro algebra in quantum charged rigid membrane: A first order formalism''}\\
{\textbf{Biswajit Paul}}\\
Published in \textbf{Phys.Rev. D 87 (2013) 045003.}\\
e-Print: arXiv:1212.5902.\\

\item \textit{``BRST symmetry and W-algebra in higher derivative models''}\\
{Rabin Banerjee, \textbf{Biswajit Paul}, Sudhaker Upadhyay}\\
Published in \textbf{Phys. Rev. D \textbf{88} (2013) 065019.}\\
e-Print: arXiv:1306.0744.\\

\item \textit{``New Hamiltonian analysis of Regge-Teitelboim minisuperspace cosmology"}\\
{Rabin Banerjee, Pradip Mukherjee, \textbf{Biswajit Paul}}\\
Published in \textbf{Phys.Rev. D 89 (2014) 043508.}\\
e-Print: arXiv:1307.4920.\\
 
 This thesis is based on the above mentioned papers.
\end{enumerate}
%%%%%%%%%%%%%%%%%%%%%%%%%%%%%%%%%%%%%%%%%%%%%%%%%%%%%%
\newpage
\thispagestyle{empty}

~~~~~~~~~~~~~~~~~ ~~~~~~~~~~~~~~~~~~~~ ~~~~~~~~~~~~~

~~~~~~~~~~~~~~~~~~~~~~~~ ~~~~~~~~~~~~~~~~~~~~~~~~~~~~~~

~~~~~~~~~~~~~~~~~~~~~~~~~~~~~~~ ~~~~~~~~~~~~~~~~~~~~~~~~

~~~~~~~~~~~~~~~~~~~~~~~~ ~~~~~~~~~~~~~~~~~~~~~~~~~~~~~~~~~

~~~~~~~~~~~~~~~~~~~~~~~~~~~~~~~~~~~~~~~~~~~~ ~~~~~~~~~~~~~~~~~~

~~~~~~~~~~~~~~~~~~~~~~~~~~~~~~~~~~~~~~~~~~~~~~~~~~~~~~~~~~~~~~

~~~~~~~~~~~~~~~~~~~~~~~~~~~~~~~~~~~~~~~~~~~ ~~~~~~~~~~~~~~~~~~~

~~~~~~~~~~~~~~~~~~~~~~~~~~~~~~~~~~~~~ ~~~~~~~~~~~~~~~~~~~~~~~~

\begin{center}

%%%%%%%%%%%%%%%%%%%%%%%%%%%%%%%%%%%%%%%%%%%%%%%%%%%%%%%%%%%%%%%%%%%%%%%%%%%%%%%%%%%%%%%%%%%
 {\bf \uppercase{\huge I\Large NVESTIGATIONS \huge I\Large N  \vspace{0.3cm}\\  \huge H\Large IGHER \huge D\Large ERIVATIVE \huge F\Large IELD \huge T\Large HEORIES   \vspace{0.3cm}\\
%\vspace{0.4 cm} \huge Q\Large uantum \huge T\Large unneling
}}
\end{center}

%%%%%%%%%%%%%%%%%%%%%%%%%%%%%%%%%%%%%%%%%%%%%%%%%%%%%%%%%%
\newpage
% Include dots between chapter name and page number
\renewcommand{\cftchapdotsep}{\cftdotsep}
%Finally, include the ToC
\tableofcontents

%%%%%%%%%%%%%%%%%%%%%%%%%%%%%%%%%%%%%%%%%%%%%%%%%%%%%%%%%%%%%

\chapter{Introduction}

  We know, in usual theories the Lagrangian is a function of fields and their first derivatives, known as first order systems.  These systems were conceptualised from the time of Newton and successfully utilised for explaining various physical phenomena. But, theories with higher time derivatives of the fields became essential in some cases to make those theories renormalisable and free from divergences \cite{stelle1, thirring1, fradkin1}. We call those theories as higher derivative theories.   The higher derivative  terms mainly occur as correction terms to the first order theories which play an  important rule during quantisation.  Lagrangian theories with higher order derivatives are interesting in their own right.  Since then, there appear numerous fields where the concepts of higher derivative theories have been used. Broadly speaking, in field theory \cite{pias1, pisarski, reyes1, podolsky}, string theory\cite{elizer}, gravity \cite{stelle1, bergshoeff, gullu, neupane} and cosmology \cite{hawking, mazitelli, nojiri1} higher derivative terms are frequently considered. Adding a higher derivative term can regularize the ultraviolet behaviour of the theory \cite{thirring1, pias1}.  These higher derivative terms can make the modified gravity renormalizable and even asymptotically free \cite{stelle1, fradkin1}. People constructed $f(R)$ gravity where higher curvature terms were added to the Einstein-Hilbert action and opened a vast sector of research \cite{stelle1, sotiriou,  paschalidis, faraoni, fabri}. For higher derivative gravity,
the list is huge \cite{stelle1, bergshoeff,  gullu, ohta, brustein, bunch, deser3, dyer}. higher derivative theories were also considered in  the most exciting fields of recent theoretical physics like AdS/CFT correspondence which indicate the importance and relevance of considering higher derivative theories \cite{nojiri1,  nojiri3, fukuma}. Due to these important applications of higher derivative theories,  a systematic analysis  of these theories  have become necessary.

% A physical system is characterised by some action principle. Varying the action and considering conditions on the field variables or their variations, one can analyse the full dynamics of the system.
	For physicists,  studies of symmetries of a dynamical system is very important as  many physical phenomena can be predicted just considering the symmetries.   Gauge symmetry plays an important role in modern theoretical physics.  Existence of gauge symmetries means not all degrees of freedoms are physically relevant.  For this reason, abstraction of gauge symmetries to make a theory physically relevant has become very important. In physics, existence of a particular symmetry in the theory always refers to some conserved quantity.  In fact, some experimental results were predicted and benefited just considering the  symmetries of the physical processes occurring  during the experiment. Recent LHC results and their consequences are nothing but finding  out the ``broken symmetries".  For this reason, the importance of finding gauge symmetries should not be overlooked. In this thesis, we  will focus on gauge symmetries and their consequences for  higher derivative theories.
 
 A systematic algorithm to extract  the symmetries of higher derivative theories can serve as a useful contribution for the future analysis of these theories. For this, we need to follow some Hamiltonian formulation.  In 1850, M. Ostrogradsky \cite{ostro} gave the Hamiltonian formulation for higher derivative theories.  But this  formalism suffers from  complications. The momenta, in this case, are defined via some non-trivial definition due to the higher derivative terms. All the higher derivatives of the fields are considered as independent variables and then Dirac's Hamiltonian formulation is followed. But, it has been noticed, in some cases, that the appearance of the number of independent primary first class constraints do not match with the number of independent gauge symmetries \footnote{For usual theories it has been shown that the no. of independent gauge parameters equals the no. of independent primary first class constraints of the theory} \cite{nesterenko}. For this reason it become necessary to have a well tested proper formulation to systematically abstract the gauge symmetries. This is the cornerstone of the present thesis.
  
 To abstract the gauge symmetries, other than the Ostrogradsy approach \cite{ostro, mulish}, we are going to follow  the first order formalism \cite{plyuschay1, plyuschay2, BMP} along with the powerful techniques developed by Banerjee et. al \cite{BRR}. According to the first order formulation, the fields and their corresponding time derivatives are considered as independent variables to convert the higher derivative Lagrangian to a first order one. From this first order Lagrangian one can construct the  Euler-Lagrange equations of motion and correspondingly develop a Hamiltonian formulation. But,  transition to Hamiltonian formulation becomes complicated when there appears constraints. Constraints are actually some functions of the phase-space variables which prohibit to have the whole access of the phase-space. For this reason, constraint systems need careful analysis. From the Hamiltonian formulation one can further proceed for quantisation of the system. This quantisation procedure is known as canonical quantisation, formerly introduced by P. A. M. Dirac \cite{dirac2}. In this seminal work by Dirac,  constrained  systems and their systematic quantisation procedure are discussed. There is other literature available for further studies on constrained systems \cite{henneaux1, gitman, hanson1, rothe1, sundermeyer1, prokhorov} particularly for the first order systems. 

% Another aspect of Hamiltonian formulation is to abstract the hidden symmetries of the system which plays a very important role to describe the quantised system. Symmetry appears in various form in nature from geometrical as well as physical point of view. The first experience of symmetry we come to experience is perhaps looking at a mirror. In terms of physics, this is known as ``mirror symmetry" where points on the either side correspond exactly to the points on the other side. Another nice and easy example can be seen from the distribution of flowers which in most cases shows the pentagonal form . The distribution or pattern of the leaves are also highly symmetric and varies among species of the  trees. It is known that there is a need for this distribution  which  stems from the need of water and food supply, geographical appearance of the species etc. What we want to emphasize is that appearance of these symmetries in nature are not arbitrary rather purposefully. Some other symmetries appearing in nature are like bilateral symmetry of the human bodies, the hexagonal symmetry of the honeycomb cells, cubic symmetry of the salt crystals etc. All these symmetries we talked about are geometrical symmetries. 

	In the first order formulation, due to redefinition of the fields, extra constraints at the Lagrangian level appear which are incorporated via Lagrange multipliers.  Now, these Lagrange multipliers are in fact treated as independent fields. Usual momenta definition is used which distinguishes this formulation from Ostrogradski method.  Due to the standard momenta definitions, the calculations become easier and transparent. 
	Constrained dynamics itself is a very interesting concept. All/some of the momenta in this case are not expressible with respect to velocities and hence the full phase space is not accessible. In this case the Lagrangian becomes singular and determinant of the Hessian matrix is equal to zero. Constraints are usually classified in the Dirac scheme as first class and second class. If their algebra closes, the set of constraints is first class, else it is second class. Consequently,  the following possibilities may arise:

	I. The original Lagrangian is singular but the additional constraints are all second class.  The reduction of phase space may be done by implementing the second class constraints strongly provided we replace all the Poisson brackets(PB) by appropriate Dirac brackets(DB).

 	II. The original Lagrangian is singular and there are both  second class and first class constraints among them. The second class constraints may be eliminated again by the DB technique. The first class constraints generate gauge transformations which are
required to be further analysed. These constraints may yield further constraints and
so on. The iterative process stops when no new constraints are generated.

   The first class constraints  generate gauge transformations. The gauge generator is defined as the linear combination of all the first class constraints. The coefficients of the constraints are called  gauge parameters. Now, all the gauge parameters may or may not be independent of each other. To identify the independent gauge symmetries we apply the effective scheme developed by Banerjee and co-authors \cite{BRR} which relates the gauge parameters and the Lagrange multipliersand suitability abstracts the independent gauge parameters. Once the independent gauge degrees of freedoms are identified the gauge generator can easily be expressed with respect to these parameters. The number of independent gauge parameters refers to the number of independent degrees of freedom. This completes the process of finding out the gauge symmetries of a higher derivative system.
  
  	 Gauge symmetric higher derivative theories appear in different contexts like the rigid relativistic particle models\cite{plyuschay1, plyuschay2}, the extended Maxwell-Chern-Simons theory \cite{reyes1}, the quantum charged rigid membrane \cite{dirac1}, the Regge-Teitelboim model in cosmology\cite{regge1} etc and many more. Applications of this first order approach in different field of physics are carried out sequentially to provide the efficacy of the first order formalism

%\section{different models}

	The first higher derivative model we are going to study is the massive relativistic particle model with curvature which is an extension of  Polyakov's string model concept \cite{polyakov}. There he added, a scale invariant  extrinsic curvature term to the usual relativistic particle model action whose influence in the infrared region determines the phase structure of the string theory. The  particle model of this theory was introduced explicitly by Pisarski \cite{pisarski} where he considered the action in which the curvature of the path appeared along with the length coupled via a dimensionless constant. The model is important for polymer physics \cite{peliti}. Also, Plyuschay has shown that  the massless case to be interpreted as the description of either bosons or fermions depending on the value of the quantized parameter \cite{plyuschay3}. The Ostrogradskian Hamiltonian analysis of this model was subsequently carried out later by many authors in \cite{nesterenko, ramos1, ramos2, capovilla1}.  M. S. Plyuschay is worthy to be mentioned \cite{plyuschay1, plyuschay2} for his contribution in relativistic particle models. The $2+1$ dimensional version for the relativistic particle model with curvature and torsion was studied  in \cite{kunzetsov}.  Ramos et al showed that the massless relativistic particle model with curvature obeys $W_3$ algebra \cite{ramos1}. More literature on $W_3$ algebra can be found here \cite{ kunzetsov, bakas, hull, gervais}. 	
	
	In the first order formalism, the relativistic particle model with curvature  is converted into a first order form by redefinition of the fields and their corresponding higher time derivatives. It is seen that the theory is constrained one. Preserving these primary constraints in time, the secondary constraints are found out for the model. After all the constraint combinatorics, we see that there are two independent primary first class constraints. Using these two first class constraints the gauge generator is constructed. It is seen that there is only one independent gauge parameter. Since there are two independent PFC we expected to have two independent gauge parameter. The mismatch  also existed previously while considering the Ostrogradski approach but remained unnoticed for a long time \cite{nesterenko}. The solution of this mismatch is clearly our finding. Also, the independent gauge symmetry is identified as reparametrisation invariance.
	
	The massless version of this model that is constructed by keeping only the curvature term is shown to have two PFC along corresponding to two gauge degrees of freedom. One of them is the usual diffeomorphism while the other shows the $W$-symmetry. It is notable that this extra symmetry is revealed by casting the equations of motion in the Bossinesq form. By examining the most general transformations that preserve the structure of the Boussinesq Lax operators it was demonstrated that the symmetry group of the model satisfies $W_3$ algebra \cite{ramos1, ramos2}. $W$-algebras are nothing but extended Virasoro algebra in conformal field theories.  
	
	For quantisation of a theory with gauge
invariance, BRST is a very powerful tool  which also helps in the proof of the renormalizability and unitarity of gauge theories. This transformation, which is characterized by an infnitesimal, global and anticommuting parameter leaves the efective action as well as path integral of the effective theory invariant. The BRST qunatisation for  usual first order theories was first introduced in \cite{becchi}. A series of works appeared thereafter in this direction \cite{brst}. The  finite field dependent BRST (FFBRST) which considers the gauge parameters being finite and dependent on the fields  have found many applications in various contexts.  In gauge field theories the usual BRST symmetry has been generalized to make it finite and field dependent.
	The implementation of BRST symmetries for higher derivative theories is quite nontrivial and poses problems. In this context, therefore, a natural question arises regarding the application of BRST formalism to
relativistic particle models. Indeed it is not surprising that in spite of a considerable volume of research on relativistic particle models, this aspect remains unstudied. 

	We consider the relavistic particle model with curvature since it possesses gauge symmetry \cite{banerjee_brst}. We construct the (anti-)BRST for the massive and massless models. The difficulties of applying BRST transformations to higher derivative theories are bypassed by working in the first order formalism developed in \cite{BMP}  instead of the conventional Ostrogradsky approach.
	
	It is shown that the (anti-)BRST symmetry
transformations for all the variables reproduce the diffeomorphism symmetry of the massive relativistic particle model including curvature. Furthermore, we also show that the
massless particle model with rigidity yields both the diffeomorphism and $W$ invariances. We explicitly demonstrate the $W_3$ algebra. For BRST transformations this algebra is
shown for all variables, excluding the antighost. Exactly the same features are revealed, but now excluding the ghost
variable instead of the antighost, for anti-BRST transformations. To get the complete picture, therefore, both BRST and
anti-BRST transformations have to be considered. Further, we implement the concept of FFBRST transformation  in the quantum mechanical relativistic particle model. The quantum mechanical version of FFBRST transformation \cite{brst} is called as finite coordinate-dependent BRST (FCBRST) transformation. We see that FCBRST transformation for the relativistic particle model is a symmetry of the action only, but not of the generating functional. Analogous to FFBRST the FCBRST transformation changes the Jacobian of path integral measure nontrivially. For an appropriate choice of finite coordinate dependent parameter FCBRST connects two different gauge-fixed
actions within functional integration.	
	
	So far we have discussed the general method of abstracting gauge symmetries  for the higher derivative theories  for particle models only. A transition to field theories bring novel features even in the first derivative systems. It is natural to ask how our method in \cite{BMP} works in the field theoretic models.  Higher derivative theories also found their place in the field of electrodynamics long ago by Podolsky \cite{podolsky}. The usual Maxwell-Chern-Simon's(MCS) theory is a first order one which also violates parity conservation principle. The  higher derivative extension of this MCS theory known as the extended Maxwell-Chern-Simon's(EMCS) theory was first introduced in \cite{deser2}. Subsequently its Hamiltonian analysis in the purterbative approach was done by \cite{reyes1}. The higher derivative version of Chern-Simons model was also considered in \cite{sarmishtha}. We take the  higher derivative model to analyse it in the first order approach \cite{mukherjee_exmcs}. Using the technique in \cite{BRR} it is shown that the number of independent gauge transformation is one. This independent gauge transformation is identified as the $U(1)$ symmetry associated with the vector fields of the usual Maxwell theory.

	The next higher derivative model we are going to consider is a membrane theory. In 1962, Dirac considered the electron to be a charged rigid membrane in \cite{dirac_membrane}.  Based on this concept, a model was proposed in \cite{onder} by adding the extrinsic curvature of the world volume swept out by the membrane. Theories with extrinsic curvatures are frequently studied, but recent inclusion of these terms in some physically interesting models added an extra urgency to revisit the symmetry features of such theories. It is shown that because of the extrinsic curvature effects, there appears geometrical frustration when nematic liquid crystals are constrained to a curved surface \cite{napoli}. Hamiltonian analysis of the membrane model of  electron   in the Ostrogradski formalism was done by Cordero et al. \cite{cordero_qcrm}. A series of work  like domain wall formation \cite{christensen}, relativistic membranes \cite{hope}, spiky membranes \cite{trzetrzelewski} and others\cite{gnadig} come into existence using this concept of membrane model.  The geometry of these deformed relativistic membranes were also discussed in \cite{capovilla}. Even the extended version of this electron model was used in general relativity  in which this particle is described as a source of the Kerr-Newman field \cite{lopez} and Reissner-Nordstrom geometry \cite{lopez}. Stability analysis of the model was studied in \cite{lopez}. We take the model in \cite{cordero_qcrm} where the Lagrangian is expressed with respect to the local coordinates in the world volume. The model is higher derivative, although a surface term can be identified. We convert this higher derivative Lagrangian into a first order form and perform the Hamiltonian analysis \cite{paul_membrane}. It is shown that the model has only one gauge symmetry viz. the reparametrization invariance. The equation of motions for  the higher derivative model matches with the first order  model both at the Lagrangian and Hamiltonian levels.

	After discussing the particle, field theoretic and membrane models  we now take a model from gravity theory to show the efficacy of this first order formalism. We know,  usually gravity is described in 4 dimensionsional pseudo Riemannian space-time. In 1977, Regge and Teitelboim introduced a model where gravity is considered to be  induced on a hypersurface embedded in a higher dimensional Minkowski space \cite{regge1}. Soon, the model was able to draw  attention  of the physics community as it is a manifestation of gravity from the string theoretic view.  The model was reviewed at various stages by many authors \cite{rt_model, davidson2, davidson3}. Unlike gravity theories where metric is considered to be an independent variable, here  the functions of the spacetime embeddings into a flat space are taken as independent variables. The canonical theory of  gravity can be developed with respect to these induced variables. The configuration space in this case is infinite dimensional, called the superspace. In  finite dimensions, the number of degrees of freedom gets restricted and the model correspondingly is described in the minisuperspace.
	
	 The Einstein Hilbert action with cosmological constant is considered to develop the miniuperspace version and apply the first order formalism \cite{cordero_rt, banerjee_rt}.  As in usual Einstein-Hilbert action a surface term can be isolated here also in this minisuperspace version. The action without the surface term was considered in \cite{rt_model, davidson2, davidson3} and Wheeler DeWitt equations \cite{WDW} were constructed. Here, on the contrary, we keep the surface term since it has relation to entropy and develop a Hamiltonian formulation. The model is a higher derivative one and we adapt the first order formalism. In this formalism we find out the gauge symmetry viz. the reparametrisation symmetry inherent in the system. The number of PFC is found out to be one. Final Dirac brackets are constructed by removing the pair of first class constraints by imposing appropriate gauge conditions. One of the gauge conditions is the usual cosmic gauge while the other one is newly proposed.  Under these gauge conditions, the existing pair of  first class constraints become second class which can be removed via the Dirac brackets. The final canonical pair is uplifted to the level of commutator. In quantum cosmology, Wheeler DeWitt equation is  analogous to the Schrodinger equation \cite{WDW}. So, to describe the quantum picture of the universe, the WDW equation is constructed and a subsequent match with existing literature is shown.

\section{Outline of the thesis}
  The thesis  starts with a review of Higher derivative theories and their canonical quantisation both in the Ostrogradski and the first order formalism. Thereafter, we will explicitly workout Hamiltonian formulation of various models e.g. the relativistic particle model with curvature \cite{BMP, banerjee_brst}, the extended Maxwell-Chern-Simons model \cite{mukherjee_exmcs}, the quantum charged rigid membrane \cite{paul_membrane}, the Regge-Teitelboim model of cosmology \cite{banerjee_rt} to show the effectiveness of this first order formalism. Also, while discussing these models we will  come across various features of these models which automatically emerge out during the course of Hamiltonian formulation.
  	\itemize
  	\item \textit{Chapter 2} discusses canonical quantisation of the constrained and unconstrained systems. This chapter  briefly describes how to extract the gauge symmetries of a first class system.  Also, transition to quantum picture is described formally. In this chapter, higher derivative systems are introduced and their canonical quantisation procedure in the Ostrogradski formalism is briefly described. Also,  first order formalism is described.
  	\item \textit{Chapter 3} considers the relativistic particle model with curvature. Both its massive and massless versions are taken to abstract the symmetries. It is shown that there appears a mismatch in the number of independent primary first class constraints and number of independent gauge symmetries. The problem is solved by taking the first order approach and additionally considering a condition which appears only due to the higher derivative nature of the theory. The massive version shows the difeomorphism while the massless version has $W_3$ symmetry along with the former one. The BRST quantisaton for both the models is also discussed.  
  	\item \textit{Chapter 4} takes the extended Maxwell-Chern-Simon theory which is a higher derivative field theoretic model. The gauge symmetry of this model is worked out. An exact mapping between the Hamiltonian gauge transformations and the $U(1)$ symmetries of the action has been established 
  	\item \textit{Chapter 5} discusses the quantum  charged rigid membrane model. The reparametrisation symmetry in the first order formalism is shown to be the symmetry of the action. The equations of motion for the higher derivative  and first order system  are shown to be equivalent to establish the validity of the first order approach.  
  	\item \textit{Chapter 6} includes the Regge Teitelboim model of cosmology. Keeping the surface term, the model is analysed in the first order approach. The reparametrisation invariances are abstracted. By choosing appropriate gauge conditions the model is formally quantised in the reduced phase space. The Wheeler DeWitt equation is constructed and matching with the existing literature is shown for this model.
  	
  	\item  \textit{Chapter 7} contains the conclusions.
  	\chapter{Canonical quantisation}
  	Information of any dynamical system can be described by real functions of time $q^i(t)$ and it's  time derivatives $\dot{q}^{i}(t)$. Using this collection which forms the configuration space, the state of any  dynamical system can be specified. The starting of our discussion should be the action principle which led to the development of the Lagrangian and Hamiltonian formulation of classical mechanics. Using the action principle, we get the equations of motion commonly known as Euler-Lagrange equations of motion. Consider the action as function of $q(t)$ with fixed end points $q_1 (t)$ and $q_2(t)$ 
  	\begin{equation}
  	S[q; t_1, t_2]=\int_{t_1}^{t_2} L(q^i,\dot{q}^i, t) dt.
\end{equation}

According to action principle, \textit{the trajectory is the extremum of this action}.  The $L(q, \dot{q})$ is a real function of time, called the Lagrangian. The extremum condition says
 
\begin{equation}
\frac{\delta{S}}{\delta{q^i(t)}} = 0.  \ \ \ \ \ \  \delta{q(t_1)} = \delta{q(t_1)} =0.
\label{boundary_cond}
\end{equation} 	

Consequently the equations of motion  can be written as 
\begin{equation}
\frac{d}{dt} \frac{\partial{L}}{\partial{\dot{q}^{i}(t)}} - \frac{\partial{L}}{\partial{q^i}} =0. \label{eulerlag_1_first order}
\end{equation}  
 These second order differential equations are  known as Euler-Lagrange(EL) equations. However, these differential equations can  be of higher order for the Lagrangians containing higher time derivatives of $q$'s. This entire chapter is devoted for the first order systems.
 
   One can convert these second order equations to first order by shifting from the configuration space ($q^i, \dot{q}^i$) to phase space ($q^i, p_i$). These $p_i$ are known as momenta and defined by
 \begin{equation}
 p_i = \frac{\partial{L}}{\partial{\dot{q}^i}} \label{momenta_def}
\end{equation}  
 Sometimes, all or some of the momenta are not expressible with respect to velocities. These equations lead to constraints. Constraints are some functions of the phase space and need special care during the process of quantisation. Below we describe quantisation of unconstrained and constrained systems separately.

  	\section{Unconstrained systems}
  	To see the evolution of the system,  we define a function using the phase space variables  called Hamiltonian  via the Legendre transformation as
  	\begin{equation}
  	H(q^i, p_i)=p_i \dot{q}^i - L(q, \dot{q}).
  	\label{hamiltonian_def}
  	\end{equation}
  	Using the same boundary conditions as in (\ref{boundary_cond}) we get the Hamilton's equations of motion
  	\begin{equation}
  	\dot{p}_i= \frac{\partial{H}}{\partial{\dot{q}^i}}; \ \ \ \ \ \dot{q}_i= \frac{\partial{H}}{\partial{\dot{p}^i}}
  	\label{hamiltonian_eom_def}
  	\end{equation}
  	These Hamilton's equations can also be easily obtained using Poisson brackets. For two functions $A(q,p)$ and $B(q,p)$ of the phase space variables the Poisson bracket(PB) is a bilinear operation defined as (see appendix for properties of Poisson Brackets)
  	\begin{equation}
  	\{A(q_i,p^i),B(q_i,p^i)\} = \frac{\partial{A(q_i,p^i)}}{\partial{q_j}} \frac{\partial{B(q_i,p^i)}}{\partial{p^j}} - \frac{\partial{A(q_i,p^i)}}{\partial{p^j}} \frac{\partial{B(q_i,p^i)}}{\partial{q_j}}.
  	\label{PB_def}
\end{equation} 
With these the PBs for the basic variables turn out to be 
\begin{equation}
\{q^i,p_j\} = \delta^i_j.
\end{equation}	 
 The Hamilton's EOM now become as
 \begin{equation}
 \dot{q}^i = \{q^i, H \}; \ \ \ \ \ \dot{p}_i = \{p_i,H \} \label{ham_eom_def}
\end{equation}  
We can move to the quantum picture of the system where the Hilbert space is spanned by these canonical variables$(q^i, p_i)$ lifted to the operators $(\hat{q}^i, \hat{p}_i)$ and the PBs are converted to commutators via the following relation
\begin{equation}
 \{ . , . \} \rightarrow \frac{1}{i \hbar}[. , .] 
 \label{quantisation_def}
\end{equation}
All the dynamical variables in the phase space are the operators in the chosen Hilbert space. This method of quantisation is known as canonical quantisation and was introduced by P. A. M. Dirac \cite{dirac2}.

  	\section{Constrained systems}
  	From (\ref{eulerlag_1_first order}) we can write 
  	\begin{equation}
  	\ddot{q}^i \frac{\partial^{2}{L}}{\partial{\dot{q}^{i}} \partial{\dot{q}^{j}}} =\frac{\partial{L}}{\partial{q^j}} - \dot{q}^i \frac {\partial^2{L}} {\partial{q^i} \partial{\dot{q}^j}}
  	\end{equation}
  	From this equation we see that if the determinant of the matrix $\frac{\partial^{2}{L}}{\partial{\dot{q}^{i}} \partial{\dot{q}^{j}}}$ is zero the accelerations $\ddot{q}^i$ are not uniquely determined which signals that all the momenta in (\ref{momenta_def}) are not expressible with respect to the velocities. This means the system is constrained one.  
  	For constrained systems the method of canonical quantisation is more complicated. We write the constraint equations in a general form as
  	\begin{equation}
  	\phi_m(q,p)\approx 0. \ \ \ \  (m = 1, .... , M)
  	\end{equation}
  	We got these equations from the definitions of the momenta itself. These constraints are known as \textit{primary constraints}.  The sign $\approx$  refers to the \textit{weak condition} which means that they should be put to zero only after evaluation of the PBs.  We can add constraints to the Hamiltonian (\ref{hamiltonian_def}) and correspondingly see the evolution. Therefore, the definition (\ref{hamiltonian_def}) is not unique for the constraint systems. For a constraint system it is better to use the total Hamiltonian defined by
\begin{equation}
H_T = H_{can} + \lambda ^m \phi_m.
\label{H_T def}
\end{equation}
Here $H_{can}$ is known as the canonical Hamiltonian defined by (\ref{hamiltonian_def}) and $\lambda^m$ are the Lagrange multipliers enforcing the primary constraints.  	

  	 Once we get the primary constraints we can see the time variation of the constraints as
  	 \begin{equation}
  	 \psi_m = \{\phi_m, H_T\}.
\end{equation}  	  
 These $\psi_m$ can be zero identically  or be linear combination of the other constraints. Another possibility can arise where they are some other functions of the phase space variables. Demanding this  to be zero we get the secondary constraints. In this way preserving the secondary constraints in time we can get the tertiary constraints. At some point, the chain terminates and we have  no more generation of new constraints. Once we have the whole list of the constraints, another important classification can be done  based on their PB structure. In this case they are defined as either first class or second class.
 \begin{itemize} 
 \item {\textit{First class} constraints are those which have a closed Poisson algebra. Effectively this implies that they have weakly vanishing PB among themselves.  First class constraints are the generators of the gauge transformation.}
 \item {\textit{Second class} constraints are those which do not fall in the above class.}
 \end{itemize}
  This classification is very important for Hamiltonian formulation. It is to be mentioned that the number of independent primary first class constraints are equal to the number on independent gauge symmetries. The existence of the first class constraints in the system means there are some redundant degrees of freedom. To quantise the theory, in the reduced phase space, we have to remove all the constraints. First we remove the second class constraints by defining the Dirac brackets(DB) analogous to Poisson brackets as
   \begin{equation}
   \{A,B\}_{D}=\{A,B\}-\{A,\chi_i\}\Delta_{ij}^{-1}\{\chi_j,B\}.
   \label{DB_def}
   \end{equation}
   Here $\chi_i$ are the set of all second class constraints and $\Delta_{ij}=\{\chi_i,\chi_j\}$. All the second class constraints can be put strongly to zero and in the quantum case they  are treated as operator identities. Also,  hereafter, whenever the PB appears they should be replaced by DB defined in (\ref{DB_def}).  With the first class constraints remaining in the system we can proceed to abstract the gauge symmetries by defining the gauge generator as the linear combination of all the first class constraints ($\varphi_a$) written as
   \begin{equation}
   \mathcal{G}=\epsilon^a \varphi_a.
   \label{gauge_gen_def}
   \end{equation}
 Here $\epsilon^a$ are known as the gauge parameters. All the gauge parameters in the system may or may not be independent. The independent gauge parameters actually signifies the  gauge symmetries in the system. Following the algorithm of \cite{BRR} we can express the dependent gauge parameters in terms of the independent set using the conditions
\begin{equation}
\delta\lambda^{a_1} = \frac{d\epsilon^{a_1}}{dt} -\epsilon^{a}\left(V_{a}^{a_1}
 +\lambda^{b_1}C_{b_1a}^{a_1}\right)
                             \label{master_eqn_1}
\end{equation}
\begin{equation}
  0 = \frac{d\epsilon^{a_2}}{dt}
 -\epsilon^{a}\left(V_{a}^{a_2}
+\lambda^{b_1}C_{b_1a}^{a_2}\right)
\label{master_eqn_2}
\end{equation}
The indices $a_1, b_1 ...$ refer to the primary first class constraints while the indices $a_2, b_2 ...$ correspond to the secondary first class constraints. Indices without any suffix like $a,b....$ comprise both the primary and secondary set. 
The coefficients $V_{a}^{a_{1}}$ and $C_{b_1a}^{a_1}$ are the structure
functions of the involutive algebra, defined as
\begin{eqnarray}
\{H_c,\varphi_{a}\} = V_{a}^b\varphi_{b} \label{structure constant_V} \\
\{\varphi_{a},\varphi_{b}\} = C_{ab}^{c}\varphi_{c}
\label{structure constant_C}
\end{eqnarray}
and $\lambda^{a_1}$ are the Lagrange multipliers(associated with the primary first class constraints)  appearing in the expression of the total Hamiltonian (\ref{H_T def}). Solving (\ref{master_eqn_1}) it is possible to choose $a_1$ independent
gauge parameters from the set $\epsilon^{a}$ and express $\mathcal{G}$ of
(\ref{gauge_gen_def}) entirely in terms of them. Once we get the independent gauge parameters ($\epsilon^{a^{\prime}}$), we can find out the gauge transformation of the fields as
\begin{equation}
\delta q^i =  \epsilon^{a} \{q^i, \varphi_{a}\}
\end{equation}
   	Now we have to get rid of these first class constraints $\varphi_a$. This can be done by introducing gauge conditions. Gauge conditions are some ad hoc relations which make the first class constraints second class. Once the first class constraints become second class they can be removed by Dirac brackets. In the reduced phase space we can  identify the canonical pairs and proceed for quantisation by promoting these BDs as commutators as in (\ref{quantisation_def}). This completes the canonical quantisation for constraint systems.

\section{Higher derivative theories}
 By higher derivative theories we specifically refer to those theories where  higher time derivative of the fields appear.  We begin with a general higher derivative theory given by the Lagrangian
\begin{equation}
L = L\left(x, \dot{x}, \ddot{x}, \cdots , x^{\left(\nu\right)}\right)
\label{originallagrangean}
\end{equation}
where $x = x_n (n = 1,2,\cdots,\nu)$ are the coordinates and $\dot{}$ means derivative with respect to time. $\nu$-th order derivative of time is denoted by $x^{\left(\nu\right)}$. Below we discuss the  canonical formulation of the higher derivative theories in the Ostrogradski and  first order formalisms. 

\subsection{Ostrogradski formalism}
In the Ostrogradski method \cite{ostro}, for the higher derivative Lagrangian (\ref{originallagrangean}), the Euler Lagrange EOM is written as
\begin{equation}
\sum_{l=0}^{\nu} (-1)^l \frac{d^l}{dt^l}(\frac{\partial{L\left(x, \dot{x}, \ddot{x}, \cdots , x^{\left(\nu\right)}\right)}}{\partial{x_{n, l}}}) =0
\end{equation}
In this approach, all the fields along with their corresponding higher derivatives are  considered as independent fields. Correspondingly, two types of momenta are defined. To begin the  Hamiltonian formulation, the momenta are defined as,
\begin{eqnarray}
P_{i} &=& \frac{\partial{L}}{\partial{q}^{(m)}_{i}}, 
\nonumber \\
 p_{i,m-1} &=&  \frac{\partial{L}}{\partial{q^{(m)}_{i}}} - \frac{d}{dt} \left(   \frac{\partial{L}}{\partial{q}_{i}^{m+1}} \right) ;  \ \ \  m= 1,2 ... n-1 
\end{eqnarray}
Here the canonical pairs are $(q_i, P_{i,m-1})$ and $(q^{m}_{i}, p_{i,m})$ and thus we get the Hamiltonian as  
 \begin{equation}
 H = \sum_{m=1}^{n-1} p_{i,m} q^{(m)} + P_i\dot{q}^{(n-1)}_i - L .
 \end{equation}
Once the Hamiltonian is defined the Hamiltonian equations of motion can be found in the usual way as (\ref{ham_eom_def}).
%\begin{equation}
%\dot{q}_{i} = \{ q_{i},H \} \ \ \ \ q^{(m)}_{i} = \{ q^{(m)}_{i},H \}.
%\end{equation}  
%In the next subsection we develop another Hamiltonian formalism to deal with the higher derivative theories. 
\subsection{The first order formalism}
The Hamiltonian formulation of the theory may be conveniently done by a variant of Ostrogradskii method commonly known as the first order formalism. The crux of the method consists in embedding  the original higher derivative theory to an effective first order theory. We define the variables $q_{n,\alpha} \left(\alpha = 1, 2, ...., \nu - 1 \right)$ as
\begin{eqnarray}
q_{n,1}   &=& x_n\nonumber\\
q_{n,\alpha} &=& \dot{q}_{n,\alpha -1}, \left(\alpha > 1 \right)
\label{newvariables}
\end{eqnarray}
This leads to the following Lagrangian constraints
\begin{eqnarray}
q_{n,\alpha} - \dot{q}_{n,\alpha -1} = 0, \left(\alpha > 1 \right)
\label{lagrangeanconstraints}
\end{eqnarray}
which must be enforced by corresponding Lagrange multipliers .   
The auxiliary Lagrange function of this extended description
of the system is given by
\begin{eqnarray}
\nonumber
L^*(q_{n,\alpha},\dot{q}_{n,\alpha},\lambda_{n,\beta})
&=& L\left(q_{n,1},q_{n,2}\cdots,q_{n,\nu-1},
\dot{q}_{n,\nu-1}\right)+ \\  &&  \sum_{\beta=2}^{\nu-1}
\left(q_{n,\beta}-\dot{q}_{n,\beta-1}\right)\lambda_{n,\beta}\ ,
\label{extendedlagrangean}
\end{eqnarray}
where $\lambda_{n,\beta} (\beta = 2,\cdots , \nu - 1)$ are the Lagrange multipliers. If we consider these multipliers as independent fields then the Lagrangian $L^*$ becomes first order to which the well known methods of Hamiltonian analysis for  first order systems apply.
The momenta canonically
conjugate to the degrees of freedom $q_{n,\alpha}$,
$(\alpha=1,2,\cdots,\nu-1)$ and
$\lambda_{n,\beta}$ $(\beta = 2, \cdots,\nu-1)$
are defined, respectively, by,
\begin{equation}
p_{n,\alpha}=\frac{\partial L^{*}}{\partial \dot{q}_{n,\alpha}}\ ,\ \
\pi_{n,\beta}=\frac{\partial L^{*}}{\partial\dot{\lambda}_{n,\beta}}\ .
\end{equation}
These immediately lead at least to the following 
primary constraints,
\begin{equation}
\Phi_{n,\beta} \approx 0\ ,
\ \ \pi_{n,\beta} \approx 0\ ,\ \
\beta = 2,\cdots,\nu -1\ ,
\label{constraints}
\end{equation}
where
\begin{equation}
\Phi_{n,\beta}\equiv p_{n,\beta -1}+\lambda_{n,\beta}\ , \ \
\beta = 2,\cdots,\nu - 1\ .
\end{equation}
Note that depending on the situation whether the original Lagrangian $L$ is singular there may be more primary constraints.\\ 
Let us first assume  that the original Lagrangian $L$ is regular. Then there are no more primary constraints. Now
the basic non-trivial Poisson brackets are
\begin{equation}
\left\{q_{n,\alpha},p_{m,\alpha '}\right\}=\delta_{nm}\delta_{\alpha \alpha '}\ ,\ \
\left\{\lambda_{n,\beta},\pi_{m,\beta '}\right\}=\delta_{nm}\delta_{\beta \beta '}
\ ,
\label{basicpbs}
\end{equation}
Consequently the primary constraints obey the algebra 
\begin{equation}
\left\{\Phi_{n,\beta},\Phi_{m,\beta '}\right\}=0\ ,\ \
\left\{\Phi_{n,\beta},\pi_{m,\beta '}\right\}=
\delta_{nm}\delta_{\beta \beta '}\ ,\ \
\left\{\pi_{n,\beta},\pi_{m,\beta '}\right\}=0\ ,
\label{algebraconstraints}
\end{equation}
implying that $\pi_{n,\beta}$ and $\Phi_{n,\beta}$ are second class constraints. Now, the 
ca\-no\-ni\-cal Hamiltonian of the modified system(\ref{extendedlagrangean}) can be written according to the usual prescription as,
\begin{equation}
H_C=\sum_n\sum_{\alpha=1}^{\nu-1}\dot{q}_{n,\alpha}p_{n,\alpha}
+\sum_n\sum_{\beta=2}^{\nu-1}\dot{\lambda}_{n,\beta}\pi_{n,\beta}-L^*\ ,
\end{equation}
We define the total Hamiltonian ($H_T$) by adding linear combinations of the primary constraints (\ref{constraints}), 
\begin{equation}
H_T = H_C + u_{n,\beta}\pi_{n,\beta} + v_{n,\beta}\Phi_{n,\beta}
\end{equation}
where $ u_{n,\beta}$ and $v_{n,\beta}$ are Lagrange multipliers. From (\ref{algebraconstraints}) we found that the constraints are second class. Thus preserving the primary constraints in time we will be able to fix the multipliers $ u_{n,\beta}$ and $v_{n,\beta}$. Fixing these and after some simplifications we find that
\begin{equation}
H_T\left(q_{n,\alpha},p_{n,\alpha};\lambda_{n,\beta}\right)=
\overline{H}_0\left(q_{n,\alpha},p_{n,\nu-1}\right)
-\sum_n\sum_{\beta=2}^{\nu-1}\lambda_{n,\beta}q_{n,\beta}\ .
\label{eq:concanH0}
\end{equation}
\begin{equation}
\overline{H}_0(q_{n,\alpha},p_{n,\nu-1})=
\sum_n\dot{q}_{n,\nu-1}p_{n,\nu-1}-
L\left(q_{n,\alpha},\dot{q}_{n,\nu-1}\right)\ .
\label{eq:H0restricted}
\end{equation}
The presence of the  second class constraints (\ref{constraints}) imply that the phase space degrees of freedom
$(q_{n,\alpha},p_{n,\alpha};\lambda_{n,\beta},\pi_{n,\beta})$ are
not all independent. If we replace the Poisson brackets by Dirac brackets the constraints (\ref{constraints}) can be strongly implemented. This enables us to eliminate the nondynamical sector $(\lambda_{n,\beta},\pi_{n,\beta})$ from the phase space variables. It is to be noted that the no. of second class constraints equals the no. of nondynamical variables eliminated from the phase space. Straightforward calculations show that the DBs between the remaining phase space variables are the same as the corresponding PBs. The total Hamiltonian now becomes
\begin{equation}
H_T\left(q_{n,\alpha},p_{n,\alpha};\lambda_{n,\beta}\right)=
\overline{H}_0\left(q_{n,\alpha},p_{n,\nu-1}\right)
+ \sum_n\sum_{\beta = 2}^{\nu-1}p_{n,\beta - 1}q_{n,\beta}\ .
\label{eq:canH0}
\end{equation}
Since the original Lagrangian system is replaced by the first order theory (\ref{extendedlagrangean}) the algorithm of \cite{BRR} can be readily applied. For the conventional first order theories this completes the picture. 
The situation for higher order theories is, however, different.
 %However unlike the first order theories this set of parameters will not be mutually independent in general.
%For the higher order theories the number of independent gauge parameters will be even less.
 This is because of the new constraints (\ref{lagrangeanconstraints}) appearing in the effective first order lagrangean (\ref{extendedlagrangean}). Owing to these  we additionally require
\begin{eqnarray}
\delta q_{n,\alpha} - \frac{d}{dt}\delta{q}_{n,\alpha -1} = 0, \left(\alpha > 1 \right)
\label{varsgauge}
\end{eqnarray}
%where $\delta A$ is the gauge variation of the phase space variable $A$.
%obtained from (\ref{lagrangeanconstraints}).
% by using (\ref{commutativity1}).
 These conditions may  reduce the number of independent gauge parameters further. Thus the number of independent gauge parameters is, in general, less than the number of independent primary first class constraints. Whereas, in usual theories the number of independent primary first class constraints is always equal to the number of independent primary first class constraints. This additional condition can play a remarkable role in determining the actual number of gauge symmetries present in the higher derivative systems making  the difference from a genuine first order theory. As is evident here for first order theory the equation is not meaningful. While considering the algorithm for abstracting the gauge symmetry we must consider this relation to get the actual realisation of the theory.
 % The conversion of the original higher derivative theory to an effective first order theory is thus a formal one containing peculiarities characteristic of higher derivative nature. 
 
 The effective first order theory, obtained from the original higher order theory, therefore contains peculiarities. In this thesis we will illustrate this peculiarity using different examples.

%Before going to the applications of the general formalism a few words about the consistency of the formalism will be appropriate. 
%Fortunately, in the algorithm of \cite{BRR, BRR1} there is a provision of checking the consistency of calculation. 
%The gauge vatriation of the Lagrange multipliers satisfy
%\begin{equation}
%\delta\lambda^{a_1} = \frac{d\epsilon^{a_1}}{dt}
%                 -\epsilon^{a}\left(V_{a}^{a_1}
%                 +\lambda^{b_1}C_{b_1a}^{a_1}\right)
%                             \label{218}
%\end{equation}
%It can be shown that the condition (\ref{218}) actually follows from (\ref{219}) \cite{BRR,BRR1}. Also the same gauge variations can be directly obtained as $$\delta\lambda^{a_1} = \{\lambda^{a_1}, G\}$$ where we assume that the generator $G$ is expressed in terms of the independent gauge parameters. If $\delta\lambda^{a_1}$ The agreement of both the expressions of $\delta\lambda^{a_1}$ will provide a nontrivial consistency check of our analysis.
  \section{Discussions}	
   	In this chapter we have shown how to canonically quantise a theory from the given action. If the theory is constrained one, then it needs special attention. To handle these theories the well known Dirac's procedure has been explained. The gauge generator is constructed from the first class constraints. It has been shown how to extract the independent gauge symmetries using the technique developed in \cite{BRR}. For the higher derivative theories, the situation however, is different where we depict the Ostrogradski formalism as well as the first order formalism. The first order formalism, which we are going to use throughout the thesis, is explained elaborately,  even in the presence of constraints. The usual momenta definition  is used to develop the Hamiltonian formulation for the first order approach.   
   	
                      \chapter {Relativistic particle model with curvature} %DONE
  	In the last chapter we  have discussed how to perform canonical formulations for the higher derivative theories. For that purpose we have developed a systematic algorithm to extract the independent gauge degrees of freedom. Our  method is an extension of similar technique for the usual first order theories \cite{BRR}. In the rest of the thesis we will illustrate our method for different particle and field theoretic models. In this chapter, we consider a particle model, known as the relativistic particle model with curvature, which is higher derivative in nature.
  	  	 The action of the relativistic particle model is defined by the arc length of the worldline \cite{polyakov}. They are important in physics in their own right and also behave as toy models to test various ideas which are useful for more general systems such as string and membranes. One can add the extrinsic curvature of the world line to this relativistic particle model which is generally known as the relativistic particle model with curvature \cite{nesterenko, plyuschay1, plyuschay2, BMP}.
  	The massive relativistic particle model with curvature\cite{pisarski}  has the action \footnote{contractions are abbreviated as $A^\mu B_\mu = AB$, $A^\mu A_\mu=A^2$. We consider the model in 3 + 1 dimensions. So $\mu$ assumes the values 0, 1, 2, 3. Also, the model is meaningful for  $\alpha < 0 $ \cite{plyuschay2}} 
\begin{equation}
S=-m \int{\sqrt{\dot{x}^{2}}}d\tau+\alpha\int{\frac{\left( \left( \dot{x}\ddot{x}\right)^{2}-\dot{x}^{2}\ddot{x}^{2} \right)^\frac{1}{2}}{\dot{x}^{2}}}d\tau\label{maction}
\end{equation} 

The Hamiltonian analysis in the Ostrogradski formalism shows that we have two independent primary first class constraints. Whereas, there is only one independent gauge degrees of freedom. Clearly, it is a mismatch with the existing literature (on usual non higher derivative theories) which says  that the number of independent gauge symmetries are always equal to the number of independent primary first class constraints. This paradox will  be solved in the following sections. Also, the gauge generator with the appropriate number of independent gauge parameters will be constructed. 
   	 
  	The action of the massless relativistic particle model can be obtained from (\ref{maction}) by putting $m=0$. This model has been investigated by many authors \cite{plyuschay2, ramos1, ramos2}. It has five first class constraints, out of which two are primary. There is no second class constraint. So the gauge generator is constructed with five gauge parameters. The method of \cite{BRR} shows that only two of them are independent. Also, there are two independent primary first class constraints. So, the case of  mismatch in the number of independent PFCs and the independent gauge transformations does not occur here. Thus in the Ostrogradski approach, the no. of independent PFCs may or may not be equal to the no. of independent  gauge degrees of freedom. In the following we will demonstrate this apparent mechanism of this arbitrariness. The massless relativistic particle model with curvature has a very interesting aspect. One of it's two gauge symmetries may easily be identified as the usual difeomorphism symmetry. The second symmetry was analysed only at the equations of motion level by introducing the Lax operators. It was demonstrated that his symmetry is nothing but $W_3$ symmetry \cite{ramos1}. In the following we will show that our method is capable of extracting both these symmetries from the action and $W_3$ algebra emerges systematically. 
  	
 We will discuss the BRST symmetries of this mmassive and massless relativistic particle model with curvature. The BRST technique is a very powerful tool to quantise a theory with gauge invariance. The BRST symmetry transformations are characterized by an infinitesimal, global and anticommuting parameter that leaves the effective action as well as the path integral of the effective theory invariant.  The BRST symmetries for higher derivative theories are still unexplored and  quite nontrivial. Since the massive as well as the massless relativistic particle model with curvature poses gauge symmetries, their BRST study is necessary. The (anti-)BRST for the above models are constructed in \cite{banerjee_brst}. For the massive relativistic particle model the (anti-)BRST transformations for all variables including ghost and anti-ghost exactly reproduce the diffeomorphism symmetry. The BRST transformation for the massless version also reproduce the $W_3$-algebra but they exclude the anti-ghost variables. Whereas, the anti-BRST transformation in this case exclude the ghost fields. Other than this,  the finite field dependent BRST is also considered for both these models. The quantum mechanical version which is known as the finite coordinate dependent BRST is a symmetry of the action only but not of the generating functional for this relativistic particle model.

  	%The Hamiltonian formulation is developed with the aim of constructing the gauge generator and abstracting the symmetries of the model. In the course of this analysis we show that the nature of the constraints is non-trivially modified at the critical point $p_{1}^{2} = m^{2}$ that corresponds to the standard dispersion relation. Hence the analysis at this point is separately given. 

%The massive relativistic point particle theory with rigidity \cite{pisarski}  has the action\footnote{contractions are abbreviated as $A^\mu B_\mu = AB$, $A^\mu A_\mu=A^2$. We consider the model in 3 + 1 dimensions. So $\mu$ assumes the values 0, 1, 2, 3. Also, the model is meaningful for  $\alpha < 0 $ \cite{plyuschay2}} 
%\begin{equation}
%S=-m \int{\sqrt{\dot{x}^{2}}}d\tau+\alpha\int{\frac{\left( \left( \dot{x}\ddot{x}\right)^{2}-\dot{x}^{2}\ddot{x}^{2} \right)^\frac{1}{2}}{\dot{x}^{2}}}d\tau\label{maction}
%\end{equation}
\section{Hamiltonian analysis and construction of the gauge generator}
To convert the Lagrangian (\ref{maction}) into a first order one we  introduce the new coordinates
\begin{equation}
q^{\mu}_1 = x^{\mu}\ \;\ \ q^{\mu}_2 = \dot{x}^{\mu}
\label{mnewcoordinate}
\end{equation}
The Lagrangian in these coordinates has a first order form given by \cite{plyuschay1}
\begin{eqnarray}
% \nonumber
%L_{0} &=& -m\sqrt{q_{2}^{2}} + \alpha \frac{\left({\left({q_{2}\dot{q}_{2}} \right)^{2} - q_{2}^{2} \dot{q}_{2}^{2} } \right)^{\frac{1}{2}} }{q_{2}^{2}}
%\nonumber \\
L &=& -m\sqrt{q_{2}^{2}} + \alpha \frac{\left({\left({q_{2}\dot{q}_{2}} \right)^{2} - q_{2}^{2} \dot{q}_{2}^{2} } \right)^{\frac{1}{2}} }{q_{2}^{2}} + q_{0}^{\mu}(\dot{q}_{1\mu} - q_{2\mu})
\label{mindependentfields}
\end{eqnarray}
where $q_{0}^{\mu}$ are the Lagrange multipliers that enforce the constraints
\begin{equation}
\dot{q}_{1\mu} - q_{2\mu} = 0
\label{27}
\end{equation}
Let $p_{0\mu}$, $p_{1\mu}$ and $p_{2\mu}$ be the canonical momenta conjugate to $q_{0\mu}$, $q_{1\mu}$ and $q_{2\mu}$ respectively. 
Then 
we immediately get the following
primary constraints
\begin{eqnarray}
%\nonumber
\Phi_{0\mu} = p_{0\mu} \approx 0 ;
\Phi_{1\mu} = p_{1\mu} - q_{0\mu} \approx 0
\label{mPFC1}
\end{eqnarray}
and
\begin{eqnarray}
%\nonumber
\Phi_{1} = p_{2}q_{2} \approx 0;
\Phi_{2} = p_{2}^{2}q_{2}^{2} + \alpha^{2} \approx 0
\label{mPFC2}
\end{eqnarray}
The first set of constraints (\ref{mPFC1}) are an outcome of our extension of the original Lagrangian. The canonical Hamiltonian following from the usual definition is given by
\begin{equation}
H_{C} =  m\sqrt{q_{2}^{2}} + q_{0\mu}q_{2}^{\mu}
 \label{CH}
\end{equation}
The total Hamiltonian is
\begin{equation}
H_{T} =  H_{C} + u_{0\mu}\Phi_{0}^{\mu} + u_{1\mu}\Phi_{1}^{\mu} + \xi^{1}\Phi_{1} + \xi^{2}\Phi_{2} 
\end{equation}
where $u_{0\mu}$, $u_{1\mu}$, $\xi^{1}$ and $\xi^{2}$ are as yet undetermined multipliers. Now, conserving the primary constraints
% we get
%\begin{eqnarray}
%\nonumber 
%\dot{\Phi}_{0\mu} &=& \{\Phi_{0\mu}, H_T\}  = -q_{2\mu} + u_{1\mu} \approx 0
%\nonumber \\
%\dot{\Phi}_{1\mu} &=& \{\Phi_{1\mu}, H_T\}  = -u_{0\mu} \approx 0
%\nonumber \\
%\dot{\Phi}_{1} &=& \{\Phi_{1},  H_T\}  = -(q_{0}q_{2} + m\sqrt{q_{2}^{2}}) \approx 0
%\nonumber \\
%\dot{\Phi}_{2} &=& \{\Phi_{2},  H_T\} = 2(q_{0}p_{2})q_{2}^{2} \approx 0
%\label{34}
%\end{eqnarray}
we find that the multipliers  $u_{0\mu}$ and $u_{1\mu}$ are fixed:
\begin{equation}
u_{0\mu} = 0 \ \ and \ \ u_{1\mu} = q_{2\mu}
\end{equation}
Also new
secondary constraints are obtained
\begin{eqnarray}
%\nonumber
\omega_{1} =  q_{0}q_{2} +  m\sqrt{q_{2}^{2}} \approx 0; \ \ \ \ 
\omega_{2} = q_{0}p_{2} \approx 0
\label{mssc}
\end{eqnarray}
The last constraint in (\ref{mssc}) is obtained by assuming $q_{2}^{2} \neq 0$ which follows from
 the structure of the Lagrangian (\ref{mindependentfields}). The total Hamiltonian now becomes
\begin{equation}
H_{T} = m\sqrt{q_{2}^{2}} + q_{0\mu}q_{2}^{\mu} + q_{2\mu}\Phi_{1}^{\mu} + \xi^{1}\Phi_{1} + \xi^{2}\Phi_{2}
\end{equation}
PBs among the constraints are given by \\
\begin{eqnarray}
\nonumber
\left\lbrace {\Phi_{0\mu}, \omega_{1}}\right\rbrace &=& -q_{2\mu}
\nonumber \\
\left\lbrace {\Phi_{0\mu}, \omega_{2}}\right\rbrace &=& -p_{2\mu}
\nonumber \\
\left\lbrace {\Phi_{1}, \omega_{1}}\right\rbrace &=& -\omega_{1} =0
\nonumber  \\
\left\lbrace {\Phi_{1}, \omega_{2}}\right\rbrace &=& \omega_{2} = 0
\nonumber \\
\left\lbrace {\Phi_{2}, \omega_{1}}\right\rbrace &=& -2( \omega_{2} + m\sqrt{q_{2}^{2}}\Phi_{2} ) = 0
\nonumber \\
\left\lbrace {\Phi_{2}, \omega_{2}}\right\rbrace &=& 2p_{2}^{2}( q_{0}q_{2} )
\nonumber \\
\left\lbrace{\omega_{1}, \omega_{2}} \right\rbrace &=& q_{0}^{2} - m^{2}
\label{consalgebra}       
\end{eqnarray}
 Conserving the secondary constraints we get
%find that 
%no tertiary constraints are obtained but
\begin{eqnarray}
\nonumber
\dot{\omega}_{1} &=& 0
\nonumber \\
\dot{\omega}_{2}  &=& -q_{0\mu}\left(  {\frac{m}{\sqrt{q_{2}^{2}}}q_{2}^{\mu} + q_{0}^{\mu}} \right) - 2\xi^{2} p_{2}^{2}(q_{0}q_{2}) = 0\label{condition1}
\end{eqnarray}
Clearly no more tertiary constraints are obtained. 
The second condition of (\ref{condition1}) fixes 
the multiplier $\xi^{2}$ 
%is fixed, so that the iterative process terminates;\\
\begin{equation}
\xi^{2} = - \frac{1}{2p_{2}^{2}(q_{0}q_{2})}\left(  {\frac{m}{\sqrt{q_{2}^{2}}}(q_{0}q_{2}) + q_{0}^{2} } \right) 
\end{equation}
Substituting this in the expression of the total Hamiltonian we get
\begin{eqnarray}
%\nonumber
H_{T} =
%  m\sqrt{q_{2}^{2}} + q_{0\mu}q_{2}^{\mu} + q_{2\mu}\Phi_{1}^{\mu} + \xi_{1}\Phi_{1}  - \frac{1}{2p_{2}^{2}(q_{0}q_{2})}\left(  {\frac{m}{\sqrt{q_{2}^{2}}}(q_{0}q_{2}) + q_{0}^{2} } \right)\Phi_{2}
%\nonumber \\
= m\sqrt{q_{2}^{2}} + q_{0\mu}q_{2}^{\mu} + \xi^{1}\Phi_{1}  - \frac{1}{2p_{2}^{2}(q_{0}q_{2})}\left(  {\frac{m}{\sqrt{q_{2}^{2}}}(q_{0}q_{2}) + q_{0}^{2} } \right)\Phi_{2}
\label{totalhamiltonian}
\end{eqnarray}
It is important to observe that though there are two primary first class constraints only one undetermined multiplier survives in the total Hamiltonian. This shows that effectively there is only one gauge degree of freedom. This feature distinguishes it from a genuine first order theory and has vital implications in the construction of the generator.
%\begin{eqnarray}
%q_{1}^{\mu}=x_{\mu}, \ q_{2}^{\mu}=\dot{x}^{\mu},\   p_{2}^{\mu}=-\frac{\partial{L}}{\partial\dot{x}^{\mu}},  \ p_{1}^{\mu}=-\frac{\partial{L}}{\partial\dot{x}^{\mu}}+\frac{d}{dt}\left( \frac{\partial{L}}{\partial{\ddot{x^{\mu}}}}\right) 
%\end{eqnarray}

 We now strongly impose the constraints (\ref{mPFC1}).  This is possible because the constraints (\ref{mPFC1}) merely eliminate the unphysical sector($q_{0\mu}$, $p_{0\mu}$) in favour of the physical variables. The constraint (\ref{mssc}) now read as
\begin{eqnarray}
%\nonumber
\omega_{1} =  p_{1}q_{2} +  m\sqrt{q_{2}^{2}} \approx 0;
\omega_{2} =  p_{1}p_{2} \approx 0
\label{mssc2}
\end{eqnarray}
The algebra of the remaining constraints can now be read off from (\ref{consalgebra}). We find
\begin{equation}
\left\lbrace \Phi_{2}, \omega_{2}\right\rbrace= 2p_{2}^{2}\left(p_{1}q_{2} \right); \ \ \ \ 
\left\lbrace \omega_{1}, \omega_{2}\right\rbrace = p_{1}^{2}-m^{2} 
\label{mconstalgb}
\end{equation}
If we take\\
\begin{equation}
\Phi_{2}^{\prime}=\left(p_{1}^{2}-m^{2} \right)\Phi_{2} - 2p_{2}^{2}\left( p_{1}q_{2}\right)\omega_{1}
\label{44}  
\end{equation}
then\\
\begin{equation}
\left\lbrace \Phi_{2}^{\prime}, \omega_{2}\right\rbrace=0
\end{equation}
If instead of the set of constraints $\Phi_{1}, \Phi_{2}, \omega_{1}, \omega_{2}$ we take their linear combinations in the form 
$\Phi_{1}, \Phi_{2}^{\prime}, \omega_{1}, \omega_{2}$ we find that only the PB between $ \omega_{1}$ and $\omega_{2}$ is non- involutive, given by the second equation in (\ref{mconstalgb}).
Clearly $\Phi_{1}, \Phi_{2}^{\prime}$ are first class and $\omega_{1}, \omega_{2}$ are second class.

 At this stage an important point is to be noticed. The canonical momentum  is set to be equal to the Lagrange multiplier $q_{0\mu}$ when we impose the constraint $\Phi_{1 \mu}$ strongly equal to zero. Thus $p_{1\mu}$ is completely arbitrary. It can be space, time or light-like. In the usual relativistic particle model $p_1^2 = m^2$ appears as a constraint of the theory. This is not the case here. In what follows we will assume that $p_1^2\ne m^2$. In other words we will consider a modified dispersion relation. 
%In the following thid modified dispersion relation will be shoen to be consistent from our analysis.
That such a modified dispersion relation is consistent will be demonstrated by our analysis. The case $p_1^2 = m^2$ will be seen as a singular point in our analysis and will be explored separately in section 3.2.

 Now the second class constraints $\omega_{1}, \omega_{2}$ may be strongly put equal to zero if the PBs $\{A, B\}$ are replaced by DBs $\{A, B\}_D$ {\footnote{Dirac brackets are denoted by $\{ , \}_D$ to distinguish them from Poisson brackets which are written as $\left\lbrace  \right\rbrace$.}}.
 The non-vanishing DBs are given by {\footnote{Note that this computation is valid only when $p_1^2 \ne m^2$. The case $p_1^2 = m^2$ is thus a singular point in our analysis which is to be treated separately. }}
\begin{eqnarray}
\nonumber
\{q_{1\mu}, q_{1\nu}\}_D
%&=&-\left\lbrace{q_{1\mu}, \omega_{1}} \right\rbrace\Delta_{12}^{-1}\left\lbrace {\omega_{2}, q_{1\nu}}\right\rbrace-\left\lbrace {q_{1\mu}, \omega_{2}}\right\rbrace\Delta_{21}^{-1}\left\lbrace {\omega_{1},q_{1\nu}}\right\rbrace
%\nonumber \\
%&=& \left( -q_{2\mu}\right)\left(-\frac{1}{p_{1}^{2}-m^{2}}\right)\left( -p_{2\nu}\right)-\left( p_{2\mu}\right)\left(\frac{1}{p_{1}^{2}-m^{2}}\right)\left( -q_{2\nu}\right)  
%\nonumber \\
&=& \frac{1}{p_{1}^{2}-m^{2}}\left({p_{2\mu}q_{2\nu}-q_{2\mu}p_{2\nu}} \right)     
\end{eqnarray}
%Now $\left\lbrace {q_{2\mu}, \omega_{1}}\right\rbrace=0, \left\lbrace {q_{2\mu}, \omega_{2}}\right\rbrace=p_{1\mu} $
%so\\
\begin{eqnarray}
\nonumber
\{q_{1\mu}, q_{2\nu}\}_D
%&=&-\left\lbrace {q_{1\mu}, \omega_{1}}\right\rbrace\Delta_{12}^{-1}\left\lbrace{\omega_{2}, q_{2\nu}} \right\rbrace
%\nonumber \\
%&=&-\left( {q_{2\mu}}\right)\left({-\frac{1}{p_{1}^{2}-m^{2}}} \right)\left({-p_{1\nu}} \right)
%\nonumber \\
&=& -\frac{q_{2\mu}p_{1\nu}}{p_{1}^{2}-m^{2}}      
%\end{eqnarray}
%\begin{equation}
%\left\lbrace{p_{1\mu}, \omega_{1}} \right\rbrace=\left\lbrace {p_{1\mu}, \omega_{2}}\right\rbrace=\left\lbrace{p_{2\mu}, \omega_{2}} \right\rbrace=0   
%\end{equation}
%\begin{eqnarray}
%\nonumber
%\left\lbrace {p_{2\mu}, \omega_{1}}\right\rbrace&=&\left\lbrace{p_{2\mu}, \left({p_{1}q_{2}-m\sqrt{q_{2}^{2}}} \right) } \right\rbrace  
%\nonumber \\
%&=& -p_{1\mu}+\frac{m}{\sqrt{q_{2}^{2}}}q_{2\mu}
%\end{eqnarray}
%Then\\
%\begin{eqnarray}
\nonumber\\
\{q_{1\mu}, p_{2\nu} \}_D
%&=& -\left\lbrace{q_{1\mu}, \omega_{2}} \right\rbrace\Delta_{21}^{-1}\left\lbrace{\omega_{1}, p_{2\nu}} \right\rbrace   
%\nonumber \\
%&=&-\left( {p_{2\mu}}\right)\left({\frac{1}{p_{1}^{2}-m^{2}}} \right)\left({p_{1\nu}-\frac{m}{\sqrt{q_{2}^{2}}}q_{2\nu}} \right) 
%\nonumber \\
&=& \frac{1}{p_{1}^{2}-m^{2}}\left( {\frac{m}{\sqrt{q_{2}^{2}}}p_{2\mu}q_{2\nu} + p_{2\mu}p_{1\nu}}\right)
\label{basicdbs1}
 \end{eqnarray}
\begin{eqnarray}
%\nonumber
%\left[ {q_{2\mu}, p_{1\nu}}\right]&=& 0, \left[ {p_{1\mu},p_{1\nu}}\right]=0
%\nonumber \\
%\left[{p_{1\mu}, p_{2\nu}} \right]&=& 0, \left[ {p_{2\mu}, p_{2\nu}}\right]=0 
\nonumber \\
\{q_{1\mu}, p_{1\nu} \}_D &=& \eta_{\mu\nu}
% \left[{q_{2\mu}, q_{2\nu}} \right]=0     
\nonumber \\
\{q_{2\mu}, p_{2\nu} \}_D &=& \eta_{\mu\nu}-\frac{1}{p_{1}^{2}-m^{2}}\left({p_{1\mu}p_{1\nu} + \frac{m}{\sqrt{q_{2}^{2}}}p_{1\mu}q_{2\nu}} \right)
\label{50} 
\end{eqnarray}
The structure of the DBs is remarkable. We find that the coordinate algebra $\left[ {q_{1\mu}, q_{1\nu}}\right]$ becomes non -- commutative. Such non -- commutativity is generally known to modify the usual dispersion relation, both effects occuring at Planck scales. Our assumption of the condition $p_1^2 \ne m^2$ is thus consistent and it exhibits the appearance of coordinate noncommutativity in a simple setting including its connection with modified dispersion relation. As shown later in section 4, the treatment of the singular case $p_1^2 = m^2$ naturally leads to a vanishing algebra among the coordinates $q_{1\mu}$

  After the substitution of the basic PBs by the DBs there is some simplification of the constraint structure.
  The second class constraints $\omega_{1}$ and $\omega_{2}$ can now be strongly set equal to zero.
%($\omega_{1}$=$\omega_{2}$=0)
% which implies  that 
%\begin{eqnarray}
%\omega_1 = p_1q_2 +m \sqrt{q_2^2} = 0\nonumber\\
%\omega_2 = p_1p_2 = 0
%\label{constrntelimination}
%\end{eqnarray} 
The constraint $\Phi_{2}^{\prime}$(see equation (\ref{44}))  then reduces to,
\begin{eqnarray}
\nonumber
\Phi_{2}^{\prime}=\left(p_{1}^{2}-m^{2} \right)\Phi_{2}   
\end{eqnarray}
%We now have $\omega_{1} = 0$. Also $\left(p_{1}^{2}-m^{2} \right)$ is nonvanishing. Thus we can replace $\Phi_{2}^{\prime}$ again by $\Phi_{2}$. The set of remaining constraints is then taken as $\Phi_{1},\Phi_{2}$.
 Let us calculate the DB between the constraints.We find,
%\begin{eqnarray}
%\left[\Phi_{1} , \Phi_{2}^{\prime}\right] = -2\left[\left(p_1p_2\right)q_2^2\left(\left(p_1p_2\right) +\frac{m}{\sqrt{q_2^2}}\left(p_1q_2\right)\right) - p_2^2\left(p_1 q_2\right)\left(p_1q_2 +m \sqrt{q_2^2}\right) \right]
%\end{eqnarray}
\begin{eqnarray}
\{\Phi_{1} , \Phi_{2}^{\prime}\}_D = -2\left[\omega_2q_2^2\left(\omega_2 +\frac{m}{\sqrt{q_2^2}}\left(p_1q_2\right)\right) - p_2^2\left(p_1 q_2\right)\omega_1 \right] = 0
\end{eqnarray}
 since $\omega_1 = \omega_2 = 0$.

We now proceed towards the discussion on gauge symmetry. There are now two  first class constraints
$\Phi_{1}$, $\Phi_{2}^{\prime}$, both of which are primary.
%, $\left[ {\Phi_{1}, \Phi_{2}}\right]= 0 $ .
 So the generator of gauge transformation is \cite{dirac2}
\begin{equation}
G=\epsilon^{1}\Phi_{1} + \epsilon^{2}\Phi^{\prime}_{2}
\label{generator}
\end{equation}
%
%$\Phi_{1}$ and $\Phi_{2}$ are first class 
%The canonical hamiltonian $H_{c}=-\omega_{1}=0$\\ 
%All the structure constants vanish.Both $\epsilon_{1}, \epsilon_{2}$ are independent gauge constants.\\
 Had it been a first order theory we would conclude that both $\epsilon^{1}, \epsilon^{2}$ are independent gauge parameters. However, due to the higher derivative nature we  have the additional  requirement 
\begin{eqnarray}
\frac{d}{d\tau}\delta{q_{1}^{\mu}} &=& \delta{q_{2}^{\mu}}\label{condition}
\end{eqnarray}
which follows from 
%(\ref{commutativity1}) and 
(\ref{varsgauge}).
Now
\begin{eqnarray}
\delta{q_{1}^{\mu}} = \left[q_{1}^{\mu}, G\right]
%&=& \epsilon_{1}\left[{q_{1}^{\mu}, \Phi_{1}}\right] + \epsilon_{2}\left[{q_{1}^{\mu}, \Phi_{2}} \right]\nonumber\\
 = 2\epsilon^{2}p_{2}^{2}q_{2}^{\mu}m\sqrt{q_{2}^{2}}
 \label{cmp1}
\end{eqnarray}
and
\begin{eqnarray}
\delta{q_{2}^{\mu}} = \left[q_{2}^{\mu}, G\right] = \epsilon^{1}q_{2}^{\mu}+2\epsilon^{2}q_{2}^{2}(p_{1}^{2} - m^{2})p_{2}^{\mu}
\label{55}
\end{eqnarray}
Hence using (\ref{condition})we get
%Here \\
%\begin{eqnarray}
%\nonumber
%\left[{q_{1}^{\mu}, \Phi_{1}}\right]=0
%\nonumber \\
%\left[{q_{1}^{\mu}, \Phi_{2}} \right]=-\frac{2p_{2}^{2}}{p_{1}^{2}-m^{2}}q_{2}^{\mu}m\sqrt{q_{2}^{2}}
%\end{eqnarray}
%$\delta{q_{2}^{\mu}} = \epsilon_{1}q_{2}^{\mu}+2\epsilon_{2}q_{2}^{2}p_{2}^{\mu}$
%now $\dot{q_{1}^{\mu}}=q_{2}^{\mu}$  so we require \\
\begin{eqnarray}
%\nonumber 
%\frac{d}{d\tau}\delta{q_{1}^{\mu}} &=& \delta{q_{2}^{\mu}}
%\nonumber \\
%or,
 \ \epsilon^{1}q_{2}^{\mu} + 2\epsilon^{2}q_{2}^{2}(p_{1}^{2} - m^{2})p_{2}^{\mu} = \frac{d}{d\tau}\left({2mp_{2}^{2}\sqrt{q_{2}^{2}}q_{2}^{\mu}\epsilon^{2}} \right)
%or, \ \epsilon_{1}q_{2}^{\mu}+ 2\epsilon_{2}q_{2}^{2}p_{2}^{\mu} &=& -\frac{2mp_{2}^{2}\sqrt{q_{2}^{2}}}{p_{1}^{2}-m^{2}}q_{2}^{\mu}\dot{\epsilon_{2}} - \epsilon_{2}\frac{d}{d\tau}\left({\frac{2mp_{2}^{2}\sqrt{q_{2}^{2}}}{p_{1}^{2}-m^{2}}q_{2}^{\mu}} \right) 
\end{eqnarray}
Taking scalar product with $q_{2\mu}$ and using $\Phi_{1} = p_{2}q_{2}\approx0$, we obtain\\
\\
\begin{eqnarray}
\epsilon^{1} = \frac{q_{2\mu}}{q_{2}^{2}}\frac{d}{d\tau}\left(2mp_{2}^{2}\sqrt{q_{2}^{2}}\epsilon^{2}q_{2}^{\mu} \right)
\label{gaugepara1}
\end{eqnarray}
Clearly, only one parameter $\epsilon^{2}$(say) is independent in the gauge generator G. This is also compatible with the observation that there is only one undetermined multiplier in the total Hamiltonian (\ref{totalhamiltonian}). The number of independent gauge parameters is thus shown to be less than the number of independent first class primary constraints.

	We next show that these findings are consistent with the reparametrization symmetry of the model to which the gauge symmetry is expected to  have a one to one correspondence. In the following we will show the mapping between the gauge parameter and the reparametrization parameter.\\

Consider the following reparametrization 
\begin{eqnarray}
\tau \to \tau + \Lambda
\label{repara_particle}
\end{eqnarray}
where $\Lambda$ is an infinitesimal reparametrization parameter. By direct substitution we can verify that (\ref{repara_particle}) is an invariance of (\ref{maction}). Now, under this reparametrization $x^{\mu}$ transforms as
\begin{equation}
x^{\prime\mu}\left(\tau\right) = x^{\mu}\left(\tau -\Lambda\right)
\end{equation}
The variation of $x^{\mu}$ is then
\begin{equation}
\delta{x^{\mu}} = x^{\prime\mu}(\tau) - x^{\mu}(\tau) =  -\Lambda\dot{x}^{\mu}\label{cmp3}
\end{equation}
%where $\Lambda$ is the reparametrization parameter.
From equation (\ref{cmp1}) we can write
\begin{equation}
\delta{x^{\mu}} = 2\epsilon^{2}p_{2}^{2}\dot{x}^{\mu}m\sqrt{q_{2}^{2}}
\label{cmp2}
\end{equation}
where we have used the identification(\ref{mnewcoordinate}). Comparing (\ref{cmp3}) and (\ref{cmp2}) we get the desired mapping
\begin{equation}
\Lambda =   - 2\epsilon^{2}p_{2}^{2}m\sqrt{q_{2}^{2}}
\label{reparavalue}
\end{equation}
Thus in our analysis an exact correspondence between the gauge and reparametrization symmetries is clearly demonstrated.
% parameter and the gauge parameter.
%\begin {appendix} 
%\renewcommand{\thesection}%{Appendix 
%\Alph{section}
%}			% redefine the command that creates the section heading.
%\setcounter{section}{0}										% redefine the command that creates the section no.

\subsection {Behaviour at the singular point ($p_1^2 = m^2$)} In the above Hamiltonian analysis of the relativistic particle model with curvature we have shown that $p_1^2$ is not constrained by the phase space structure and may be space, time or light-like. In our analysis we have assumed that $p_1^2$ is not equal to $m^2$. This is because the condition $p_1^2 = m^2$  is a singular point and must be treated separately. 
In the following we will discuss the construction of the gauge generator when the singular limit is assumed. This will also reveal the versatality of our method.

 The set of constraints after removal of  the unphysical sector $q_{0} = p_{1}$ is now given by,
 \begin{eqnarray}
\Phi_{1} &=& p_{2}q_{2} \approx 0; \ \ \ \Phi_{2} =p_{2}^{2}q_{2}^{2} + \alpha^{2} \approx 0; \nonumber \\ 
 \omega_{1} &=& p_{1}q_{2} + m\sqrt{q_{2}^{2}} \approx 0 ; \ \ \  \omega_{2} = p_{1}p_{2} \approx  0\label{singularconstraints}
\end{eqnarray}
Here $\Phi_1$  and $\Phi_2$ are primary while $\omega_{1}$ and $\omega_{2}$ are secondary constraints. The constraint algebra shows that $\Phi_1$  and $\omega_{1}$ are first class while the rest are second class. So the constraint structure in the singular limit is different from that in the general case discussed above. Specifically the appearance of a secondary first class constraint is to be noted. 
%Also, the multiplier  $\xi_{2}$ in the the expreturns out to be vanishing. Hence again there is only one undetermined multiplier in 
The total Hamiltonian reads as
\begin{equation}
H_{T} = H_{c} + \xi^{1}\Phi_{1}
\label{singtotalH}
\end{equation}
where $H_{c}$ is the canonical Hamiltonian, given by 
\begin{equation}
H_{c} = m \sqrt{q_{2}^{2}} + p_{1}q_{2}
\end{equation}
Note that, as before, the total Hamiltonian consists of one undetermined multiplier
indicating one independent gauge degree of freedom.

    The construction of the  generator of the gauge transformations follow the course outlined in section 2. At first we strongly impose the second class  constraints $\left( {\Phi_{2}, \omega_{2}}\right)$ using the Dirac bracket formalism.
The non-trivial Dirac brackets among the phase-space variables are\\
\begin{eqnarray}
\nonumber \\
\{ q_{1\mu}, q_{2\nu}\}_D &=& \frac{q_{2}^{2}}{p_{2}^{2}} \frac{p_{2\mu}p_{2\nu}}{p_{1}q_{2}} 
\nonumber \\
\{q_{1\mu}, p_{2\nu} \}_D &=& - \frac{p_{2\mu}q_{2\nu}}{p_{1}q_{2}} 
\nonumber \\
\{{q_{1\mu}, p_{1\nu}} \}_D  &=& \eta_{\mu \nu}
\nonumber \\
\{q_{2\mu}, p_{2\nu} \}_D &=& \eta_{\mu \nu} - \frac{p_{1\mu} q_{2\nu}}{p_{1}q_{2}}
\nonumber \\
\{q_{2\mu}, q_{2\nu}\}_D &=& \frac{q_{2}^{2}}{p_{2}^{2}} \frac{\left( {p_{1\mu}p_{2\nu} - p_{2\mu}p_{1\nu}}\right) }{p_{1}q_{2}}
\label{singDBs}
\end{eqnarray}
 Note that the coordinate algebra $\{q_{1\mu}, q_{1\nu}\}_D$ becomes commutative as a result of usual dispersion relation. This may be contrasted with the general case ($p_{1}^{2} \neq m^{2}$) leading to a noncommutative coordinate algebra (first equation in (\ref{basicdbs1})).
The  generator  is given by\\
\begin{equation}
G^{\prime} = \epsilon^{1} \Phi_{1} + \epsilon^{2}\omega_{1}
\label{ncgg}
\end{equation} 
Due to the presence of a secondary first class constraint($\omega_{1}$), $\epsilon^{1}$ and $\epsilon^{2}$ are not independent. There is a restriction \cite{BRR}
\begin{eqnarray}
\frac{d\epsilon^2}{d\tau} - \sum_{a=1,2} \epsilon^a\left(V_a^{\ 2} + \xi^1C_{1 a}^{\ \ 2}\right)=0
\end{eqnarray} The other restriction(\ref{condition}) follows from the higher derivative nature . Interestingly, both the restrictions lead to the same condition 
\begin{equation}
\epsilon^{1} = \dot{\epsilon}^{2} + \xi^{1} \epsilon^{2}
\label{nc12}
\end{equation}
 To proceed further we require to express the Lagrange multiplier $\xi^1$ in terms of the phase space variables.  To this end we calculate $\dot{q}_2^\mu$ as
\begin{equation}
\dot{q}_2^\mu = \{ {q}_2^\mu, H_T\}_D 
\end{equation}
where $H_T$ is the total Hamiltonian given by equation (\ref{singtotalH}). Using the basic brackets (\ref{singDBs}) we get
%\begin{array}{rcl}
\begin{eqnarray}
%\nonumber
\dot{q}_2^\mu &=& \xi^1\left[{q}_2^\mu - \frac{p_1^\mu}{p_1^2 - m^2}\left(p_1q_2+m\sqrt{q_2^2}\right)\right] - \frac{q_2^2}{p_2^2p_1q_2}\left(\frac{m}{\sqrt{q_2^2}}(p_1q_2)+p_1^2\right)\nonumber\\
&&\left[p_2^\mu -\frac{p_1^\mu}{p_1^2 - m^2}\left(p_1p_2+\frac{m}{\sqrt{q_2^2}}p_2q_2\right)\right]
%\nonumber \\
\label{qdot}
\end{eqnarray}
%From equation (\ref{constrntelimination}) we find that $\omega_1 = p_1q_2 + m\sqrt{q_2^2} = 0$ and $\omega_2 = p_1p_2 = 0$. These constraints are strongly imposed. Also note the constraint algebra between the first class constraints $\Phi_1$ and $\Phi^{\prime}_2$ given by equation (\ref{53}). Since the bracket between these first class constraints vanishes strongly we can use $\Phi_1 = 0$ in (\ref{qdot}). Using these simplifications we get 
   \  \  Implementing the constraints (\ref{singularconstraints})  and simplifying, yields,
\begin{eqnarray}
\dot{q}_2^\mu = \xi^1{q}_2^\mu - \left(p_1^2 - m^2\right)\frac{q_2^2p_2^\mu}{p_2^2p_1q_2} ,
\end{eqnarray}
which immediately gives, on contraction with $q_{2\mu}$,
\begin{equation}
\xi^1 = \frac{1}{2q_2^2}\frac{d}{d\tau}\left(q_2^2\right)\label{ncvarxi1}
%\label{lagrange}
\end{equation}
\\

%\begin{equation}
%\xi^{1} = \frac{q_{2}\dot{q}_{2}}{q_{2}^{2}} 
%\label{ncvarxi1}
%\end{equation}\\
Using (\ref{nc12}, \ref{ncvarxi1}) we can express one of the gauge parameters appearing in $G^{\prime}$(\ref{ncgg}) in terms of the other. 
%The consistency of the scheme may again be checked by the method outlined above.
 We observe that in spite of the modified constraint structure corresponding to the singular point, the gauge generator may be consistently constructed by our general method. 
\section{A consistency check}
An important relation to check the consistency of our scheme is given by \cite{BRR}
\begin{eqnarray}
\delta\xi^1 = \frac{d\epsilon^1}{dt} - \sum_{a= 1,2} \epsilon^a\left(V_a^{\  1} + \xi^1C_{1 a}^{\ \  1}\right)\label{parametervariation}
\end{eqnarray}
where the coefficients $V_a^{\  b}$ and $C_{ab}^{\ \ c}$ are defined by (\ref{structure constant_V}, \ref{structure constant_C}).
%\begin{eqnarray}
%\left[\Phi_a, H_C\right] = V_a^{\ b}\Phi_b ; \left[\Phi_a, \Phi_b\right] =C_{ab}^{\ \ c}\Phi_c
%\end{eqnarray}
Here both $V_a^{\ b}$ and $C_{ab}^{\ \ c}$ vanish.

   Now we can independently calculate $\delta\xi^1 $ for our theory and an exact agreement with (\ref{parametervariation}) will be demonstrated. 

	Taking the gauge variation on both sides of (\ref{ncvarxi1}) and substituting $\delta q_2^\mu$ from (\ref{55}) we arrive at,
\begin{eqnarray}
\delta\xi^1 = &-&\frac{1}{(q_2^2)^2}q_{2\mu}\left(\epsilon^1 q_2^\mu + 2\epsilon^2q_2^2(p_1^2-m^2)p_{2}^{\mu}\right)\frac{d}{d\tau}q_2^2\nonumber\\  &+& \frac{1}{q_2^2}\frac{d}{d\tau}\left[q_{2\mu}\left(\epsilon^1 q_2^\mu + 2\epsilon^2q_2^2(p_{1}^{2}-m^{2})p_{2}^{\mu}\right)\right]
\end{eqnarray}
Imposing the constraints (\ref{mPFC2}, \ref{mssc2}) and simplifying the ensuing algebra immediately reproduces(\ref{parametervariation}).This completes our consistency check.
%\begin{equation}
%\delta\xi_1 = \frac{d\epsilon_1}{d\tau}
%\end{equation}
%which is just the same as (\ref{parametervariation}), following from the general formalism.

\section{The rigid relativistic particle model}

	 The massless version of the model  known as `rigid relativistic particle', is obtained by setting $m=0$ in (\ref{maction}). It presents some unique features. We perform a detailed Hamiltonian analysis which is quite distinct from the earlier(massive) model due to a modified symplectic structure.  It is interesting to point out in this case that we will be able to find out an extra symmetry apart from the expected diffeomorphism  symmetry. This  is the $W_{3}$-symmetry \cite{ramos1}.
\subsection{Hamiltonian analysis}
The relativistic point particle theory with rigidity only has the action \cite{BMP, plyuschay2}
%\footnote{contractions are abbreviated as $A^\mu B_\mu = AB$, $A^\mu A_\mu=A^2$} 
\begin{equation}
S=
%-m \int{\sqrt{\dot{x}^{2}}}d\tau+
\alpha\int{\frac{\left( \left( \dot{x}\ddot{x}\right)^{2}-\dot{x}^{2}\ddot{x}^{2} \right)^\frac{1}{2}}{\dot{x}^{2}}}d\tau\label{action}
\end{equation}
We introduce the new coordinates
\begin{equation}
q^{\mu}_1 = x^{\mu}\ \;\ \ q^{\mu}_2 = \dot{x}^{\mu}
\label{newcoordinate}
\end{equation}
The Lagrangian in these coordinates has a first order form given by
\begin{eqnarray}
% \nonumber
%L_{0} &=& -m\sqrt{q_{2}^{2}} + \alpha \frac{\left({\left({q_{2}\dot{q}_{2}} \right)^{2} - q_{2}^{2} \dot{q}_{2}^{2} } \right)^{\frac{1}{2}} }{q_{2}^{2}}
%\nonumber \\
L &=& \alpha \frac{\left({\left({q_{2}\dot{q}_{2}} \right)^{2} - q_{2}^{2} \dot{q}_{2}^{2} } \right)^{\frac{1}{2}} }{q_{2}^{2}} + q_{0}^{\mu}(\dot{q}_{1\mu} - q_{2\mu})
\label{independentfields}
\end{eqnarray}
where $q_{0}^{\mu}$ are the Lagrange multipliers that enforce the constraints
\begin{equation}
\dot{q}_{1\mu} - q_{2\mu} = 0
\label{27}
\end{equation}
Let $p_{0\mu}$, $p_{1\mu}$ and $p_{2\mu}$ be the canonical momenta conjugate to $q_{0\mu}$, $q_{1\mu}$ and $q_{2\mu}$ respectively. Then
\begin{eqnarray}
\nonumber
p_{0\mu} &=& \frac{\partial{L}}{\partial{\dot{q}_{0}^{\mu}}} = 0
\nonumber \\
p_{1\mu} &=& q_{0\mu}
\nonumber \\
p_{2\mu} &=& \frac{\alpha}{q_{2}^{2}\sqrt{g}} l_{\mu}
\label{regidmomenta}
\end{eqnarray}
 where,\\
\begin{eqnarray}
\nonumber
g &=& \left({\left({q_{2}\dot{q}_{2}} \right)^{2} - q_{2}^{2} \dot{q}_{2}^{2} } \right)
\nonumber \\
l_{\mu} &=& (q_{2}\dot{q}_{2})q_{2\mu} - q_{2}^{2}\dot{q}_{2\mu}
\end{eqnarray}
We immediately get the following
primary constraints
\begin{eqnarray}
\nonumber
\Phi_{0\mu} = p_{0\mu} \approx 0
\nonumber \\
\Phi_{1\mu} = p_{1\mu} - q_{0\mu} \approx 0
\label{PFC1}
\end{eqnarray}
and
\begin{eqnarray}
\nonumber
\Phi_{1} &=& p_{2}q_{2} \approx 0
\nonumber \\
\Phi_{2} &=& p_{2}^{2}q_{2}^{2} + \alpha^{2} \approx 0
\label{PFC2new}
\end{eqnarray}
The first set of constraints (\ref{PFC1}) are an outcome of our extension of the original Lagrangian. The canonical Hamiltonian following from the usual definition is given by
\begin{equation}
H_{C} =  q_{0\mu}q_{2}^{\mu}
 \label{CH}
\end{equation}
The total Hamiltonian is
\begin{equation}
H_{T} =  H_{C} + u_{0\mu}\Phi_{0}^{\mu} + u_{1\mu}\Phi_{1}^{\mu} + \lambda^{1}\Phi_{1} + \lambda^{2}\Phi_{2} 
\end{equation}
where $u_{0\mu}$, $u_{1\mu}$, $\lambda^{1}$ and $\lambda^{2}$ are as yet undetermined multipliers. Now, conserving the primary constraints we get
\begin{eqnarray}
\nonumber 
\dot{\Phi}_{0\mu} &=& \{\Phi_{0\mu}, H_T\}  = -q_{2\mu} + u_{1\mu} \approx 0
\nonumber \\
\dot{\Phi}_{1\mu} &=& \{\Phi_{1\mu}, H_T\}  = -u_{0\mu} \approx 0
\nonumber \\
\dot{\Phi}_{1} &=& \{\Phi_{1},  H_T\}  = -q_{0}q_{2}  \approx 0
\nonumber \\
\dot{\Phi}_{2} &=& \{\Phi_{2},  H_T\} = - 2(q_{0}p_{2})q_{2}^{2} \approx 0
\label{34}
\end{eqnarray}
We find that the multipliers  $u_{0\mu}$ and $u_{1\mu}$ are fixed
\begin{equation}
u_{0\mu} = 0 \ \ and \ \ u_{1\mu} = q_{2\mu}
\end{equation}
while new
secondary constraints are obtained
\begin{eqnarray}
\nonumber
\omega_{1} =  q_{0}q_{2}  \approx 0
\nonumber \\
\omega_{2} = q_{0}p_{2} \approx 0
\label{36new}
\end{eqnarray}
The last constraint in (\ref{34}) simplifies to $\omega_{2}$ since $q_{2}^{2} \neq 0$ as may be observed from the structure of the Lagrangian (\ref{independentfields}). The total Hamiltonian now becomes
\begin{equation}
H_{T} =  q_{0\mu}q_{2}^{\mu} + q_{2\mu}\Phi_{1}^{\mu} + \lambda^{1}\Phi_{1} + \lambda^{2}\Phi_{2} \label{tothamfin}
\end{equation}
	Preserving the constraints\ref{36new} we get $\dot{\omega}_{1} \approx 0$ ; $\dot{\omega}_{2} \approx -q_{0}^{2}$ thereby yielding a tertiary constraint $\Phi_{3} = q_{0}^{2} \approx 0$. This terminates the iterative process of obtaining constraints. The complete set of constraints is now given by,
\begin{eqnarray}
\Phi_{0\mu} &=& p_{0\mu} \approx 0\nonumber\\
\Phi_{1\mu} &=& p_{1\mu} - q_{0\mu} \approx 0\nonumber\\
\Phi_{1} &=& p_{2}q_{2} \approx 0\nonumber\\
\Phi_{2} &=& p_{2}^2q_{2}^2 + \alpha^2 \approx 0\nonumber\\
\omega_1 &=& q_0q_2\approx 0\nonumber\\
\omega_2 &=& q_0p_2\approx 0\nonumber\\
\Phi_3 &=& q_0^2 \approx 0
\end{eqnarray}
PBs among the constraints are given by \\
\begin{eqnarray}
\nonumber
\left\lbrace {\Phi_{0\mu}, \Phi_{1\nu}}\right\rbrace &=& \eta_{\mu\nu}
\nonumber \\
\nonumber
\left\lbrace {\Phi_{0\mu}, \omega_{1}}\right\rbrace &=& -q_{2\mu}
\nonumber \\
\left\lbrace {\Phi_{0\mu}, \omega_{2}}\right\rbrace &=& -p_{2\mu}
\nonumber \\
\left\lbrace {\Phi_{0\mu}, \Phi_{3}}\right\rbrace &=& -2q_{0\mu}
\nonumber \\
\left\lbrace {\Phi_{1}, \omega_{1}}\right\rbrace &=& -\omega_{1} =0
\nonumber  \\
\left\lbrace {\Phi_{1}, \omega_{2}}\right\rbrace &=& \omega_{2} = 0
\nonumber \\
\nonumber  \\
\left\lbrace {\Phi_{1}, \Phi_{2}}\right\rbrace&=& 0
\nonumber \\
\left\lbrace {\Phi_{2}, \omega_{1}}\right\rbrace &=& -2q_{2}^{2}\omega_2 = 0
\nonumber \\
\left\lbrace {\Phi_{2}, \omega_{2}}\right\rbrace &=& 2p_{2}^{2}( q_{0}q_{2} )=0
\nonumber \\
\left\lbrace{\omega_{1}, \omega_{2}} \right\rbrace &=& q_{0}^{2} =0
\label{algebra}       
\end{eqnarray}
Apparently, $\Phi_{0\mu}$, $\Phi_{1\mu}$, $\omega_1$, $\omega_2$ and $\Phi_3$ are second class. However we may substitute $\Phi_3$ by $\Phi_3^{\prime}$ where
\begin{equation} 
\Phi_3^{\prime} = \Phi_3 - 2q_{0}^{\nu} \Phi_{1\nu}\label{consnew}
\end{equation} 
One can easily verify that $\Phi_3^{\prime}$ commutes with all the constraints. The set of constraints are taken to be $\Phi_{0\mu}$, $\Phi_{1\mu}$, $\Phi_{1}$, $\Phi_{2}$, $\omega_1$, $\omega_2$ and $\Phi_3^{\prime}$ which are may be classified in  table 1.
\begin{table}[h]
%\begin{table}
\label{table:constraints}
\caption{Classification of Constraints}
\centering
\begin{tabular}{l  c  c}
\\[0.5ex]
\hline
\hline\\[-2ex]
& First class & Second class \\[0.5ex]
\hline\\[-2ex]
Primary &\ \ $\Phi_1, \Phi_2$ &\ \ $\Phi_{0\mu}, \Phi_{1\nu}$\\[0.5ex]
\hline\\[-2ex]
Secondary &\ $\Phi_3^{\prime}$ &\ \ $\omega_1, \omega_2$\\[0.5ex]
\hline
\hline
\end{tabular}
\end{table}
\subsection {Gauge symmetries and the emergence of the $W_3$ algebra}
 In the above we have considered the hamiltonian formulation of the  model (\ref{action}) in the first order approach. The expression of the total Hamiltonian (\ref{tothamfin}) suggests two independent gauge symmetries of the model. In the following we will construct the gauge generator and interpret the different gauge symmetries physically. 

 The variables $q_{0\mu}$ and their associated momenta $p_{0\mu}$ comprise the unphysical sector of the phase space. This is characterised by the second class pair $\Phi_{0\mu}$ and $\Phi_{1\mu}$. To find the gauge symmetries we have to eliminate these constraints. Considering their unphysical nature it will be appropriate to work in the reduced phase by putting
%In the following it will be convenient to work in the reduced phase space.
$\Phi_{0\mu}$ and $\Phi_{1\mu}$ equal to zero.
% and substituting the Poisson brackets (PB) by the corresponding Dirac Brackets (DB). It turns out that the DBs between the remaining phase space variables are equal to the corresponding PBs. 
The set of remaining constraints in the reduced phase space become 
\begin{eqnarray}
\Phi_{1} &=& p_{2}q_{2} \approx 0\nonumber\\
\Phi_{2} &=& p_{2}^2q_{2}^2 + \alpha^2 \approx 0\nonumber\\
\omega_1 &=& p_1q_2\approx 0\nonumber\\
\omega_2 &=& p_1p_2\approx 0\nonumber\\
\Phi_3 &=& p_1^2 \approx 0
\end{eqnarray}
Inspection of the algebra (\ref{algebra}) shows that in the reduced phase space
$ \Phi_{1}$, $\Phi_{2}$, $\omega_1$, $\omega_2$ and $\Phi_3^{\prime}$ form a first class set. In what follows it will be advantageous to rename the constraints as
\begin{eqnarray}
%\Phi_{0\mu} &=& p_{0\mu} \approx 0\nonumber\\
%\Phi_{1\mu} &=& p_{1\mu} - q_{0\mu} \approx 0\nonumber\\
\Omega_1 = \Phi_{1} &=& p_{2}q_{2} \approx 0\nonumber\\
\Omega_2 = \Phi_{2} &=& p_{2}^2q_{2}^2 + \alpha^2 \approx 0\nonumber\\
\Omega_3 = \omega_1 &=& p_1q_2\approx 0\nonumber\\
\Omega_4 = \omega_2 &=& p_1p_2\approx 0\nonumber\\
\Omega_5 = \Phi_3 &=& p_1^2 \approx 0
\end{eqnarray}
Also the canonical Hamiltonian is obtained from (\ref{CH}) as
\begin{equation}
H_c = \Omega_3\label{canham}
\end{equation}
The corresponding total Hamiltonian is
%The total Hamiltonian is
\begin{equation}
H_{T} =  H_{C}  + \lambda^{1}\Phi_{1} + \lambda^{2}\Phi_{2} 
\label{totham}
\end{equation}

%  As has been mentioned earlier we will follow the canonical algorithm of \cite{BRR} to construct the gauge generator. According to Dirac's conjecture the gauge generator $G$ is given by
%\begin{equation}
%G = \epsilon^a\Omega_a\label{gauge}
%\end{equation}
%Note that all first class constraints appear in the expression multiplied by the corresponding gauge parameters $\epsilon^a$. However not all of these parameters are independent. For the usual first order lagrangean theories the number of independent gauge parameters is equal to the number of primary first class constraints \cite{BRR}. The dependent gauge parameters are expressed in terms of the independent ones from the following set of equations:
%\begin{equation}
%  0 = \frac{d\epsilon^{a_2}}{dt}
% -\epsilon^{a}\left(V_{a}^{a_2}
%+\lambda^{b_1}C_{b_1a}^{a_2}\right)
%\label{219}
%\end{equation}
%Here the coefficients $V_{a}^{a_{1}}$ and $C_{b_1a}^{a_1}$ are the structure
%functions of the involutive algebra, defined as
%\begin{eqnarray}
%\{H_c,\Omega_{a}\} = V_{a}^b\Omega_{b}\nonumber\\
%\{\Omega_{a},\Omega_{b}\} = C_{ab}^{c}\Omega_{c}
%\label{2110}
%\end{eqnarray}
%Solving (\ref{219}) it is possible to choose $a_1$ independent
%gauge parameters from the set $\epsilon^{a}$ and express $G$ of
%(\ref{gauge}) entirely in terms of them. The relations (\ref{219}) are obtained 
%from the commutativity of the gauge transformation.
 
 As done previously, the gauge generator is written as a combination of all the first class constraints,
\begin{equation}
G = \sum_{a=1}^{5} \epsilon^{a}\Omega_{a}.
\label{rigidsumG}
\end{equation} 
 However,due to the presence of secondary first-class constraints, the parameters of gauge transformation($\epsilon_{a}$) are not independent. The independent parameters will be isolated by using \ref{master_eqn_2}. The first step in this direction is to calculate the structure functions $V_{a}^b$ and $C_{ab}^{c}$ using their definitions (\ref{structure constant_V}, \ref{structure constant_C}). A straightforward calculation gives the following nonzero values:
\begin{eqnarray}
V_1^{\ 3} &=& 1,\hspace{1cm} V_2^{\ 4} = 2q_2^{\ 2}\nonumber\\
V_4^{\ 5} &=& 1
 \label{v's}
\end{eqnarray}
and
\begin{eqnarray}
C_{13}^{\  3} &=& -1 ,\hspace{1cm} C_{14}^{\  4} = 1 \nonumber\\
C_{23}^{\   4} &=& -2q_2^{\ 2},\hspace{1cm} C_{24}^{\  3} = 2p_2^{ 2}\nonumber\\
C_{34}^{\ \ 5} &=& 1 
\label{c's}
\end{eqnarray}

Using these(\ref{v's}, \ref{c's})  in the master equation (\ref{master_eqn_2}) we arrive at the following equations
\begin{eqnarray}
{\dot{\epsilon}}^3 - \epsilon^1 + \lambda^1\epsilon^3 - 2\lambda^2\epsilon^4p_2^{\ 2} = 0\nonumber\\
{\dot{\epsilon}}^4 - 2q_2^{\ 2}\epsilon^2 - \lambda^1\epsilon^4 + 2\lambda^2\epsilon^3q_2^{\ 2} = 0\nonumber\\
{\dot{\epsilon}}^5 = \epsilon^4\label{gaugerelations}
\end{eqnarray}
Certain points are immediately apparent from the above equations. Out of the five gauge parameters $\epsilon^a$ three parameters may be expressed in terms of the remaining two using the equations(\ref{gaugerelations}). It is most convenient to take $\epsilon^3$
and $\epsilon^5$ as independent. 

To proceed further we require to work out the Lagrange multipliers in (\ref{totham}). For that purpose we first compute $\dot{q}_{2\mu} = \left\lbrace {q_{2\mu}, H_{T}}\right\rbrace $. Using the expression for total Hamiltonian given in(\ref{totham})we get 
   \begin{equation}
\dot{q}_{2\mu} = \lambda^{1}q_{2\mu} + 2\lambda^{2}q_{2}^{2}p_{2\mu}
\label{rigidq2dot}
\end{equation}  
  Now scalar multiplying the above equation(\ref{rigidq2dot}) by $q_{2}^{\mu}$ and $p_{2}^{\mu}$ we obtain respectively solutions for $\lambda^1$ and $\lambda^2$ as
\begin{eqnarray}
\lambda^1 &=& \frac{q_2{\dot{q}}_2}{q_2^{\ 2}}\nonumber\\
\lambda^2 &=& 
\frac{\alpha
\sqrt{g}}
{2p_2^{2}q_2^{\ 4}}
\end{eqnarray}
Substituting these results for $\lambda^{1}$, $\lambda^{2}$ and $\epsilon^{4}=\dot{\epsilon}^{5}$ in the first two equations of (\ref{gaugerelations}) one can express 
\begin{eqnarray}
 \epsilon^1 &=& {\dot{\epsilon}}^3  + \frac{q_2{\dot{q}}_2}{q_2^{\ 2}}\epsilon^3 - \frac{\alpha\sqrt{g}}{q_2^{\ 4}}\dot{\epsilon}^{5} \nonumber\\
\epsilon^2 &=& \frac{1}{2q_2^{\ 2}}
\left({\ddot{\epsilon}}^5 - \frac{q_2{\dot{q}}_2}{q_2^{\ 2}}{\dot{\epsilon^5}} + \frac{\alpha\sqrt{g}}{p_2^{\ 2}q_2^{\ 2}}\epsilon^3\right)% \nonumber\\
%\epsilon^4 &=& {\dot{\epsilon}}^5 
 \label{gaugerelations1}
\end{eqnarray}
Note that the theory under investigation is a higher derivative theory. So the gauge transformations are additionally subject to the condition (\ref{varsgauge}). After simplification this condition reduces to
\begin{eqnarray}
2q_2{\dot{p}}_1\epsilon^5 = \left(q_2{\dot{p}}_2 + \frac{\alpha
\sqrt{g}}
{q_2^{\ 2}}\right)
\end{eqnarray}
Using the constraints of the theory this condition reduces to a tautological statement $0 = 0$. So, for this particular model, (\ref{varsgauge}) does not impose any new conditions on the gauge parameters. We thus find that there are two independent gauge parameters in the expression of the gauge generator $G$.Taking $\epsilon^{3}$ and $\epsilon^{5}$ to be independent, the expression for the gauge generator becomes(\ref{rigidsumG})
\begin{eqnarray}
G &=& \left( {\dot{\epsilon}}^3  + \frac{q_2{\dot{q}}_2}{q_2^{\ 2}}\epsilon^3 - \frac{\alpha\sqrt{g}}{q_2^{\ 4}}\dot{\epsilon}^{5}\right) \Omega_{1}+ \nonumber\\
&&  \frac{1}{2q_2^{\ 2}}
\left({\ddot{\epsilon}}^5 - \frac{q_2{\dot{q}}_2}{q_2^{\ 2}}{\dot{\epsilon^5}} + \frac{\alpha\sqrt{g}}{p_2^{\ 2}q_2^{\ 2}}\epsilon^3\right)\Omega_{2}+\epsilon^{3}\Omega_{3}+\dot{\epsilon}^{5}\Omega_{4}+\epsilon^{5}\Omega_{5}
\end{eqnarray}
that there are only two independent parameters  is consistent with the fact that there were two independent Lagrange multipliers in the expression of the total hamiltonian (\ref{totham}). Note, however, the distinction from the massive model discussed in the previous section. There we found only one independent gauge degree of freedom which was shown to have a one to one correspondence with the diffeomorhism invariance of the model. Clearly, the rigid relativistic particle is endowed with more general symmetries as is indicated by its gauge generator. 
	
	In order to unravel the meaning of the additional gauge symmetry we calculate the gauge variations of the dynamical variables, defined as $\delta{q} = \left\lbrace {q, G}\right\rbrace $. These are given by,
\begin{eqnarray}
\nonumber
\delta{q_{1}^{\mu}} &=& \epsilon^{3}q_{2}^{\mu}+ \dot{\epsilon}^{5}p_{2}^{\mu} + 2\epsilon^{5}p_{1}^{\mu}\\
\nonumber
\delta{q_{2}^{\mu}} &=& \left( {{\dot{\epsilon}}^3  + \frac{q_2{\dot{q}}_2}{q_2^{\ 2}}\epsilon^3 - \frac{\alpha\sqrt{g}}{q_2^{\ 4}}\dot{\epsilon}^5 } \right)  q_{2}^{\mu} + \left({\ddot{\epsilon}}^5 - \frac{q_2{\dot{q}}_2}{q_2^{\ 2}} {\dot{\epsilon^5}} + \frac{\alpha\sqrt{g}} {p_2^{\ 2}q_2^{\ 2}}\epsilon^3\right)p_{2}^{\mu} + \dot{\epsilon}^{5}p_{1}^{\mu}\\
\nonumber
\delta{p_{1}^{ \mu}} &=& 0\\
\nonumber
 \delta{p_{2}^{\mu}} &=& -\left( {{\dot{\epsilon}}^3  + \frac{q_2{\dot{q}}_2}{q_2^{\ 2}}\epsilon^3 - \frac{\alpha\sqrt{g}}{q_2^{\ 4}}\dot{\epsilon}^5} \right)  p_{2}^{\mu} \nonumber\\
 &&
 -\frac{p_{2}^{2}} {q_{2}^{2}} \left({\ddot{\epsilon}}^5 - \frac{q_2{\dot{q}}_2}{q_2^{\ 2}} {\dot{\epsilon^5}} + \frac{\alpha\sqrt{g}} {p_2^{\ 2}q_2^{\ 2}}\epsilon^3\right)q_{2}^{\mu}  -\epsilon^{3}p_{1}^{\mu}
%\nonumber
\label{dwinvariance}
\end{eqnarray}
%Let us consider that $\epsilon^{3}$ corresponds to reparametrization parameter and $\epsilon^{5}$ corresponds to w-symmetry which we shall show in the next section. Justification about $\epsilon^{3}$ is given bellow.
	
%Let us consider a reparametrization $\tau \rightarrow \tau + \Lambda $, where $\Lambda$ is an infinitesimal reparamtrization parameter then 
%	\begin{equation}
%\delta{x^{\mu}} = x^{\mu}(\tau - \Lambda) -  x^{\mu}(\tau ) = - \Lambda\dot{x}^{\mu}\label{repinva}
%\end{equation}
%Using our definition (\ref{newcoordinate}) this becomes
%\begin{equation}
%\delta{q_1^{\ \mu}} = - \Lambda q_2^{\ \mu}\label{repinv1}
%\end{equation}
\begin{figure}[th]
\centering
\includegraphics{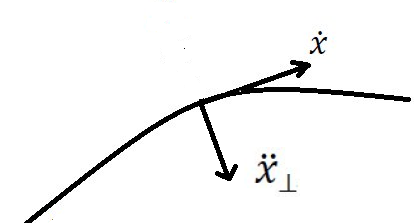}
\caption{	orthogonal frame attached with particle world trajectory}
\label{fig:figure1}
\end{figure}
Now it would be customary to identify the geometrical origin of these independent gauge transformations. It is well known  that  the most general deformation of the rigid particle trajectory may be resolved as
%%by an orthogonal coordinate system written as
\cite{ramos2}
\begin{equation}
\delta_{\beta, \eta}{x} = \beta(t)\dot{x} + \eta(t)\ddot{x}_{\perp}
\label{deformation}
\end{equation}  
where $\ddot{x}_{\perp}^{\mu} = \frac{l^{\mu}}{\dot{x}^{2}}$ is orthogonal to the tangent space mapped by $\dot{x}$(see figure.$1$). The coefficients $\beta$, $\eta $ are 
%some terms containing 
respectively the diffeo-morphism and w-morphism parameters. Now we rewrite the first equation of (\ref{dwinvariance}) as
\begin{equation}
\delta{q_{1}^{\mu}} = \epsilon^{3}{\dot{x}}^{\mu} + \dot{\epsilon}^{5}\frac{\alpha}{\sqrt{g}}\ddot{x}^{\mu}_{\perp} + 2\epsilon^{5}p_{1}^{\mu}
\label{dwvariance1}
\end{equation}
As we have noted earlier, the gauge variations (\ref{dwinvariance}) contain two independent gauge parameters $\epsilon^{3}$ and $\epsilon^{5}$. Let us first assume 
\begin{equation}
\epsilon^{3} \ne  0\ \ \  {\rm{and}}  \ \ \ \epsilon^{5} = 0
\label{D}
\end{equation}
 Substituting this condition in  (\ref{dwvariance1}) we get 
\begin{equation}
\delta{q_{1}^{\mu}} = \epsilon^{3}{\dot{x}}^{\mu} 
\end{equation}
Comparing the above with (\ref{deformation}) we can easily see that the gauge generator subject to the limiting condition (\ref{D}) generates diffeomorphism(with $\epsilon^{3}=\beta$) and the corresponding gauge symmetry is identified with diffeomorphism invariance.
% which is reconfirmed from (\ref{cmp3}).
  This identification may be confirmed by working out the variations of other phase space variables under reparametrization and comparing them with the gauge variation generated by $G$, subject to the same condition (\ref{D}).

       It is now clear that the gauge generator subject to the other extreme condition
\begin{equation}
\epsilon^{3} =  0\ \ \  {\rm{and}}  \ \ \ \epsilon^{5} \ne 0 
\label{W}
\end{equation}
 generates symmetry transformations 
other than diffeos. Substituting (\ref{W}) in (\ref{dwvariance1}) we now get
\begin{equation}
\delta{q_{1}^{\mu}} =  \dot{\epsilon}^{5}\frac{\alpha}{\sqrt{g}}\ddot{x}^{\mu}_{\perp} + 2\epsilon^{5}p_{1}^{\mu}
\label{dwvariance12}
\end{equation}
Looking back at the second equation of (\ref{regidmomenta})  we find that the linear term containing $p_{1}^{\mu}$ in the variation (\ref{dwvariance12}) can be neglected, as $p_{1}^{\mu}$ is a Lagrange multiplier. Comparing with (\ref{deformation}) we can then associate (\ref{dwvariance12}) with w - transformation, by identifying $ \eta = \dot{\epsilon}^{5}\frac{\alpha}{\sqrt{g}}$ . Thus the additional symmetry corresponding to (\ref{W}) is nothing but
 W - symmetry. It is now straightforward to show that the two different symmetries together satisfy the ${\rm{W}}_3$ algebra.
% again comparing (\ref{deformation}) where the linear term containing $p_{1}^{\mu}$ can be neglected from its own definition(second equation of (\ref{regidmomenta}) ) as it is a Lagrange multiplier, so arbitrary. We again can reconfirm the identification of $\epsilon^{5}$ as the gauge parameter for w-symmetry as they obeys the following algebra \cite{RR0, RR}. 
 Let us denote the  transformations of category 1 (diffeomorphisms) by the superscript `$D$' and the category 2  transformations by `$W$'. Detailed calculations on all the phase-space variables show that
 \begin{eqnarray}
 \nonumber
\left\lbrace { \delta^{(D)}_{\epsilon^{3}_{1}} , \delta^{(D)}_{\epsilon^{3}_{2}} }\right\rbrace  &=& \delta^{(D)}_{\epsilon^{3}}; \ \ \ with  \ \ \ \ \epsilon^{3}= \dot{\epsilon}_{1}^{3} \epsilon_{2}^{3} - \epsilon_{1}^{3} \dot{\epsilon}_{2}^{3}  \\
 \nonumber 
\left\lbrace { \delta^{(D)}_{\epsilon^{3}} , \delta^{(W)}_{\epsilon^{5}}}\right\rbrace  &=& \delta^{(W)}_{\epsilon^{\prime 5}}; \ \ \ with  \ \ \ \ \epsilon^{\prime5}= -\epsilon^{3}\dot{\epsilon}^{5}\\ 
 \nonumber 
\left\lbrace { \delta^{(W)}_{\epsilon^{5}_{1}} , \delta^{(W)}_{\epsilon^{5}_{2}} }\right\rbrace  &=& \nonumber \delta^{(W)}_{\epsilon^{5}}; \ \ \ with  \ \ \ \ \epsilon^{5}= \frac{p_{2}^{2}}{q_{2}^{2}}\left( \dot{\epsilon}_{2}^{5} \epsilon_{1}^{5} - \epsilon_{2}^{5} \dot{\epsilon}_{1}^{5}  \right) 
\end{eqnarray}
%This is  
which is nothing but the  $W_3$ algebra. 
%So the category 2 transformations correspond to the $W$ -- symmetry.
\section{(Anti-)BRST symmetries and $W_3$-algebra }
 In this section we construct the    nilpotent BRST  and anti-BRST  symmetries for the theory. For this purpose
we  need to fix a gauge before the quantization of the theory as the theory is gauge invariant and therefore has some redundant degrees of freedom. The
general gauge condition in this case is chosen as:
\begin{equation}
 F_1 [f(q)] =0, \label{gauge1}
\end{equation}
where $f(q)$ is a general function of all the generic variables $q$. Some 
explicit
examples of gauge  conditions corresponding to
relativistic particle models 
 are \cite{plyuschay2}.
 \begin{eqnarray}
q_1^0 -\tau =0,\ \  q_2^0 -1 =0,\ \  p_2^0 =0,\ \ q_2^2 =0.
 \end{eqnarray}
The general gauge condition (\ref{gauge1}) can be incorporated at a quantum level by adding the 
appropriate gauge-fixing term to  classical action. 
 
The linearised gauge-fixing term  using
Nakanishi-Lautrup auxiliary variable $  B ( q) $ is given by
\begin{equation}
 S_{gf} = \int d\tau    \left [ \frac{1}{2} B^2 +BF_1 [f(q) ] \right].
\end{equation}
To complete the effective theory we need a further Faddeev-Popov ghost term in the action. The ghost term in this  case is constructed as    
\begin{eqnarray}
  S_{gh} &=&  \int d\tau    \left [  \bar c s F_1 [f(q) ] \right],\nonumber\\
 &=& - \int d\tau    \left [    c \bar s F_1 [f(q) ] \right], 
\end{eqnarray}
where 
  $c$ and $ \bar{c}  $ are ghost and  anti-ghost variables.  
 Now the effective action can be written as
 \begin{equation}
 S_{eff} =S+S_{gf} +S_{gh}.\label{seff}
 \end{equation}
 The source free generating functional for this theory is defined as
 \begin{equation}
 Z[0] =\int {\cal D} q\ e^{iS_{eff}},\label{zfun}
 \end{equation}
 where ${\cal D} q$ is the path integral measure. 
 The nilpotent BRST symmetry of the  effective action in the case of relativistic 
 particle model with curvature 
is defined  by  replacing the infinitesimal reparametrisation parameter ($\Lambda$) to ghost 
variable $c$ in the gauge transformation given in equation (\ref{cmp1}) as 
 \begin{eqnarray}
&& s^{D} q_1^\mu =-c  q_2^\mu,\ \ \ s^{D} q_2^\mu =-\dot c   q_2^\mu -c  \dot q_2^\mu,\nonumber\\
&&  s^{D} c =0,\ \ \ s^{D} \bar c  =B, \ \ \ s^{D} B  =0,
\label{brs1}
 \end{eqnarray}
 where $c$, $\bar c $ and $B $ are ghost, anti-ghost and auxiliary variables  respectively 
 for relativistic  particle model with curvature.  This  BRST transformation, corresponding to
 gauge symmetry identified with
 the diffeomorphism invariance, leaves both  the effective action as well as 
 generating functional, invariant. 
 Similarly, we construct the anti-BRST symmetry
 transformation, where the roles of ghost and anti-ghosts are interchanged with some coefficients,
 as 
  \begin{eqnarray}
&&\bar s^{D} q_1^\mu =-\bar c  q_2^\mu,\ \ \ \bar  s^{D} q_2^\mu =-\dot {\bar c}   q_2^\mu -\bar c  \dot q_2^\mu,
\nonumber\\
&& \bar s^{D} 
\bar c =0,\ \ \ \bar s^{D}  c  =-B, \ \ \ \bar s^{D} B  =0.\label{antibrs1}
 \end{eqnarray}
 These transformations are nilpotent and absolutely anticommuting in nature i.e.
 \begin{equation}
 (s^{D})^{2}=0,\ \ ( {\bar s^D})^2=0,\ \ s^{D}\bar s^{D}+\bar s^{D} s^{D} =0.
 \end{equation}
 The above (anti-)BRST transformations are  valid for both the models. On the other hand, the nilpotent BRST and anti-BRST symmetry transformations,  identified with
 $W$-symmetry (with $\epsilon^3 =0$) in (\ref{dwinvariance}), for relativistic massless particle model with rigidity  only, are constructed as  
  \begin{eqnarray}
 s^{W} q_1^\mu &=& \dot \eta p_2^\mu +2 \eta p_1^\mu,\nonumber\\
 s^{W} {q_{2}^{\mu}} &=&  - \frac{\alpha\sqrt{g}}{q_2^{\ 4}}\dot \eta   q_{2}^{\mu} +\left({\ddot \eta}  - \frac{q_2{\dot{q}}_2}{q_2^{\ 2}} {\dot \eta} \right)p_{2}^{\mu} + \dot \eta p_{1}^{\mu},
 \nonumber\\
 s^{W} p_1^\mu &=&0, \ \ s^{W} p_2^\mu =    \frac{\alpha\sqrt{g}}{q_2^{\ 4}}\dot \eta  p_{2}^{\mu} -\frac{p_{2}^{2}} {q_{2}^{2}}\left({\ddot \eta}  - \frac{q_2{\dot{q}}_2}{q_2^{\ 2}} {\dot \eta} \right)q_{2}^{\mu},\nonumber\\
      s^{W} \eta &=&0,\ \ s^{W} \bar \eta =B, \ \ s^{W} B =0,\label{brs2}
 \end{eqnarray} and
  \begin{eqnarray}
\bar s^{W} q_1^\mu &=& \dot {\bar \eta} p_2^\mu +2 {\bar \eta} p_1^\mu,\nonumber\\
 \bar s^{W} {q_{2}^{\mu}} &=&  - \frac{\alpha\sqrt{g}}{q_2^{\ 4}}\dot {\bar \eta}   q_{2}^{\mu} +\left({\ddot{\bar\eta}}  - \frac{q_2{\dot{q}}_2}{q_2^{\ 2}} {\dot {\bar\eta}} \right)p_{2}^{\mu} + \dot {\bar \eta} p_{1}^{\mu},
 \nonumber\\
\bar s^{W} p_1^\mu &=&0, \ \ \bar s^{W} p_2^\mu =    \frac{\alpha\sqrt{g}}{q_2^{\ 4}}\dot{\bar \eta}  p_{2}^{\mu} -\frac{p_{2}^{2}} {q_{2}^{2}}\left({\ddot {\bar \eta}}  - \frac{q_2{\dot{q}}_2}{q_2^{\ 2}} {\dot{\bar \eta}} \right)q_{2}^{\mu},\nonumber\\
  \bar    s^{W} {\bar \eta}&=&0,\ \ \bar s^{W} \eta =-B, \ \ \bar s^{W} B =0,
  \label{antibrs2}
 \end{eqnarray}
  where $\eta, \bar \eta$ and $B $ are ghost, anti-ghost and auxiliary variables,  respectively, 
 for  relativistic massless particle model with rigidity. 
 
 Here we observe interestingly  that the BRST symmetry transformations of all variables 
(excluding the anti-ghost  variable) given in equations (\ref{brs1}) and (\ref{brs2}) also satisfy the $W_3$-algebra as
  \begin{eqnarray}
 \nonumber
\left[ {s_{c_1}^{D} } , s_{c_2}^{D}  \right]   &=& s_{c_3}^{D};\ \ \ \mbox{with}  \ \ \ \ c_3= c_2 \dot c_1 - \dot c_2 c_1\\ 
 \nonumber
\left[ { s_c^{D}, s_\eta^{W}}\right]     &=& s_{\eta'}^{W};\ \ \ \mbox{with}  \ \ \ \ \eta' =\dot \eta c\\ 
 \nonumber
\left[ { s_{\eta_1}^{W}, s_{\eta_2}^{W} }\right]     &=& \nonumber s_{\eta_3}^{W};
\ \ \ \mbox{with}  \ \ \ \ \eta_3= \frac{p_{2}^{2}}{q_{2}^{2}}\left( \eta_2\dot\eta_1 -\eta_1\dot \eta_2 \right),
\end{eqnarray} 
and the anti-BRST symmetry transformations of all variables 
(excluding  ghost  variable) given in equations (\ref{antibrs1}) and (\ref{antibrs2}) also satisfy the $W_3$-algebra as
  \begin{eqnarray}
 \nonumber
\left[ {\bar s_{\bar c_1}^{D} } , \bar s_{\bar c_2}^{D}  \right]    &=& \bar s_{\bar c_3}^{D};\ \ \ \mbox{with}  \ \ \ \ \bar c_3= \bar c_2 \dot {\bar c}_1 - \dot {\bar c}_2 \bar c_1 \\
 \nonumber
\left[ {\bar s_{\bar c}^{D}, \bar s_{\bar \eta}^{W}}\right]     &=&\bar  s_{\bar \eta'}^{W};\ \ \ \mbox{with}  \ \ \ \ \bar\eta' =\dot {\bar\eta}\bar c\\ 
 \nonumber
\left[ {\bar s_{\bar\eta_1}^{W}; \bar s_{\bar\eta_2}^{W} }\right]     &=& \nonumber \bar s_{\bar\eta_3}^{W};\ \ \ \mbox{with}  \ \ \ \ \bar\eta_3= \frac{p_{2}^{2}}{q_{2}^{2}}\left(\bar \eta_2\dot{\bar\eta}_1 -\bar\eta_1\dot {\bar\eta}_2 \right).
\end{eqnarray} 
This completes our analysis of the  connection  between the (anti-)BRST symmetries  and  $W_3$-algebra.
 \section{FCBRST formulation for higher derivative theory}
In this section we investigate the finite coordinate-dependent BRST (FCBRST) formulation for general higher derivative theory. To do so, 
 we first define the infinitesimal BRST symmetry transformation 
 with Grassmannian constant parameter $\delta\rho$ as

 \begin{equation}
 \delta_b q =s q\ \delta\rho,\label{brst}
 \end{equation}
 where $s q$ is the BRST variation of generic variables $q$  in the higher derivative theories. 
The properties of the usual BRST transformation in equation (\ref{brst})  do not depend on whether 
the parameter $\delta\rho$  is (i) finite or infinitesimal, (ii) variable-dependent or not, as long 
as it is anticommuting and global in nature. These observations give us a freedom to 
generalize the BRST transformation by making the parameter $\delta\rho$ finite and coordinate-dependent without
 affecting its properties. We call such generalized BRST transformation in quantum mechanical systems
  as FCBRST transformation.
 In the field theory such generalization is known as FFBRST  transformation \cite{brst}. 
 Here we adopt a similar technique to generalize the BRST transformation in quantum mechanical theory.
 We start 
by making the  infinitesimal parameter coordinate-dependent with introduction of an arbitrary parameter $\kappa\ 
(0\leq \kappa\leq 1)$.
We allow the generalized coordinates, $q( \kappa)$, to depend on  $\kappa$  in such a way that $q( \kappa =0)=q $ and $q( \kappa 
=1)=q^\prime $, the transformed coordinate.

The usual infinitesimal BRST transformation, thus can be written generically as 
\begin{equation}
{dq( \kappa)}=s  [q  ]\Theta^\prime [q ( \kappa ) ]{d\kappa},
\label{diff}
\end{equation}
where the $\Theta^\prime [q ( \kappa ) ]{d\kappa}$ is the infinitesimal but coordinate-dependent parameter.
The FCBRST transformation with the finite coordinate-dependent parameter then can be 
constructed by integrating such infinitesimal transformation from $\kappa =0$ to $\kappa= 1$, to obtain
\cite{brst}
\begin{equation}
q^\prime\equiv q (\kappa =1)=q( \kappa=0)+s  (q  )\Theta[q  ],
\label{kdep}
\end{equation}
where 
\begin{equation}
\Theta[q ]=\int_0^1 d\kappa^\prime\Theta^\prime [q( \kappa^\prime)],\label{fin}
\end{equation}
 is the finite coordinate-dependent parameter. 

Such a generalized BRST transformation with finite coordinate-dependent
 parameter is the symmetry  of the effective action in equation (\ref{seff}). However, the 
path integral measure in equation (\ref{zfun}) is not invariant under such transformation as the 
BRST parameter is finite in nature. 
The Jacobian of the path integral measure for such transformations is then evaluated for some 
particular choices of the finite coordinate-dependent parameter, $\Theta[q(x)]$, as
\begin{eqnarray}
{\cal D}q^\prime &=&J(
\kappa) {\cal D}q(\kappa).
\end{eqnarray}
The Jacobian, $J(\kappa )$ can be replaced (within the functional integral) as
\begin{equation}
J(\kappa )\rightarrow \exp[iS_1[q( \kappa) ]],
\end{equation}
 iff the following condition is satisfied \cite{brst} 
\begin{eqnarray}
 \int {\cal{D}}q   \;  \left [ \frac{1}{J}\frac{dJ}{d\kappa}-i\frac
{dS_1[q (x,\kappa )]}{d\kappa}\right ] 
 \exp{[i(S_{eff}+S_1)]}=0, \label{mcond}
\end{eqnarray}
where $ S_1[q ]$ is local functional of variables such that at $\kappa =0$ it must vanish.

The infinitesimal change in the $J(\kappa)$ is written as \cite{brst},
\begin{equation}
\frac{1}{J}\frac{dJ}{d\kappa}=-\int d\tau \left [\pm s q ( \kappa )\frac{
\partial\Theta^\prime [q ( \kappa )]}{\partial q ( \kappa )}\right],\label{jac}
\end{equation}
where $\pm$ sign refers to whether $q$ is a bosonic or a fermionic variable.

Thus, the FCBRST transformation with appropriate $\Theta$, changes the
effective action $S_{eff} $ to a  new effective 
action  $S_{eff}+S_1(\kappa=1)$ within the functional integration.  
\section{Connecting different gauges in relativistic particle models}
Here we will exploit the general FCBRST formulation developed in the previous section to connect the path integral of relativistic particle models with different gauge conditions. The FCBRST transformations $(f_b)$ for the relativistic particle model with 
curvature are constructed   as follows:
 \begin{eqnarray}
 \nonumber
 f_b q_1^\mu =-c  q_2^\mu \Theta[q],\ \  f_b q_2^\mu =(-\dot c   q_2^\mu -c  \dot q_2^\mu)\Theta[q],\ \  \\  f_b c =0,\ \  f_b\bar c  =B \Theta[q], \ \  f_b B  =0,\label{ffbrst}
 \end{eqnarray}
 where $\Theta[q]$ is an arbitrary finite coordinate-dependent parameter.  
 Now, we show how two different gauges (say $F_1(q) = 0$ and $F_2(q) = 0$) in the relativistic particle model may be connected by such transformations. For this purpose, let us choose the following infinitesimal coordinate dependent parameter (through equation (\ref{fin}))
\begin{equation}
\Theta'[q] =-i\int d\tau\ \bar c (F_1 -F_2 ).
\end{equation}
Let us first calculate the  infinitesimal change in the Jacobian $J(\kappa) $ for above $\Theta'[q]$  using the relation (\ref{jac})  as
\begin{eqnarray}
 \frac{1}{J}\frac{dJ}{d\kappa} &=&i\int d\tau [-B(F_1 -F_2) +s  (F_1 -F_2 ) \bar c ],\nonumber\\
  &=&-i\int d\tau [ B(F_1 -F_2) + \bar c\ s (F_1 -F_2 ) ].\label{jac1}
\end{eqnarray}
To express the Jacobian as  $e^{iS_1}$ \cite{brst}, we take the ansatz, 
\begin{eqnarray}
S_1 [   \kappa ]=\int d\tau [\zeta_1(\kappa )  B F_1  +\zeta_2 (\kappa ) BF_2    +\zeta_3(\kappa ) \bar c\ s  F_1 +
\zeta_4(\kappa )\bar c\ s  F_2],\label{s1}
\end{eqnarray}
where $\zeta_i(\kappa ) (i =1,...4)$ are constant parameters satisfying the boundary conditions
\begin{equation}
\zeta_i(\kappa =0 ) =0.\label{con}
\end{equation} 
 
To satisfy the crucial condition  (\ref{mcond}), we calculate the
infinitesimal change in $S_1$ with respect to $\kappa$ using the relation (\ref{diff}) as
\begin{eqnarray}
\frac{ dS_1 [ q, \kappa ]}{d\kappa} &=&\int d\tau [\zeta_1'  B F_1  +\zeta_2' BF_2    +\zeta_3' \bar c\ s  F_1 +
\zeta_4'\bar c\ s  F_2\nonumber\\
& +&(\zeta_1 -\zeta_2 )B(s  F_1) \Theta' +(\zeta_2 -\zeta_4 )B  (s  F_2) \Theta'],\label{dis}
\end{eqnarray}
where prime denotes the differentiation with respect to $\kappa$.
Exploiting equations  (\ref{jac1}) and (\ref{dis}), the condition  (\ref{mcond}) simplifies to, 
\begin{eqnarray}
&&\int {\cal{D}}q   \;  \left [ (\zeta_1' +1)  BF_1  +(\zeta_2' -1) BF_2    +(\zeta_3' +1) \bar c\ s  F_1 +
(\zeta_4' -1)\bar c\ s  F_2\right.\nonumber\\
& &+\left.(\zeta_1 -\zeta_3 )B (s  F_1) \Theta' +(\zeta_2 -\zeta_4 )B (s  F_2) \Theta'\right ] 
e^{ i(S_{eff}+S_1) }=0.  
\end{eqnarray}
The comparison of coefficients from the terms of the above equation   
gives the following constraints on the parameters $\zeta_i$
\begin{eqnarray}
&&\zeta_1' +1 = 0,\ \ \zeta_2' -1 =0,\ \ \zeta_3' +1 =0,\ \ \zeta_4' -1 =0,\nonumber\\
&&\zeta_1 -\zeta_3=0,\ \ \zeta_2 -\zeta_4 =0.
\end{eqnarray}
The solutions of the above equations satisfying the boundary conditions (\ref{con}) are
\begin{eqnarray}
 \zeta_1 =-\kappa,\ \ \zeta_2 =\kappa,\ \ \zeta_3 =-\kappa,\ \ \zeta_4 =\kappa.
\end{eqnarray}
With these values of $\zeta_i$  the expression of $S_1[\kappa]$ given in equation (\ref{s1}) becomes
\begin{eqnarray}
S_1 [\kappa ]=\int d\tau [ -\kappa   B F_1  + \kappa  BF_2    - \kappa  \bar c\ s F_1 +
 \kappa  \bar c\ s  F_2],
\end{eqnarray}
which vanishes at $\kappa=0$. 
Now, by adding   $S_1(\kappa =1)$ to the effective action ($S_{eff}$) given in equation (\ref{seff}) we get
\begin{eqnarray}
S_{eff}+S_1(\kappa =1) = S+ \int d\tau \left[ \frac{1}{2} B^2 +BF_2 [f(q) ] +\bar c s F_2 [f(q) ] \right],
\end{eqnarray}
which is nothing but the effective action for relativistic particle models satisfying the
different gauge condition $F_2[f(q)] =0$.
Thus, under FCBRST transformation, the generating functional of higher derivative models
changes from one gauge condition ($F_1 [f(q) ]=0$) to another gauge ($F_2 [f(q) ]=0$) as 
\begin{eqnarray}
\int d\tau e^{iS_{eff}}\stackrel{FCBRST} { ----\longrightarrow}\left( \int d\tau e^{i[S_{eff}+S_1(\kappa =1)]} 
\right).
\end{eqnarray}
We end  this section by noting that the FCBRST transformation with appropriate 
finite coordinate-dependent parameter is able to connect two different (arbitrary) gauges of the relativistic particle model.  
\section{Discussions}
In this chapter we have taken the relativistic particle model with curvature. Both the massive and massless versions were considered in the first order formalism to explore their gauge symmetries. We have used an extended version of the algorithm of Banerjee et. al.\cite{BRR} to construct the gauge generators. The apparent mismatch in counting the gauge degrees of freedom fromthe no, of PFCs \cite{nesterenko} has been explained. Also, from the methodology we have shown why the massless version has two gauge degrees of freedom. Moreover, we have extracted both these symmetries using the gauge generator. One of the symmetries was identified with the usual diffeomorphism invariance whereas the other has been shown to be $W_3$ symmetry. In this connection, it may be noted that this result was earlier obtained only at equations of motion level. The BRST symmetries were  explored for both the massive and massless models. The massless version in this case exhibited the $W_3$ algebra also for the (anti-)BRST. 

	\chapter{Extended Maxwell-Chern-Simons model} %DONE
  	In the previous chapter we have demonstrated the application of our method of canonical analysis for particle models. In this chapter, we extend the application of our method to field theories. We consider the action of the extended-Maxwell-Chern-Simon's (EMCS) in $2+1$ dimension which is given by

\begin{equation}
S = \int d^{3}x  \left({ -\frac{1}{4}F_{\mu\nu}F^{\mu\nu} + \frac{g}{2} \epsilon^{ \alpha \beta \gamma }(\partial^{\rho}\partial_{\rho} A_{\alpha})(\partial_{\beta}A_{\gamma})}\right) 
\label{mcslag}
\end{equation}
where $F_{\mu \nu} =  \partial_{\mu}A_{\nu} - \partial_{\nu}A_{\mu}$. Note that the model is endowed with the $U(1)$ symmetry
\begin{equation}
  	A_\mu \rightarrow A_\mu + \partial_{\mu} \Lambda. \label{gengaugeinv}
\end{equation}
and is an extension of the usual Maxwell Chern Simons theory \cite{deser2}. The second term in the action contains higher derivative terms and may be viewed as the higher derivative version of the Chern-Simons piece. The model actually derive its physical significance from the fact that it give rise to the topological gravity in the higher dimensions. The fields in this case are the vector potential $A_\mu$ and it's higher time derivatives.  The complete Hamiltonian analysis is done in the first order formalism. It is shown that there is only one primary first class constraint along with two secondary first class constraints. Consequently, the gauge generator has three gauge parameters. The exact mapping of this independent gauge symmetry with the $U(1)$ symmetry has been established. 
  	
  	   \section{Hamiltonian analysis of the model in the equivalent first order formalism}
 In our approach the time derivative of the field $ A_{\mu} $ will be considered as additional fields. Thus, it will be convenient to expand the Lagrangian of the model (\ref{mcslag})
in space and time parts. Using the mostly positive metric ($ \eta_{\mu\nu} $ = -, +, +) the Lagrangian is written as
 \begin{eqnarray}
 \nonumber
\mathcal{L} =  &=& \frac{1}{2} ( \dot{A}_{i}^{2} + (\partial_{i}A_{0})^{2} - (\partial_{i}A_{j})^{2} -2 \dot{A}_{i}\partial_{i}A_{0} + \partial_{i}A_{j}\partial_{j}A_{i} ) + \frac{g}{2} \epsilon_{ i j } ( -\ddot{A}_{0}     \\ \nonumber
&& +\nabla^{2} A_{0} )\partial_{i}A_{j}- \frac{g}{2} \epsilon_{ i j } ( -\ddot{A}_{i} + \nabla^{2} A_{i} )\dot{A}_{j}+ \\
&&   \frac{g}{2} \epsilon_{ i j } ( -\ddot{A}_{i} +  \nabla^{2} A_{i} )\partial_{j}A_{0} 
\label{mcslag1}
\end{eqnarray}
Here the fields are referred to their covariant components and dot represents derivative with respect to time. Note that the effect of the relativistic metric $\eta_{\mu\nu}$ has been taken care of explicitly in writing (\ref{mcslag1}). In the following subscripts from the middle of the Greek alphabet $ \mu $, $ \nu $ assume the values 0,1, 2 and those from the middle of the Latin alphabet $i$, $j$ take values 1 and 2. In any case, they just label the components and no further reference to the relativistic metric is implied.\\

 To analyze the model in the equivalent first order formalism we define the new coordinates
\begin{equation}
\xi_{1\mu} = A_{\mu}  \ \ \ {\rm{and}} \ \ \xi_{2\mu} = \dot{A}_{\mu}
\label{fielddef}
\end{equation}
This immediately imposes the constraint 
\begin{equation}
\xi_{2\mu} = \dot{\xi}_{1\mu} 
\label{lagconst}
\end{equation}
 %Enforcing this constraint by the Lagrange multiplier $ \xi_{0\mu} $ 
 The equivalent first order Lagrangian is obtained from (\ref{mcslag1}) using the definitions (\ref{fielddef}) as 

%Decomposing into space and time parts and taking $\xi_{2\mu} = \dot{A}_{\mu}$ we get the auxilliary lagrangian. Here , $\epsilon^{012} = \epsilon^{12} = 1$ and the extended phase space is spanned by ($\xi_{1\mu}$,$\Pi_{1\mu}$),($\xi_{2\mu}$,$\Pi_{2\mu}$), ($\xi_{0\mu}$,$\Pi_{0\mu}$) .
\begin{eqnarray}
\nonumber
\mathcal{L}^{\prime} &=& \frac{1}{2} ( \xi_{2i}\xi_{2i} + \partial_{i}\xi_{10}\partial_{i}\xi_{10} - \partial_{i}\xi_{1j}\partial_{i}\xi_{1j} -2 \xi_{2j}\partial_{j}\xi_{10} + \partial_{i}\xi_{1j}\partial_{j}\xi_{1i} ) \\ \nonumber
&& + \frac{g}{2} \epsilon_{ i j } ( -\dot{\xi}_{20} +\nabla^{2} \xi_{10} )\partial_{i}\xi_{1j}- \frac{g}{2} \epsilon_{ i j } ( -\dot{\xi}_{2i} + \nabla^{2} \xi_{1i} )\xi_{2j}  \\
&& +  \frac{g}{2} \epsilon_{ i j } ( -\dot{\xi}_{2i} + \nabla^{2} \xi_{1i} )\partial_{j}\xi_{10} + \xi_{0\mu} (\xi_{2\mu} - \dot{\xi}_{1\mu})
\label{auxilag}
\end{eqnarray}
where the constraint (\ref{lagconst}) is enforced by the Lagrange multiplier $ \xi_{0\mu} $.
Henceforth, $ \xi_{0\mu} $ will be considered as independent fields.\\

  To proceed with the canonical analysis we define the momenta $ \Pi_{0\mu} $,$ \Pi_{1\mu} $, $ \Pi_{2\mu} $ conjugate to the fields $ \xi_{0\mu} $, $ \xi_{1\mu} $, $ \xi_{2\mu} $ respectively in the usual way : 
\begin{eqnarray}
\Pi_{\alpha\mu} &=& \frac{\partial \mathcal{L}^{\prime}}{\partial \dot{\xi}_{\alpha\mu}} \ \ \ \ ; \ \ \ \ \ \ \ \ \ \ \  \alpha = 0, 1, 2
\end{eqnarray}  
    As a result the following primary constraints emerge: 
\begin{eqnarray}
\nonumber\\
\Phi_{0\mu} &=& \Pi_{0\mu}  \approx 0
\nonumber\\
\Phi_{1\mu} &=& \Pi_{1\mu} + \xi_{0\mu} \approx 0
\nonumber\\
\Phi_{20} &=& \Pi_{20} +  \frac{g}{2} \epsilon_{  i j }\partial_{i}\xi_{1j} \approx 0
\nonumber\\
\Phi_{2i} &=& \Pi_{2i} -  \frac{g}{2} \epsilon_{ i j }\xi_{2j} + \frac{g}{2} \epsilon_{ i j }\partial_{j}\xi_{10} \approx 0
%\nonumber\\
%\Phi_{3}^{\mu} &=& h^{\mu} \approx 0
\end{eqnarray}
The basic  Poisson brackets are 
\begin{eqnarray}
\nonumber\\
\lbrace{ \xi_{\alpha\mu}(\textbf{x}), \Pi_{\beta \nu}(\textbf{x}^{\prime}) }\rbrace &=& \delta_{\alpha\beta}\delta_{\mu\nu} \delta^{2}( \textbf{x} - \textbf{x}^{\prime} )
\label{basicpoisbarc}
\end{eqnarray}
where $ \alpha $, $ \beta $  = 0, 1, 2. This leads to the following  algebra of the primary constraint,
\begin{eqnarray}
\nonumber\\
\lbrace{ \Phi_{10}(\textbf{x}), \Phi_{2i}(\textbf{x}^{\prime}) }\rbrace &=&   -\frac{g}{2} \epsilon_{ij} \partial^{\prime }_{j}\delta^{2}( \textbf{x}-\textbf{x}^{\prime} )
\nonumber\\
\left\lbrace {\Phi_{1i}(\textbf{x}), \Phi_{20}(\textbf{x}^{\prime})}\right\rbrace &=& \frac{g}{2} \epsilon_{ij}\partial_{j}^{\prime} \delta^{2}(\textbf{x} - \textbf{x}^{\prime}) 
\nonumber\\
\lbrace{ \Phi_{2 i}(\textbf{x}), \Phi_{2j}(\textbf{x}^{\prime}) }\rbrace &=& - g \epsilon_{ i j}  \delta^{2}( \textbf{x}-\textbf{x}^{\prime} )
\nonumber\\
\lbrace{ \Phi_{0\mu}(\textbf{x}), \Phi_{1\nu}(\textbf{x}^{\prime}) }\rbrace &=& - \delta_{\mu\nu} \delta^{2}( \textbf{x}-\textbf{x}^{\prime} )
\label{constalgebra}
\end{eqnarray}
All other brackets between the constraints vanish. Apparently all the primary constraints have non trivial brackets among themselves. However, we can make the following linear combinations of the primary constraints 
\begin{eqnarray}
\nonumber
\Phi^{\prime}_{20} = \Phi_{20} + \frac{g}{2}\epsilon_{ij}\partial_{i}\Phi_{0j}  \approx 0
\nonumber\\
\Phi^{\prime}_{2i} = \Phi_{2i} + \frac{g}{2}\epsilon_{ij}\partial_{j}\Phi_{00}  \approx 0
\end{eqnarray}
%If we replace
Using the algebra of the primary constraints (\ref{constalgebra}) we find that the constraint algebra simplifies to   
\begin{eqnarray}
%\nonumber\\
%\lbrace{ \Phi_{0\mu}(x), \Phi_{2\nu}(x^{\prime}) }\rbrace &=&   -\delta_{\mu\nu}\delta^2( x-x^{\prime} )
\nonumber\\
\lbrace{ \Phi_{0\mu}(\textbf{x}), \Phi_{1\nu}(\textbf{x}^{\prime}) }\rbrace &=& - \delta_{\mu\nu} \delta^{2}( \textbf{x}-\textbf{x}^{\prime} )
\nonumber\\
\lbrace{ \Phi_{2 i}^{\prime}(\textbf{x}), \Phi_{2j}^{\prime}(\textbf{x}^{\prime}) }\rbrace &=& - g \epsilon_{ i j}  \delta^{2}( \textbf{x}-\textbf{x}^{\prime} )
\end{eqnarray}
It will thus be convenient to replace  the original set of primary constraints  \{$ \Phi_{0\mu} $ , $ \Phi_{10} $, $ \Phi_{1i} $, $ \Phi_{20} $, $ \Phi_{2i} $\}  by \{$ \Phi_{0\mu} $, $ \Phi_{10} $, $ \Phi_{1i} $, $ \Phi_{20}^{\prime} $, $\Phi_{2i}^{\prime} $\}. % the constraint algebra simplifies to
Explicitly, the new set of primary constraints are 
\begin{eqnarray}
\nonumber\\
\Phi_{0\mu} &=& \Pi_{0\mu}  \approx 0
\nonumber\\
\Phi_{1\mu} &=& \Pi_{1\mu} + \xi_{0\mu} \approx 0
\nonumber\\
\Phi_{20}^{\prime} &=&  \Pi_{20}+ \frac{g}{2} \epsilon_{  i j }\partial_{i}\xi_{1j} + \frac{g}{2} \epsilon_{  i j }\partial_{i}\Pi_{0j} \approx 0
\nonumber\\
\Phi_{2i}^{\prime} &=& \Pi_{2i} -  \frac{g}{2} \epsilon_{ i j }\xi_{2j} + \frac{g}{2} \epsilon_{ i j }\partial_{j}\xi_{10} + \frac{g}{2} \epsilon_{  i j }\partial_{j}\Pi_{00} \approx 0
%\nonumber\\
%\Phi_{3}^{\mu} &=& h^{\mu} \approx 0
\label{primarycons}
\end{eqnarray}

%\begin{equation}
%\Phi^{\prime}_{2i} = \Phi_{2i} + \frac{g}{2}\epsilon_{ij}\partial_{j}\Phi_{00}  \approx 0
%\end{equation}
%So the modified brackets are
%\begin{eqnarray}
%\nonumber\\
%\lbrace{ \Phi_{2 i}^{\prime}(x), \Phi_{2j}^{\prime}(x^{\prime}) }\rbrace &=& - g \epsilon_{ i j}  \delta^{2}( x-x^{\prime} )
%\nonumber\\
%\lbrace{ \Phi_{0\mu}(x), \Phi_{1\nu}(x^{\prime}) }\rbrace &=& - \delta_{\mu\nu} \delta^{2}( x-x^{\prime} )
%\nonumber\\
%\end{eqnarray}
 The canonical Hamiltonian is obtained by Legendre transformation as
\begin{eqnarray}
\nonumber\\
H_{can} &=& \int \mathcal{H}_{can} d^{2}\textbf{x}
\end{eqnarray}
where $\mathcal{H}_{can}$ is the canonical Hamiltonian density, given by:
\begin{eqnarray}
\nonumber
\mathcal{H}_{can}&=& -\frac{1}{2} ( \xi_{2i}\xi_{2i} + \partial_{i}\xi_{10}\partial_{i}\xi_{10} - \partial_{i}\xi_{1j}\partial_{i}\xi_{1j} -2 \xi_{2j}\partial_{j}\xi_{10} +  \partial_{i}\xi_{1j}\partial_{j}\xi_{1i} )-  \\
&& \frac{g}{2} \epsilon_{ i j } \nabla^{2} \xi_{10} \partial_{i}\xi_{1j}+ \frac{g}{2} \epsilon_{ i j } \nabla^{2} \xi_{1i} .\xi_{2j}- \frac{g}{2} \epsilon_{ i j } \nabla^{2} \xi_{1i} \partial_{j}\xi_{10} - \xi_{0\mu} \xi_{2\mu} 
\end{eqnarray}
The total Hamiltonian is
\begin{eqnarray}
H_{T} &=& \int d^{2}\textbf{x}( \mathcal{H}_{can} + \Lambda_{0\mu} \Phi_{0\mu} + \Lambda_{1\mu} \Phi_{1\mu} + \Lambda_{20} \Phi_{20}^{\prime} +\Lambda_{2i} \Phi_{2i}^{\prime}) 
\end{eqnarray}
The multipliers $ \Lambda_{0\mu} $, $ \Lambda_{1\mu} $, and $ \Lambda_{2\mu} $ are arbitrary at this stage.\\

The primary constraints (\ref{primarycons}) should be conserved in time, i.e. there Poisson bracket with $ H_{T} $ should vanish. Conserving $ \Phi_{0\mu}$, $ \Phi_{1\mu}$, $ \Phi_{2i}^{\prime}$ in time the following multipliers are fixed, \\
%constraint evolutions
\begin{eqnarray}
  \nonumber
  \Lambda_{00}  &=& \nabla^{2}\xi_{10} - \partial_{i}\xi_{2i} - g\epsilon_{ij}\nabla^{2}\partial_{i}\xi_{1j}
\nonumber \\
 \Lambda_{0i} &=& -\nabla^{2}\xi_{1i} + \partial_{i}\partial_{j}\xi_{1j} - g\epsilon_{ij}\nabla^{2}\partial_{j}\xi_{10} + \frac{g}{2}\epsilon_{ij}\xi_{2j}
 \nonumber\\
  \Lambda_{1\mu} &=& \xi_{2\mu}
 \nonumber\\
 \Lambda_{2i} &=& \frac{1}{2}( \nabla^{2}\xi_{1i} + \partial_{i}\xi_{20} ) + \frac{1}{g}\epsilon_{ij}( \partial_{j}\xi_{10} - \xi_{0j} - \xi_{2j} ) 
%\Lambda_{2i} = \frac{1}{g}\left({ \epsilon_{ij}\partial_{j}\xi_{10} -\epsilon_{ij}\xi_{2j} + \frac{g}{2}\nabla^{2}\xi_{1i}  -\epsilon_{ij}\xi_{0j} - \frac{g}{2}\partial_{i}\xi_{20}} \right)\\
%\dot{\Psi}_{2} &=& 0 \Rightarrow 0 = 0
\label{multilpliers}
\end{eqnarray}
Only $ \Lambda_{20} $ remains arbitrary. Substituting these in the total Hamiltonian we find that it contains only one arbitrary multiplier $ \Lambda_{20} $. This shows that there is only one gauge degree of freedom, a result consistent with the gauge transformation of the original field $A_\mu$.\\

 Conserving $ \Phi_{20}^{\prime} $ in time,  a secondary constraint emerges.
\begin{equation}
\Psi_{1} = \xi_{00} + \frac{g}{2}\epsilon_{ij}\partial_{i}\xi_{2j} \approx 0
\end{equation}
 From $ \dot{\Psi}_{1} = 0 $ we get 
 \begin{equation}
\left\lbrace{\Psi_{1}, H_{T}} \right\rbrace = 0
\end{equation} 
A straightforward calculation gives 
\begin{equation}
\Lambda_{00} - g \epsilon_{ij} \partial_{j} \Lambda_{2i} =0
\end{equation}
Using the values of $ \Lambda_{00} $ and $ \Lambda_{2i} $ from (\ref{multilpliers}) and simplifying we get
\begin{equation}
\partial_{i}\xi_{0i} - \frac{g}{2} \epsilon_{ij} \nabla^{2}\partial_{i}\xi_{1j} = 0
\end{equation}
  %This constraint must also hold in all time. From its conservation we get another secondary constraint.
%Nonzero brackets for the secondary constraints
%\begin{eqnarray}
%\nonumber\\
%\left\lbrace{\Psi_{1}(x), \Phi_{0\mu}(x^{\prime})} \right\rbrace &=& \delta_{0\mu} \delta^{2}(x-x^{\prime})
%\nonumber\\
 %\left\lbrace{\Psi_{1}(\textbf{x}), \Phi_{2i}^{\prime}(\textbf{x}^{\prime})} \right\rbrace &=& -g\epsilon_{ij}\partial_{j}\delta^{2}(\textbf{x}-\textbf{x}^{\prime})
%\end{eqnarray}
%Evolution of the secondary constraint results a tertiary constraint\\
which is a new secondary constraint
\begin{equation}
 \Psi_{2} = \partial_{i}\xi_{0i} - \frac{g}{2} \epsilon_{ij} \nabla^{2}\partial_{i}\xi_{1j} \approx 0
\end{equation}
The condition $ \left\lbrace {\Psi_{2}, H_{T}}\right\rbrace = 0 $ gives 
\begin{equation}
\partial_{i} \Lambda_{0i} - \frac{g}{2} \epsilon_{ij} \nabla^{2} \partial_{i} \Lambda_{1j} = 0
\end{equation}
Substituting the values of $ \Lambda_{0i}$ and $  \Lambda_{1j}$ the above equation reduces to the form 0 = 0. Hence the iterative process stops here giving no further constraints. The primary constraints of the theory are \{$ \Phi_{0\mu} $, $ \Phi_{1\mu} $, $ \Phi_{20}^{\prime} $, $ \Phi_{2i}^{\prime} $\} while the secondary constraints are $ \Psi_{1} $ and $ \Psi_{2} $.\\

 Using the Poisson brackets (\ref{basicpoisbarc}) the complete algebra of constraints can be worked out  as \\
\begin{eqnarray}
\nonumber\\
\left\lbrace{\Phi_{0\mu}(\textbf{x}), \Phi_{1\nu}(\textbf{x}^{\prime})} \right\rbrace &=& -\delta_{\mu \nu}\delta^{2}(\textbf{x}-\textbf{x}^{\prime})
\nonumber\\
\left\lbrace{\Phi_{2i}^{\prime}(\textbf{x}), \Phi_{2j}^{\prime}(\textbf{x}^{\prime})} \right\rbrace &=& -g \epsilon_{ij} \delta^{2}(\textbf{x}-\textbf{x}^{\prime})
\nonumber\\
\left\lbrace{\Psi_{1}(\textbf{x}), \Phi_{0\nu}(\textbf{x}^{\prime})} \right\rbrace &=& \delta_{0\nu }\delta^{2}(\textbf{x}-\textbf{x}^{\prime})
\nonumber\\
\left\lbrace{\Psi_{1}(\textbf{x}), \Phi_{2i}^{\prime}(\textbf{x}^{\prime})} \right\rbrace &=& -g \epsilon_{ij}\partial_{j}\delta^{2}(\textbf{x}-\textbf{x}^{\prime})
\nonumber\\
\left\lbrace{\Psi_{2}(\textbf{x}), \Phi_{0\mu}(\textbf{x}^{\prime})} \right\rbrace &=& \delta_{\mu i}\partial_{i}\delta^{2}(\textbf{x}-\textbf{x}^{\prime})
\nonumber\\
\left\lbrace{\Psi_{2}(\textbf{x}), \Phi_{1\mu}(\textbf{x}^{\prime})} \right\rbrace &=& -\frac{g}{2}\epsilon_{ij}    \delta_{j \mu} \nabla^{2}\partial_{i}\delta^{2}(\textbf{x}-\textbf{x}^{\prime})
\label{brackets3}
\end{eqnarray}
The constraint algebra appears to be complicated but new linear combinations will simplify the algebra. Before going into that discussion it is time to get rid of the unphysical variables $ \xi_{0\mu} $ and $ \Pi_{0\mu}  $.
\subsection{Calculation in reduced phase space}
The fields $ \xi_{0\mu} $ and  $ \Pi_{0\mu} $  can be eliminated by strongly imposing the constraints $\Phi_{0\mu}$ and $\Phi_{1\mu}$\footnote{Technically this should be done by replacing the  Poisson brackets by the corresponding Dirac brackets. However, the Dirac brackets here are trivial i.e. the Dirac brackets  between the remaining phase space variables are the same as the Poisson brackets .}. The remaining constraints of the theory can now be rewritten as 
\begin{eqnarray}
\nonumber\\
\Phi_{20} &=& \Pi_{20} +  \frac{g}{2} \epsilon_{  i j }\partial_{i}\xi_{1j}  \approx 0
\nonumber\\
\Phi_{2i} &=& \Pi_{2i} -  \frac{g}{2} \epsilon_{ i j }( \xi_{2j} - \partial_{j}\xi_{10} ) \approx 0
\nonumber\\
\Psi_{1} &=& -\Pi_{10} + \frac{g}{2}\epsilon_{ij}\partial_{i}\xi_{2j} \approx 0
\nonumber\\
\Psi_{2} &=& -\partial_{i}\Pi_{1i} - \frac{g}{2} \epsilon_{ij} \nabla^{2}\partial_{i}\xi_{1j} \approx 0
\end{eqnarray} 
The Poisson brackets between these constraints can be read from  (\ref{brackets3}). The nontrivial brackets are
\begin{eqnarray}
\nonumber\\
\left\lbrace{\Phi_{2i}(\textbf{x}), \Phi_{2j}(\textbf{x})} \right\rbrace &=& -g \epsilon_{ij}\delta^{2}(\textbf{x}-\textbf{x}^{\prime})
\nonumber\\
\left\lbrace{\Psi_{1}(\textbf{x}), \Phi_{2i}(\textbf{x}^{\prime})} \right\rbrace &=& -g \epsilon_{ij}\partial_{j}\delta^{2}(\textbf{x}-\textbf{x}^{\prime})
\end{eqnarray}
 We can form the linear combination  \\
\begin{equation}
\Psi_{1}^{\prime} = \Psi_{1} + \partial_{i}\Phi_{2i}
\end{equation}
It can be easily checked that $\Psi_{1}^{\prime} $ has vanishing brackets with all other constraints. Replacing the set of constraints \{$ \Phi_{20} $, $ \Phi_{2i} $ ,$ \Psi_{1} $, and  $ \Psi_{2} $\} by the new set \{$ \Phi_{20} $, $ \Phi_{2i} $ , $ \Psi_{1}^{\prime} $, and $ \Psi_{2} $\}, we find that there are three first class constraints: $  \Phi_{20} $,  $ \Psi_{1}^{\prime} $, and $ \Psi_{2} $ and two second-class constraints $ \Phi_{2i} $ . The classification of the constraints of the theory is tabulated in Table 1.\\
\begin{table}[h]
%\begin{table}
\label{table:constraints}
\caption{Classification of Constraints of the model (\ref{mcslag})}
\centering
\begin{tabular}{l  c  c}
\\[0.5ex]
\hline
\hline\\[-2ex]
& First class & Second class \\[0.5ex]
\hline\\[-2ex]
Primary &\ \ $\Phi_{20}$ &\ \ $\Phi_{2i}$\\[0.5ex]
\hline\\[-2ex]
Secondary &\ $\Psi_1^{\prime}$, $ \Psi_{2} $ &\ \ \\[0.5ex]
\hline
\hline
\end{tabular}
\end{table} 

Before proceeding further, a degrees of freedom count will be instructive. The total number of phase space variables is 12. There are three first class constraints and two second-class constraints. Hence, the no of degrees of freedom is
\begin{equation}
\nonumber
12- (2 \times 3 + 2) = 4 
\end{equation}
We find that the number of degrees of freedom is doubled compared with the Maxwell theory, which is expected due to the higher derivative nature \cite{ostro}. 
\subsection{Reduction of second class constraints}
After the elimination of the unphysical sector ( $ \xi_{0\mu} $, $ \Pi_{0\mu} $ ), the total Hamiltonian becomes \\
 %Using the definition of $\Phi_{2i}^{\prime}$ and after some algebra we get
%\begin{equation}
%\Psi_{1}^{\prime} = -\Pi_{10} + \partial_{i}\Pi_{2i}
%\end{equation}
%So finally for this system we got one PFC $ \Phi_{20}^{\prime} $ , one SFC $ \Psi_{1}^{\prime} $ and one tertiary first class constraint $ \Psi_{2} $. In the reduced phase space the canonical Hamiltonian
%density is given by
%\begin{eqnarray}
%\nonumber\\
%\mathcal{H}_{c} &=& -\frac{1}{2} ( \xi_{2i}\xi_{2i} + \partial_{i}\xi_{10}\partial_{i}\xi_{10} - \partial_{i}\xi_{1j}\partial_{i}\xi_{1j} -2 \xi_{2j}\partial_{j}\xi_{10} +  \partial_{i}\xi_{1j}\partial_{j}\xi_{1i} )- \frac{g}{2} \epsilon_{ i j } \nabla^{2} \xi_{10} \partial_{i}\xi_{1j} \\
% && + \frac{g}{2} \epsilon_{ i j } \nabla^{2} \xi_{1i} ( \xi_{2j} - \partial_{j}\xi_{10}) + \Pi_{1\mu} \xi_{2\mu} 
 %\nonumber
%\end{eqnarray}
%And total hamiltonian is written as
\begin{equation}
H_{T}(\textbf{x}) = \int d^{2}\textbf{x} ( \mathcal{H}_{can}(\textbf{x}) + \Lambda_{20}(\textbf{x}) \Phi_{20}(\textbf{x}) + \Lambda_{2i}(\textbf{x}) \Phi_{2i}(\textbf{x}))
\end{equation}
where $ \mathcal{H}_{can} $ is the canonical Hamiltonian density given by
\begin{eqnarray}
\nonumber
\mathcal{H}_{can} &=& -\frac{1}{2} ( \xi_{2i}\xi_{2i} + \partial_{i}\xi_{10}\partial_{i}\xi_{10} - \partial_{i}\xi_{1j}\partial_{i}\xi_{1j} -2 \xi_{2j}\partial_{j}\xi_{10} +  \partial_{i}\xi_{1j}\partial_{j}\xi_{1i} ) \\
 && - \frac{g}{2} \epsilon_{ i j } \nabla^{2} \xi_{10} \partial_{i}\xi_{1j} + \frac{g}{2} \epsilon_{ i j } \nabla^{2} \xi_{1i} ( \xi_{2j} - \partial_{j}\xi_{10}) + \Pi_{1\mu} \xi_{2\mu} 
\end{eqnarray}
 and
 \begin{equation}
 \nonumber
 \Lambda_{2i} = \frac{1}{2}( \nabla^{2}\xi_{1i} + \partial_{i}\xi_{20} ) + \frac{1}{g}\epsilon_{ij}( \partial_{j}\xi_{10}+ \Pi_{1j}  - \xi_{2j} )  
  \end{equation}
$ \Lambda_{20} $ is arbitrary. It signifies that there is one continuous gauge degree of freedom. \\

	In the next section we will explicitly construct the gauge generator using the method given in \cite{BMP}. Since the method is directly applicable to theories with first class constraint only, we have to eliminate the second class constraints of our theory. Following Dirac's method of constraint Hamiltonian analysis we can strongly put the second class constraints to be zero if the Poisson brackets are replaced by the corresponding Dirac brackets.\\
	
	  The field theoretic version of the Dirac brackets \ref{DB_def}  between two phase space variables A and B is can be written as
  \begin{eqnarray}
  \nonumber
\{A(\textbf{x}), B(\textbf{x}^{\prime})\}_D   &=& \left\lbrace {A(\textbf{x}), B(\textbf{x}^{\prime})}\right\rbrace     - \int \left\lbrace {A(\textbf{x}), \Phi_{2i}(\textbf{y})} \right\rbrace \\ && \Delta_{ij}^{-1}(\textbf{y},\textbf{z})\left\lbrace {\Phi_{2j}(\textbf{z}), B(\textbf{x}^{\prime}) } \right\rbrace  d^{2}\textbf{y} d^{2}\textbf{z}  
  \end{eqnarray}
where   $\Delta_{ij}^{-1}(\textbf{x} , \textbf{x}^{\prime})$ is the inverse of the matrix 
\begin{equation}
\Delta_{ij}(\textbf{x} , \textbf{x}^{\prime}) = \lbrace{ \Phi_{2 i}(\textbf{x}), \Phi_{2j}(\textbf{x}^{\prime}) }\rbrace %= - g \epsilon_{ i j}  \delta^{2}( \textbf{x}-\textbf{x}^{\prime} )
\end{equation} 
%\begin{eqnarray}
%\Delta_{ij}^{-1}(\textbf{x} , \textbf{x}^{\prime})=  \frac{1}{g} \epsilon_{ i j}  \delta^{2}( \textbf{x}-\textbf{x}^{\prime} )
%\end{eqnarray} 
%Dirac brackets are defined as\\
%\begin{equation}
%\left[ {A(\textbf{x}), B(\textbf{x}^{\prime})}\right]   = \left\lbrace {A(\textbf{x}), B(\textbf{x}^{\prime})}\right\rbrace - \int \left\lbrace {A(\textbf{x}), \Phi_{2i}(\textbf{y})} \right\rbrace \Delta_{ij}^{-1}(\textbf{y},\textbf{z})\left\lbrace {\Phi_{2j}(\textbf{z}), B(\textbf{x}^{\prime}) } \right\rbrace  d^{2}\textbf{y} d^{2}\textbf{z}  
%\end{equation}
The nontrivial Dirac brackets between the phase space variables are calculated as
\begin{eqnarray}
\nonumber
\{\xi_{1\mu}(\textbf{x}), \Pi_{1\nu}(\textbf{x}^{\prime})\}_D &=& \delta_{\mu\nu} \delta^{2}(\textbf{x}-\textbf{x}^{\prime})
\nonumber\\
\{\xi_{2i}(\textbf{x}), \xi_{2j}(\textbf{x}^{\prime})\}_D &=& \frac{1}{g} \epsilon_{ij} \delta^{2}(\textbf{x}-\textbf{x}^{\prime})
\nonumber\\
\{\xi_{2i}(\textbf{x}), \Pi_{2j}(\textbf{x}^{\prime})\}_D &=& \frac{1}{2} \delta_{ij} \delta^{2}(\textbf{x}-\textbf{x}^{\prime})
\nonumber\\
\{\xi_{2i}(\textbf{x}), \Pi_{10}(\textbf{x}^{\prime})\}_D &=& -\frac{1}{2} \partial^{\prime}_{i} \delta^{2}(\textbf{x}-\textbf{x}^{\prime})
\nonumber\\
\{\Pi_{2i}(\textbf{x}), \Pi_{10}(\textbf{x}^{\prime})\}_D &=& \frac{g}{4} \epsilon_{ij} \partial^{\prime}_{j} \delta^{2}(\textbf{x}-\textbf{x}^{\prime})
\nonumber\\
\{\Pi_{2i}(\textbf{x}), \Pi_{2j}(\textbf{x}^{\prime})\}_D &=& \frac{g}{4} \epsilon_{ij} \delta^{2}(\textbf{x}-\textbf{x}^{\prime})
\label{basicdirac}
\end{eqnarray}
All other Dirac brackets are the same as the corresponding Poisson brackets.

\section{Gauge symmetry in EMCS}
As has been mentioned earlier we will follow the method of \cite{BMP} to construct the gauge generator containing the exact number of independent gauge parameters. The essence of the method has been reviewed in Sec. 2. Accordingly, we rename the constraints as $ \Omega_{1}=\Phi_{20} $, $ \Omega_{2}=\Psi_{1}^{\prime} $ and $ \Omega_{3}=\Psi_{2} $. The gauge generator is \\
\begin{equation}
G = \int\epsilon_{a} \Omega_{a} d^{2}\textbf{x}
\end{equation}
which is a field theoretic extension of (\ref{gauge_gen_def}).  These structure functions are now defined by  
\begin{eqnarray}
\nonumber
\{H_{can}, \Omega_{a}(\textbf{x})\}_D  &=& \int d^{2}\textbf{y}  V_{ab}(\textbf{y}, \textbf{x}) \Omega_{b}(\textbf{y})
\nonumber\\
\{\Omega_{a}(\textbf{x}), \Omega_{b}(\textbf{y}) \}_D &=& \int d^{2}\textbf{z}  C_{abc}(\textbf{z}, \textbf{x}, \textbf{y}) \Omega_{c}(\textbf{z})
\label{fieldstruc} 
\end{eqnarray}
 and the master equation (\ref{master_eqn_2}) takes the form 
\begin{eqnarray}
\nonumber
0 &=& \frac{d\epsilon_{a_{1}}(\textbf{x})}{dt} - \int d^{2}\textbf{y} \epsilon_{b}(\textbf{y}) V_{ba_{1}}(\textbf{x}, \textbf{y}) - \\ && \int d^{2}\textbf{y} d^{2}\textbf{z} \epsilon_{b}(\textbf{y}) \Lambda_{c_{1}}(\textbf{z}) C_{c_{1}ba_{1}}(\textbf{z}, \textbf{y}, \textbf{x}) 
\label{mastereqn3}
\end{eqnarray}
Note that Dirac brackets appear on the left hand sides of Eq. (\ref{fieldstruc}). This is because there were second class constraints in our theory which have been eliminated by the Dirac bracket formalism.\\

 Using the defining relations (\ref{fieldstruc}) and the Dirac brackets (\ref{basicdirac}) we find that the only nonvanishing $V_{ab}$ are given by 
\begin{eqnarray}
\nonumber
V_{12}(\textbf{x}, \textbf{y}) &=& - \delta^{2}(\textbf{x}-\textbf{y})
\nonumber\\
V_{23}({\textbf{x}, \textbf{y}}) &=& - \delta^{2}(\textbf{x}-\textbf{y})
\end{eqnarray}
Similarly, from the algebra of the constraints  we find all $C_{abc}=0$. Substituting these values in the Eq.(\ref{mastereqn3}) we get the following conditions on the gauge parameters $\epsilon_{a}$:
%The master equation for the SFCs is given by
%\begin{eqnarray}
%\frac{d\epsilon_{a_{2}}(\textbf{x})}{dt} - \int d^{2}\textbf{y} \epsilon_{a}(\textbf{y}) V_{ba_{2}}(\textbf{x}, \textbf{y}) = 0
%\end{eqnarray}
%which yields
\begin{eqnarray}
\nonumber
\dot{\epsilon}_{2} + \epsilon_{1} &=& 0
\nonumber\\
\dot{\epsilon_{3}} + \epsilon_{2} &=& 0
\end{eqnarray}
Solving these, we find 
\begin{eqnarray}
\nonumber
  \epsilon_{1} &=& \ddot{\epsilon}_{3}
\nonumber\\
  \epsilon_{2} &=& -\dot{\epsilon_{3}}
\end{eqnarray}
Hence the desired gauge generator assumes the form
%So, finally there is only one independent gauge parameter in the generator. Considering $ \epsilon_{3} $ to be independent we rewrite the gauge generator 
\begin{equation}
G = \int d^{2}{x} (\ddot{\epsilon}_{3} \Omega_{1} - \dot{\epsilon}_{3} \Omega_{2} +\epsilon_{3}\Omega_{3})
\label{generator3}
\end{equation}
It is immediately observed that G contains one arbitrary gauge parameter, namely $ \epsilon_{3} $.\\

 We still have the additional restrictions (\ref{varsgauge}). In our case this leads to the condition 
 \begin{equation}
\delta\xi_{2\mu} = \frac{d}{dt} \delta \xi_{1\mu}
\label{addrestriction2}
\end{equation} 
where $  \delta \xi_{1\mu} $, $  \delta \xi_{2\mu} $ are the gauge variations of $ \xi_{1\mu} $ and $ \xi_{2\mu} $ respectively. Using the generator G (\ref{generator3}) we get
\begin{equation}
\delta \xi_{2\mu} = \{\xi_{2\mu}, G\}_D   =  \partial_{\mu} \dot{\epsilon}_{3}
\end{equation}
Similarly 
\begin{equation}
\delta \xi_{1\mu} = \{\xi_{1\mu}, G\}_D   =  \partial_{\mu} \epsilon_{3}
\label{gaugevarxi1}
\end{equation}
Clearly the additional restriction (\ref{addrestriction2}) is identically satisfied. Thus, no more restriction is imposed on the gauge parameters. \\%Hence \ref{generator3}.\\

 Finally we look at the comparison of the transformations generated by the Hamiltonian gauge generator with Lagrangian gauge symmetry  (\ref{gengaugeinv}). Since $ \xi_{1\mu} = A_{\mu} $ we have
 \begin{equation}
\delta A_{\mu} = \partial_{\mu} \epsilon_{3}
\end{equation}
from (\ref{gaugevarxi1}). This is the same transformation as (\ref{gengaugeinv}) if we put $ \epsilon_{3} $ = $ \lambda $.
\section{Discussions}
In this chapter, we have considered the extended Maxwell Chern Simons (EMCS) model as an example to illustrate the application of the algorithm developed in this thesis to a higher derivative field theoretic model. The EMCS model was introduced by Deser and is an extension of the usual Maxwell Chern Simons model\cite{deser2}. The first order formalism which we are using singles out the time derivative of the fields as additional `coordinate'. Due to this, a manifest Lorentz covariance is lost. We have therefore, converted the entire theory in terms of covariant components and the effect of the relativistic metric was taken care of by explicitly writing the effective first rder Lagrangian.

	A detailed constraint analysis from the first order Lagrangian has been performed. We obtained three first class constraints and a pair of second class constraint. An usual degree of freedom count demonstrated that there was a doubling of freedom, compared with the usual MCS theory. This result is consistent with the higher derivative nature of the model.
	
	As has been discussed earlier, the algorithm for constructing independent gauge generator of the model developed in this thesis is applicable for theories with first class constraints only. We have therefore reduced the phase space by replacing the Poisson brackets by the corresponding Dirac brackets. This enables one to put the second class constraints strongly equal to zero.  The model has been treated in the first order formalism. Since it is a higher derivative theory the degrees of freedom count shows that the phase space has been doubled. Finally, we show that the gauge generator obtained by us  generate $U(1)$ gauge transformation of the phase space variables. 
  	\chapter{The quantum charged rigid membrane}
  	We are continuing the application of our method to different types of systems. In this chapter we will consider the example of charged rigid membrane. The action for this  model will be constructed from the extrinsic curvature of the membrane evolving in a Minkowski space time.   
  	
  	Theories with extrinsic curvatures are frequently studied, especially in string theory. Although this concept is not new, but recent inclusion of these theories in some physically interesting models added an extra
urgency to revisit their symmetry features. As proposed by Dirac in 1962,  the membrane model of the electron is a charged bubble, evolving with time.  The model is not a physical one but has important consequences on  brane related theories.
               To have a basic knowledge of the membrane we have taken, consider the evolving surface $\Sigma$ in a background  Minkowski spacetime $\eta^{\mu\nu}$ \footnote{with $\mu, \nu=0,1,2,3$ and $a,b=0,1,2$} . The surface is described by the local coordinate $x^{\mu}$  of the background spacetime. The embedding function $  X^{\mu}(\xi^{a})= x^{\mu}$ is a function of the local coordinates of the world volume m, swept out by the surface. We consider the following effective action underlying the dynamics of the surface $\Sigma$\cite{cordero}:
\begin{equation}
S[X^{\mu}] = \int_{m} d^{3}\xi (-\alpha K + \beta j^{a} e^{\mu}_{ \ a} A_{\mu}),
\label{action1}
\end{equation}
where $K=g^{ab} K_{ab}$ being the extrinsic curvature \footnote{$g_{ab}$ is the worldvolume metric and $e^{\mu}_{ \ a} = X^{\mu}_{ \ ,a}$ are tangent vectors to the worldvolume} and $\alpha, \beta$  are constant related to the rigidity parameter and form factor respectively. On the other hand, $j^{a}$ which minimally couples the charged surface  and the electromagnetic field $A_{\mu}$ \cite{barut}, is a constant electric current density distributed over the world volume and is locally conserved on m with $\partial_{a}j^{a}=0$. Variation of the action with respect to the embedding function $X^{\mu}(\xi^{a})$ leads to the equation of motion 
\begin{equation}
\alpha \mathcal{R} = \frac{\beta}{\sqrt{-g}} j^{a}n^{\mu} e^{\nu}_{ \ a} F_{\mu\nu}.
\label{eom1}
\end{equation}
The above equation (\ref{eom1}) can be thought as a Lorentz force equation with $\mathcal{R}$  being the Gaussian curvature and $F_{\mu\nu} = 2 \partial_{[\mu}A_{\nu]}$ the electromagnetic field tensor. Under suitable choice of the embedding functions ($X^{\mu}(\tau, \theta, \varphi)=(t(\tau), r(\tau), \theta, \varphi)$) equation (\ref{action1}) boils down to\cite{cordero}
\begin{equation}
S = 4\pi \int d\tau L(r, \dot{r}, \ddot{r}, \dot{t}, \ddot{t}) \label{action2}
\end{equation}
where the Lagrangian L, which is higher derivative in nature is given by,
\begin{eqnarray}
L = -\frac{\alpha r^{2}}{\dot{t}^{2} - \dot{r}^{2}}(\ddot{r}\dot{t}- \dot{r}\ddot{t}) -2\alpha r \dot{t} - \frac{\beta q^{2} \dot{t}}{r}. 
\label{hdlagrangian}
\end{eqnarray}
 Promptly, we can write down the equation of motion for the higher derivative Lagrangian:
\begin{equation}
\frac{d}{d\tau}\left({\frac{\dot{r}}{\dot{t}}} \right) = - \frac{\dot{t}^{2} - \dot{r}^{2}}{2r \dot{t}^{3}}\left({\dot{t}^{2} - \frac{\beta(\dot{t}^{2} - \dot{r}^{2})^{2}q^{2}}{2 \alpha r^{2}}} \right). 
\label{eom2}
\end{equation} 
 Lagrangian  (\ref{hdlagrangian}) will be our sole interest. The model is reparametrisation invariant under the parameter $\tau \rightarrow \tau + \sigma $, with $\sigma$ being infinitesimal. In the proceeding sections we will develop the Hamiltonian formulation  for this model.  
\section{Hamiltonian analysis }
Before we start the Hamiltonian analysis we need to convert the higher derivative Lagrangian (\ref{hdlagrangian}) to a first order lagrangian, named as the auxiliary lagrangian, by introduction  of the new fields 
\begin{eqnarray}
\nonumber
\dot{r} &=& R\\
\dot{t} &=& T
\end{eqnarray}
So, we write down the auxiliary Lagrangian as \footnote{ consider $N^{2} = T^{2} - R^{2}$, for convenience}
\begin{equation}
L^{\prime} = -\frac{\alpha r^{2}}{N^{2}}(\dot{R}T -R \dot{T}) - 2 \alpha r T - \frac{\beta q^{2}T}{r} + \lambda_{1}(R-\dot{r}) + \lambda_{2}(T- \dot{t})
\end{equation}
 Inclusion of  new fields  impose  constraints
 \begin{equation}
  R -\dot{r} \approx 0,  \ \ \ \ \ T- \dot{t} \approx 0
 \label{lagcons} 
 \end{equation}
  which are taken care of via the  multipliers $\lambda_{1}$ and $\lambda_{2}$.  Variation of $L^{\prime}$ with respect to $r, R, t, T, \lambda_{1}$ and $\lambda_{2}$ give rise to the following equation of motions:
 \begin{eqnarray}
- \frac{2 \alpha r}{N^{2}} ( \dot{R} T - R \dot{T}) - 2 \alpha T + \frac{\beta q^{2} T}{r^{2}} + \dot{\lambda}_{1} = 0 \\ \label{eom_r}
 -\frac{2 \alpha r^{2}}{N^{4}}R ( \dot{R} T - R \dot{T}) + \frac{d}{d\tau} \left( {\frac{\alpha r^{2}}{N^{2}}T}\right)+ \frac{\alpha r^{2}}{N^{2}}\dot{T} + \lambda_{1} = 0 \\ \label{eom_R} 
 \dot{\lambda}_{2}=0 \\ \label{eom_t}
  \frac{2 \alpha r^{2}}{N^{4}}T ( \dot{R} T - R \dot{T}) - \frac{d}{d\tau} \left( {\frac{\alpha r^{2}}{N^{2}}R}\right)- \frac{\alpha r^{2}}{N^{2}}\dot{R} -2 \alpha r - \frac{\beta q^{2}}{r} + \lambda_{2} = 0 \\ \label{eom_T} 
  R-\dot{r}=0 \\ \label{eom_lambda1}
  T- \dot{t} = 0 \label{eom_lambda2}
\end{eqnarray}  
(\ref{eom_lambda1}), (\ref{eom_lambda2}) are obvious since they correspond to (\ref{lagcons}).\\

Before proceeding for Hamiltonian formulation, we identify  the new phase space which is constituted of the variables are $(r, \Pi_{r})$, $(t, \Pi_{t})$, $(R, \Pi_{R})$, $(T, \Pi_{T})$,$ ( \lambda_{1}, \Pi_{\lambda_{1}})$, $(\lambda_{2}, \Pi_{\lambda_{2}})$. Here $\Pi_{x^{\mu}}=\frac{\partial{L^{\prime}}}{\partial{\dot{x}^{\mu}}}$, are the momenta corresponding to $x^{\mu}$ which generically stands for the variables $r, R, t, T, \lambda_{1}, \lambda_{2} $. We immediately obtain the  primary constraints as listed bellow  
\begin{eqnarray}
\nonumber
\Phi_{1} &=& \Pi_{r} + \lambda_{1} \approx 0
\nonumber \\
\Phi_{2} &=& \Pi_{t} + \lambda_{2} \approx 0
\nonumber\\
\Phi_{3} &=& \Pi_{R} + \frac{\alpha r^{2}}{N^{2} } T \approx 0
\nonumber \\
\Phi_{4} &=& \Pi_{T} - \frac{\alpha r^{2}}{N^{2} } R \approx 0
\nonumber \\
\Phi_{5} &=& \Pi_{\lambda_{1}} \approx 0
\nonumber \\
\Phi_{6} &=& \Pi_{\lambda_{2}} \approx 0
\end{eqnarray}
The poisson brackets between the field variables are defined as:
\begin{eqnarray}
\nonumber
\left\lbrace{x^{\mu}, \Pi_{x^{\nu}}} \right\rbrace &=& \delta_{\mu\nu}
\nonumber\\
\left\lbrace{x^{\mu}, x^{\nu}} \right\rbrace&=& \left\lbrace{\Pi_{x^{\mu}}, \Pi_{x^{\nu}}} \right\rbrace=0
\label{pbs}
\end{eqnarray}
With the aid of (\ref{pbs})  the  non zero Poisson brackets between the primary constraints can be written down
\begin{eqnarray}
\nonumber
\left\lbrace{\Phi_{1}, \Phi_{3}} \right\rbrace &=& - \frac{2 \alpha r}{N^{2}}T
\nonumber \\
\left\lbrace{\Phi_{1}, \Phi_{4}} \right\rbrace &=& \frac{2 \alpha r}{N^{2}}R
\nonumber \\
\left\lbrace{\Phi_{1}, \Phi_{5}} \right\rbrace &=& 1
\nonumber \\
\left\lbrace{\Phi_{2}, \Phi_{6}} \right\rbrace &=& 1
\label{pbs2}
\end{eqnarray}
 We can take the following combination of the constraints  
\begin{eqnarray}
\Phi_{3}^{\prime} &=& R \Phi_{3} + T \Phi_{4} \approx 0 \\
\Phi_{4}^{\prime} &=& \Phi_{4} - \frac{2\alpha r R}{N^{2}} \Phi_{5} \approx 0 
\end{eqnarray}
so that the new set  of primary constraints are $\Phi_{1}, \Phi_{2}, \Phi_{3}^{\prime}, \Phi_{4}^{\prime}, \Phi_{5}, \Phi_{6}$. The complete algebra of primary constraints is now given by (only the nonzero brackets are listed),
\begin{eqnarray}
\left\lbrace{\Phi_{1}, \Phi_{5}} \right\rbrace &=&\left\lbrace{\Phi_{2}, \Phi_{6}} \right\rbrace =1
\label{pbs3}
\end{eqnarray}
 The  canonical Hamiltonian  via Legendre transformation takes the form as
\begin{equation}
H_{can} = 2 \alpha r T + \frac{\beta q^{2} T}{r} -\lambda_{1} R - \lambda_{2}T.
\label{H_{can}}
\end{equation}
 The total Hamiltonian is
\begin{equation}
H_{T} =H_{can} +  \Lambda_{1}\Phi_{1}+  \Lambda_{2}\Phi_{2}+  \Lambda_{3}\Phi_{3}^{\prime}+  \Lambda_{4}\Phi_{4}^{\prime}+  \Lambda_{5}\Phi_{5} +  \Lambda_{6}\Phi_{6}
\end{equation}
Here $\Lambda_{1},\Lambda_{2},\Lambda_{3}, \Lambda_{4},\Lambda_{5}, \Lambda_{6} $ are the Lagrange multipliers which are arbitrary at this stage. Only those multipliers which are attached to the primary second-class constraints will be determined, others corresponding to primary first class constraints will remain undetermined (although they can be determined too via equation of motion). At this level, loosely speaking  $\Phi_{3}^{\prime}$ and $\Phi_{4}^{\prime} $ are first class constraints (this classification may be changed after we get the full list of constraints).  These two may provide us  two new secondary constraints and the list can still keep increasing until we get all the constraints. Now, we move towards extracting all constraints of this system. This can be done by demanding  that Poisson brackets of the constraints with the total Hamiltonian(time evolution) of the constraints is zero. Preserving $\Phi_{1}, \Phi_{2}, \Phi_{5}, \Phi_{6}$ in time solves the following multipliers respectively 
\begin{eqnarray}
\nonumber
    \Lambda_{5} &=&  2\alpha T - \frac{\beta q^{2} T}{r^{2}} 
    \nonumber\\
 \Lambda_{6} &=& 0
 \nonumber\\
 \Lambda_{1} &=& R 
 \nonumber\\
 \Lambda_{2} &=& T.
\end{eqnarray}
Whereas, time conservation of the primary constraints $\Phi_{3}^{\prime}$ and  $\Phi_{4}^{\prime}$ leads to the secondary constraints $\Psi_{1}$ and $\Psi_{2}$ respectively given by
\begin{eqnarray}
\nonumber
\Psi_{1}&=&  -2\alpha r T - \frac{\beta q^{2} T}{r} + \lambda_{1}R  + \lambda_{2}T  \approx 0
\nonumber\\
 \Psi_{2} &=& -2 \alpha r - \frac{\beta q^{2}}{r} + \lambda_{2} - \frac{2 \alpha r}{N^{2}} R^{2} \approx 0
\end{eqnarray}
Before proceeding further we list below all  the nonzero Poisson brackets  of the secondary constraints  $\Psi_{1}$, $\Psi_{2}$  with other constraints:
\begin{eqnarray}
\nonumber
\left\lbrace {\Phi_{1}, \Psi_{1}}\right\rbrace &=& 2 \alpha T - \frac{\beta q^{2} T}{r^{2}} 
%\nonumber\\
%\left\lbrace {\Phi_{2}, \Psi_{1}}\right\rbrace &=& 0
%\nonumber\\ \left\lbrace {\Phi_{3}^{\prime}, \Psi_{1}}\right\rbrace &=& -\Psi_{1} \approx 0
%\nonumber\\ \left\lbrace {\Phi_{4}^{\prime}, \Psi_{1}}\right\rbrace &=&  -\Psi_{2} \approx 0
\nonumber\\ 
\left\lbrace {\Phi_{5}, \Psi_{1}}\right\rbrace &=&  -R
\nonumber\\
\left\lbrace {\Phi_{6}, \Psi_{1}}\right\rbrace &=& -T
%\nonumber\\ \left\lbrace {\Psi_{1}, \Psi_{1}}\right\rbrace &=& 0
%\nonumber\\ \left\lbrace {\Psi_{1}, \Psi_{2}}\right\rbrace &=& 0
\nonumber\\ 
\left\lbrace {\Phi_{1}, \Psi_{2}}\right\rbrace &=& \frac{2 \alpha}{N^{2}} T^{2} - \frac{\beta q^{2}}{r^{2}}
%\nonumber\\ \left\lbrace {\Phi_{2}, \Psi_{2}}\right\rbrace &=& 0
%\nonumber\\ \left\lbrace {\Phi_{3}^{\prime}, \Psi_{2}}\right\rbrace &=& 0
\nonumber\\
\left\lbrace {\Phi_{4}^{\prime}, \Psi_{2}}\right\rbrace &=& - \frac{4 \alpha r}{N^{4}} T R^{2}
%\nonumber\\ \left\lbrace {\Phi_{5}, \Psi_{2}}\right\rbrace &=& 0
\nonumber\\
\left\lbrace {\Phi_{6}, \Psi_{2}}\right\rbrace &=& -1
%\nonumber\\ \left\lbrace {\Psi_{2}, \Psi_{2}}\right\rbrace &=& 0
%\nonumber\\ \left\lbrace {\Psi_{1}, \Psi_{1}}\right\rbrace &=& 0
\label{pbs4}
\end{eqnarray}
Now, time preservation of the secondary constraint $\Psi_{1}$ gives identically $0=0$. And requirement of   $\dot{\Psi}_{2}= 0$ solves the Lagrange multiplier $\Lambda_{4} = -\frac{A}{B} R$, with  $A = \frac{2 \alpha T^{2}}{N^{2}}  - \frac{\beta q^{2}}{r^{2}}$ and $B= -\frac{4 \alpha rT R^{2} }{N^{4}}$.\\

%And from (10) we get $\Lambda_{5} = 2 \alpha T - \frac{\beta q^{2} T}{r^{2}}$.
 From the constraint algebra (\ref{pbs3}) and (\ref{pbs4}) one can clearly assert that there is only one first class constraint $\Phi_{3}^{\prime}$ with seven other second class constraints $\Phi_{1}, \Phi_{2}, \Phi_{4}^{\prime}, \Phi_{5}, \Phi_{6}, \Psi_{1}, \Psi_{2}$. One point worth noting since there are odd number of second class constraints, it indicate there might be some other first class constraint to make the pair of second class constraints even. Judiciously, we can choose a combination  $\Psi_{1}^{\prime} = \Psi_{1}  -  \Lambda_{1}\Phi_{1}-  \Lambda_{2}\Phi_{2}-    \Lambda_{4}\Phi_{4}^{\prime}- \Lambda_{5}\Phi_{5} -  \Lambda_{6}\Phi_{6} $ so that the pair  ($\Phi_{3}^{\prime}, \Psi_{1}^{\prime}$)  becomes first-class. This completes our constraint classification.% We list below the constraints with for a clear understanding:

 Having completed the constraint classification, its time to get rid of the unphysical sector $ (\lambda_{1}, \Pi_{\lambda_{1}})$           and $(\lambda_{2}, \Pi_{\lambda_{2}})$ by imposing the primary second class constraints  $\Phi_{1}, \Phi_{2}, \Phi_{5}, \Phi_{6}$  strongly zero. This can be done by replacing all Poisson brackets by Dirac brackets for rest of the calculations. Surprisingly, Dirac brackets between the basic fields remain same as their corresponding Poisson brackets. So, now our phase space is spanned by $\{ r,\Pi_{r}, t, \Pi_{t}, R, \Pi_{R}, T, \Pi_{T}\}$.
 For convenience of future calculations we rename the constraints as 
 \begin{eqnarray}
 F_{1} &=& \Phi_{3}^{\prime} = R\Phi_{3} + T \Phi_{4} \approx 0
 \\
 F_{2} &=&  \Psi_{1} - \Lambda_{4} \Phi_{4} \approx0
 \\
 S_{1}&=& \Phi_{4}  \approx 0 \\
 S_{2}&=& \Psi_{2}= - \Pi_{t} -2 \alpha r - \frac{\beta q^{2}}{r} - \frac{2 \alpha r R^{2}}{N^{2}} \approx 0.
 \end{eqnarray}\\
 
Here,  ${F_{1}, F_{2}}$ is the first class pair with  $F_{1}$ as primary first class constraint.
So far we observed that  in this theory, there is only one primary first class constraint with one undetermined multiplier which clearly indicate existence of gauge symmetry(s) in the system. In the next section we will extract the gauge symmetries of this quantum charged rigid membrane.
\section{Gauge symmetry and Virasoro algebra}
To study gauge symmetry we need to remove all the second class constraint from the system by setting them strongly zero and performing Dirac bracket defined by (\ref{DB_def})
%\begin{equation}
%\left\lbrace {f, g}\right\rbrace _{D} = \{f,g\} - \sum_{i,j = 1,2}\{ f,S_{i}\} \triangle^{-1}_{ij} \{ S_{j},g\} 
%\end{equation}
 . To compute $\triangle^{-1}_{ij}$ for the set of  of second class constraints, we have  $ \{ S_{1}, S_{2} \} = - \frac{4 \alpha rt R^{2}}{N^{4}}$. So, we can compute the  Dirac Brackets between the basic fields. The  nonzero DBs are \footnote{with $A= \frac{2 \alpha T^2}{N^2} - \frac{\beta q^2}{r^2}, B= \frac{\alpha r T R^2}{N^4}$}:
 \begin{eqnarray}
 \nonumber
 \left\lbrace {r, \Pi_{r}} \right\rbrace_{D} &=& 1
 \nonumber\\
 \left\lbrace {\Pi_{r}, t} \right\rbrace_{D} &=& - \frac{N^{2}}{2TR}
 \nonumber\\
 \left\lbrace {\Pi_{r}, T} \right\rbrace_{D} &=& -\frac{A}{B}
 \nonumber\\
 \left\lbrace {\Pi_{r}, \Pi_{T}} \right\rbrace_{D} &=& -\frac{2 \alpha r R}{N^{2}}+ \frac{Ar}{2R}
 \nonumber\\
 \left\lbrace { t, \Pi_{t}} \right\rbrace_{D} &=& 1
 \nonumber\\
 \left\lbrace { t, \Pi_{R}} \right\rbrace_{D} &=& \frac{r (T^{2} + R^{2})}{4 T R^{2}}
 \nonumber\\
 \left\lbrace { t, T} \right\rbrace_{D} &=& - \frac{1}{B}
 \nonumber\\
 \left\lbrace { t, \Pi_{T}} \right\rbrace_{D} &=& - \frac{r}{2 R}
 \nonumber\\
 \left\lbrace { R, \Pi_{R}} \right\rbrace_{D} &=&1
 \nonumber\\
 \left\lbrace {  \Pi_{R}, T} \right\rbrace_{D} &=& - \frac{T}{R}
 \nonumber\\
 \left\lbrace { \Pi_{r}, \Pi_{R}} \right\rbrace_{D} &=& \frac{2 \alpha r T}{N^{2}} + \frac{A}{B}  \frac{\alpha r^{2}(T^{2}+R^{2})}{N^{4}}
 \nonumber\\
 \left\lbrace { \Pi_{R}, \Pi_{T}} \right\rbrace_{D} &=& \frac{\alpha r^{2}}{N^{2}}
 \label{dbs2}
\end{eqnarray}

The  generator of the gauge transformation is  given by a linear combination of all first class constraints,
 \begin{equation}
 G= \epsilon_{1} F_{1} + \epsilon_{2} F_{2}
 \label{gaugegenerator11}
 \end{equation}
where   $\epsilon_{1}$ and  $\epsilon_{2}$ are gauge parameters. We need to find out whether these gauge parameters are independent or not.\\
The Dirac brackets between the first class constraints are given by
\begin{equation}
\{F_{i}, F_{j}\}_{D} = - \epsilon_{ij}F_{2} \ \ \  ;\ \ \ \ \ i, j=1, 2
\label{frstconsbrak}
\end{equation}
Using a suggestive notation we rename the constraints $F_{1}$ and $F_{2}$ as 
\begin{eqnarray}
 L_{0} &=&   F_{1}\\
 L_{1} &=&  F_{2}
\end{eqnarray} 
We can easily identify a sort of truncated Virasoro algebra of the form 
\begin{equation}
\{ L_{m}, L_{n}\}_{D} = (m-n) L_{m+n}
\end{equation} 
 with $m=0$, $n=1$  as proposed in \cite {ho} for higher derivative cases. 

 Now,  using  equations (\ref{structure constant_V}, \ref{structure constant_C}) we compute the structure constants as $C_{122} = -1 = -C_{212}$ and $V_{12} =1$ (other structure constants are zero). Exploiting the master equations (\ref{master_eqn_2}) we find the the following relation between the gauge parameters  
\begin{equation}
 \epsilon_{1}= - \Lambda_{3}\epsilon_{2} - \dot{\epsilon}_{2}.
 \label{parameterrelation1}
 \end{equation}
  
 %\begin{equation}
 %\delta{\Lambda_{3}} = \dot{\epsilon}_{2}
  %\label{parameterrelation2}
% \end{equation}
It is clear that we have only one independent gauge  symmetry in this system which is supported by the fact that there is only one undetermined multiplier. We consider $\epsilon_{2}$ to be independent and compute the gauge transformation of the fields
 \begin{eqnarray}
 \delta{r} &=& -  \epsilon_{2} R \label{gaugetrans_r}\\
 \delta{t} &=& -  \epsilon_{2} T \label{gaugetrans_t}\\
 \delta{R} &=&   \epsilon_{1} R \label{gaugetrans_R}\\
 \delta{T} &=&   \epsilon_{1} T + \epsilon_{2} \frac{A}{B} R 
 \label{gaugetrans}
 \end{eqnarray}
 
 We can identify this gauge symmetry as reparametrisation symmetry in the following manner. Consider an infinitesimal transformation of r and t on the worldvolume as $\tau \rightarrow \tau + \sigma$. For some infinitesimal $\sigma$,  we can write 
\begin{eqnarray}
\nonumber
\delta{r} = -\sigma r\\
\delta{t} = -\sigma t
\label{repara}
\end{eqnarray}

 Clearly, a comparison between (\ref{gaugetrans_r}, \ref{gaugetrans_t}) and both equations of  (\ref{repara}) shows that the reparametrisation parameter is given by $\sigma = \epsilon_{2}$. Using (\ref{repara}) we  compute of Gauge variation of the Lagrangian (\ref{hdlagrangian})which simplifies to
\begin{equation}
\delta{L} = \frac{d}{d \tau} (\sigma L) 
\end{equation}
and ensure the invariance of the action under (\ref{repara}).
\section{Consistency check} 

  It would be worth to find out the  Hamiltonian equations of motion which are given by  
  \begin{eqnarray}
  \dot{r}&=& R  \label{eom_rH}\\
  \dot{t}&=& T \label{eom_tH}\\
   \dot{R}&=& \Lambda_{3} R \label{eom_RH} \\
  \dot{T}&=& - \frac{A}{B} R + \frac{\dot{R}}{R} T  \label{eom3}
  \end{eqnarray}
Equations (\ref{eom_rH}) and (\ref{eom_tH}) are obvious as they arise as constraints at the Lagrangian level and  agrees with (\ref{eom_lambda1}) and (\ref{eom_lambda2}). Taking time derivative of (\ref{gaugetrans_r}) and (\ref{gaugetrans_t}) we get
\begin{eqnarray}
\frac{d}{d \tau} \delta{r}&=& - \dot{\epsilon}_{2} R - \epsilon_{2} \dot{R} \label{commuta1}\\
\frac{d}{d \tau} \delta{t}&=& - \dot{\epsilon}_{2} T - \epsilon_{2} \dot{T} \label{commuta2}
\end{eqnarray}
Using  equation (\ref{parameterrelation1}) alongwith (\ref{eom_RH}, \ref{eom3}) the above equations (\ref{commuta1}, \ref{commuta2}) simplify to
\begin{eqnarray}
\frac{d}{d\tau} \delta{r} &=& \delta{R} \\
\frac{d}{d\tau} \delta{t} &=& \delta{T}
\end{eqnarray}
which is a direct verification for (\ref{varsgauge}). 
Whereas, (\ref{eom3}) along with the trivial equation of motions (\ref{eom_rH}) and (\ref{eom_tH})  can be cast into the form so that it verify (\ref{eom2}). This indeed is  an important outcome of this analysis which agrees the validity  of this first order formalism via matching the equation of motion at higher derivative and first order level.\\

Taking gauge variation of the  equation of  (\ref{eom_RH}) and using (\ref{gaugetrans_R})we get 
\begin{equation}
\delta{\Lambda_{3}} = \dot{\epsilon_{1}}
\end{equation}
which in turn verifies the first master equation (\ref{master_eqn_1}).

\section{Discussions}
In this chapter, we considered the quantum charged rigid membrane,  which actually was proposed by Dirac \cite{dirac_membrane}.  The Lagrangian for this model contains  higher derivatives. The model is important because it can serve as a toy model for brane related theories like the geodetic brane gravity. We have taken this model to apply the first order formalism as it belongs to a very important class of physics. For Hamiltonian formulation, the Lagrangian is  converted into the first order form and we obtained the  Lagrangian equations of motion with respect to the dynamical variables. We followed the programme developed in \cite{BMP} to unveil the gauge symmetries of this model.  We found  that there is only one independent primary first class constraint which is consistent with the fact that there is only one independent gauge degree of freedom. The gauge degree of freedom is found out to be nothing but the reparametrisation invariance. Also, we matched the equation of motions at the first order and higher derivative level to show that even in the first order approach we are dealing, the same dynamics of the system persists. 
  \chapter{The Regge-Teitelboim model of Cosmology}
  After discussing the models from various fields of physics, the final model to test the first order formalism of this thesis will be from gravity.  Specifically, we will consider the  Regge-Teitelboim (RT) model. The RT model can be thought as an implementation of a gravity theory from the perspective of string theory. Gravity, in this case, is considered to be induced on a hypersurface which is evolving in a higher dimensional Minkowski space.  According to the embedding theorem, the dimension of the embedding space should be atleast $\frac{N(N+1)}{2}$, where $N$ is the dimension of the hypersurface. So, the Einstein Hilbert action on a 4D hypersurface should be embedded in 10D Minkoswski background \cite{banerjee_rt}.  The whole gravity theory, thereafter, is expressible with respect to the the local coordinates on the hypersurface. 
  
  These local coordinates, induced on the hypersurface, are considered to form a Minisuperspace in the higher dimensional embedding space. Unlike usual gravity theories where the metric is considered to be fundamental variable, these induced coordinates in this case are considered to be dynamical variables. Considering the FriedmannLemaîtreRobertsonWalker (FRLW) metric induced on the hypersurface, the Einstein Hilbert action is computed. Apparently, the model is  higher derivative  but there appears a surface term  which can be neglected as it has no effect on the equations of motion and the model becomes first order. But, in gravity theories these surface terms cannot be neglected so naively as they carry the information for the entropy of the system. So, keeping the surface term intact we consider  the theory higher derivative version of the model. 
  
   For canonical analysis and hence exploring the gauge symmetries of this minisuperspace model we follow the first order formalism. It is found that there appears two first class and two second class constraints. Out of the two first class constraints, one is primary first class constraint.  The method of Banerjee et. al. shows that there is one independent gauge parameter \cite{BRR}. The independent gauge symmetry is identified as the reparametrisation invariance. 
   
  	Now, the next step is to quantise this model. The formal quantisation is performed by  promoting the Poisson/Dirac brackets between the canonical variable to the level of commutator. Proper identification of the  phase space is crucial to begin the quantisation procedure and it is done by removing all the constraints from the theory. The first class constraints are converted into second class by introduction of some ad hoc relations among the phasespace variables, known as gauge conditions. The number of gauge conditions are always equal to the number of independent first class constraints.  The canonical variables, in the final reduced phase space, are identified. These canonical variables are the operators which  span in some appropriate Hilbert space.  The analogous Schroedinger equation  in cosmology is known as the Wheeler-Dewitt (WDW) equation. The WDW equation in the fully reduced phase space is constructed and the conserved energy is identified.  
  
  \section{Introduction of the model}
The RT model considers a d-dimensional brane $\Sigma$ which evolves in a $N$ dimensional bulk space time with fixed Minkowski metric $\eta_{\mu\nu}$. The world volume swept out by the brane is a $d+1$ dimensional manifold $m$ defined by the embedding $x^\mu = X^\mu(\xi^a)$ where $x^\mu$ are the local coordinates of the background spacetime and $\xi^a$ are local coordinates for $m$. The theory is given by the action functional
\begin{equation}
S[X] = \int_{m} d^{d+1} \xi  \sqrt{-g} (\frac{\beta}{2}\mathcal{R} - \Lambda), \label{main_lag}
\end{equation}
where $\beta$ has the dimension $[L]^{1-d}$ and $g$ is the determinant of the induced metric $g_{ab}$. $\Lambda$ denotes cosmological constant and  $\mathcal{R}$ is the Ricci scalar. As has been already stated above, we will be confined to the minisuperspace cosmological model following from the RT model. 

 The standard procedure in cosmology is to assume that on the large scale the universe is homogeneous and isotropic. These special symmetries enable the $4$ dimensional world volume representing the evolving universe to be embedded in a 5-dimensional Minkowski space time 
 \begin{equation}
 ds^{2} = - dt^{2} + da^{2} + a^{2}d\Omega_{3}^{2},
\end{equation}  
where $d\Omega_{3}^{2}$ is the metric for unit 3 sphere. To ensure the FRW case we take the following parametric representation for the brane
\begin{eqnarray}
x^{\mu} = X^{\mu}(\xi^{a}) = \left( t(\tau),a(\tau), \chi, \theta, \phi \right),
\end{eqnarray}
 $a(\tau)$ is known as the scale factor.

 After ADM decomposition  with space like unit normals ($N=\sqrt{\dot{t}^{2}-\dot{a}^{2}}$ is the lapse function)
 \begin{equation}
 n_{\mu} = \frac{1}{N}(-\dot{a}, \dot{t}, 0,  0,0),
 \end{equation}
   the induced metric on the world volume is given by,
  \begin{equation}
  ds^{2} = -N^{2} d\tau^{2} + a^{2} d \Omega_{3}^{2}.
\end{equation}
Now, one can compute the Ricci scalar which is given by
\begin{equation}
\mathcal{R} = \frac{6 \dot{t}}{a^{2} N^{4}}(a \ddot{a}\dot{t} - a\dot{a}\ddot{t} + N^{2}\dot{t}).
\end{equation}
With these functions we can easily construct the Lagrangian density as
\begin{equation}
\mathcal{L} = \sqrt{-g} \left(\frac{\beta}{2}\mathcal{R} - \Lambda\right).
\end{equation} 
The Lagrangian in terms of arbitrary parameter $\tau$ can be written as \cite{cordero_rt}\footnote{here $H^2=\frac{\Lambda}{3\beta}$, a constant quantity}
\begin{equation}
L(a, \dot{a}, \ddot{a}, \dot{t}, \ddot{t}) = \frac{a\dot{t}}{N^{3}} \left({a \ddot{a} \dot{t}-a \dot{a} \ddot{t} + N^{2}\dot{t}} \right) - N a^{3} H^{2}.
\label{hdlag}
\end{equation}
Varying the action with respect to the field $a(\tau)$ we get 
 the corresponding Euler Lagrange equation as 
\begin{equation}
\frac{d}{d \tau}\left( \frac{\dot{a}}{\dot{t}} \right) = - \frac{N^{2}}{a \dot{t}}\frac{(\dot{t}^{2}- 3N^{2}a^{2}H^{2})}{(3\dot{t}^{2}- N^{2}a^{2}H^{2})}.
\label{ELeqn}
\end{equation} 
Note that the Lagrangian (\ref{hdlag}) contains higher derivative terms of the field $a$. However we can write it as \cite{cordero_rt}
\begin{equation}
L= -\frac{a{\dot {a}}^2}{N} + aN\left(1 - a^2H^2\right) + \frac{d}{d\tau}\left(\frac{a^2{\dot{a}}}{N}\right).
\label{orglag}
\end{equation}
If we neglect the boundary term the resulting Lagrangian becomes usual first order one. As is well known the equation of motion is still given by (\ref{ELeqn}). However the Hamiltonian analysis is facilitated if we retain the higher derivative term. Thus our Hamiltonian analysis will proceed from 
(\ref{hdlag}). 

\section{Hamiltonian analysis }
This section contains the main results of this present chapter. As stated above our aim is to develop a new Hamiltonian analysis following from the Lagrangian (\ref{hdlag}) which is a second order theory. A Hamiltonian analysis of the same model has been discussed in \cite{cordero_rt} from the Ostrogradsky approach. We on the other hand adopt the equivalent first order formalism which has been demonstrated to be useful, specifically in treating the gauge invariances from the Hamiltonian point of view \cite{BMP, mukherjee_exmcs, paul_membrane}. The point of departure is to convert (\ref{hdlag}) to a first order theory by defining the first derivative of $a$ and $t$ as additional fields and including the following constraints into the Lagrangian with the help of undetermined multipliers. These multipliers are then treated as new fields and the phase space is constructed by the entire set of fields along with their conjugate momenta defined in the usual way as is done for first order theories. Automatically primary constraints arise. The constraint analysis is then presented in detail. In addition to first class constraints the model also has second class constraints. The second class constraints are then strongly implemented by substituting the Poisson brackets by the corresponding Dirac brackets. Effectively the theory becomes a first class system with the symplectic algebra given by these Dirac brackets of which a complete list has been given.

 The results derived so far are then used in two ways. First an analysis of the gauge invariances of the model has been done and its connection with the reparametrization invariance of the action has been discussed. Secondly, the gauge redundancy of the model has been eliminated by choosing an appropriate gauge. The final Dirac brackets have been used to reduce the phase space and indicate a formal quantization of the model.

  In the equivalent first order formalism, we define the new  fields as,  
\begin{eqnarray}
\nonumber
\dot{a} &=& A\\
\dot{t} &=& T,
\end{eqnarray}
which also introduce new constraints in the system given by 
 \begin{eqnarray}
\nonumber
A -\dot{a} \approx 0 \\
 T - \dot{t}  \approx 0.
\label{hdconst}
\end{eqnarray}
 Now the higher derivative Lagrangian (\ref{hdlag}) is transformed to the first order Lagrangian  where the constraints (\ref{hdconst}) are enforced through the Lagrange multipliers $\lambda_{a}$, and $\lambda_{t}$ as 
\begin{eqnarray}
\nonumber
L^{\prime}&=&\frac{aT}{\left(T^{2}-A^{2}\right)^{\frac{3}{2}}}\left( {aT\dot{A}-aA\dot{T}+\left(T^{2}-A^{2}\right)T}\right) - \left( T^{2}-A^{2}\right) ^{\frac{1}{2}}a^{3}H^{2} \\ && + \lambda_{a}\left(A-\dot{a}\right) + \lambda_{t}\left(T-\dot{t}\right).
\label{auxlag} 
\end{eqnarray}
The Euler Lagrange equation of motion, obtained    from the first order Lagrangian (\ref{auxlag}),  by varying w.r.t. a, A, t, T, $\lambda_{a}$ and $\lambda_{t}$, are respectively given  by
\begin{eqnarray}
&& \frac{2a(\dot{A}T^{2} - AT\dot{T})}{(T^{2}-A^{2})^{\frac{3}{2}}} + \frac{T^{2}}{(T^{2}-A^{2})^{\frac{1}{2}}} -3 a^{2}H^{2}(T^{2}-A^{2})^{\frac{1}{2}} + \dot{\lambda}_{a} =0
  \label{lag_eom_a} \\
%\end{eqnarray}
%\begin{eqnarray}
  && \frac{3a^{2}A(\dot{A}T^{2} - AT\dot{T})}{(T^{2}-A^{2})^{\frac{5}{2}}} - \frac{d}{d \tau}    \left( \frac{a^{2}T^{2}}{(T^{2}-A^{2})^{\frac{3}{2}}}  \right) - \frac{a^{2}T \dot{T}}{(T^{2}-A^{2})^{\frac{3}{2}}} + \frac{aAT^{2}}{(T^{2}-A^{2})^{\frac{3}{2}}} + \nonumber \\ &&  \frac{a^{3}AH^{2}}{(T^{2}-A^{2})^{\frac{1}{2}}} + \lambda_{a}  =0 \label{lag_eom_A}\\ 
 %\end{eqnarray}
 %\begin{eqnarray}
 &&\dot{\lambda}_{t} =0  \label{lag_eom_t} \\
%\end{eqnarray}
\nonumber\\ \nonumber\\
%\begin{eqnarray}
&& \frac{3a^{2}T(\dot{A}T^{2} - AT\dot{T})}{(T^{2}-A^{2})^{\frac{5}{2}}} + \frac{2a^{2} \dot{A}T}{(T^{2}-A^{2})^{\frac{3}{2}}} - \frac{d}{d \tau} \left( \frac{a^{2} A T}{(T^{2}-A^{2})^{\frac{3}{2}}}\right) -\frac{a^{2} A \dot{T}}{(T^{2}-A^{2})^{\frac{3}{2}}} \nonumber\\ && + \frac{2aT}{(T^{2}-A^{2})^{\frac{1}{2}}}   - \frac{aT^{3}}{(T^{2}-A^{2})^{\frac{1}{2}}} + \lambda_{t} =0 \label{lag_eom_T}\\
&& A-\dot{a} =0  \label{lag_eom_lambda_a}\\
&& T-\dot{t}=0. \label{lag_eom_lambda_t}
\end{eqnarray} 
Eliminating the multipliers $\lambda_{a}$, and $\lambda_{t}$ from the above equations we get back equation (\ref{ELeqn})

In the Hamiltonian formulation adopted here, the Lagrange multipliers are considered formally as independent fields and the momenta corresponding to them are introduced in the usual way. Here we denote  the phase space coordinates by $q_{\mu}=a, t, A, T, \lambda_{a},\lambda_{t}$ and their corresponding momenta  as $\Pi_{q_\mu}=\Pi_{a},\Pi_{t}, \Pi_{A}, \Pi_{T}, \Pi_{\lambda_{a}}, \Pi_{\lambda_{t}}$  with $\mu=0, 1, 2, 3, 4, 5$. We adopt the usual definition 
%Since the Lagrangian (\ref{auxlag}) is first order so momenta are defined trivially by $
\begin{equation}
\Pi_{q_{\mu}}  = \frac{\partial{L}^{\prime}}{\partial{\dot{q}_{\mu}}},
\end{equation} 
since the Lagrangian (\ref{auxlag}) is in the first order form. This is the point of departure of our Hamiltonian formulation from the Ostrogradsky formulation of \cite{cordero_rt}.
%We get
%\begin{eqnarray}
%\nonumber\\
%\Pi_{t}&=& -\lambda_{t}
%\nonumber\\
%\Pi_{a}&=& -\lambda_{a}
%\nonumber\\
%\Pi_{T}&=& - \frac{a^{2}TA}{\left(T^{2}-%A^{2}\right)^{\frac{3}{2}}}
%\nonumber\\
%\Pi_{A} &=& \frac{a^{2}T^{2}}{\left(T^{2}-A^{2}\right)^{\frac{3}{2}}}
%\nonumber\\
%\Pi_{\lambda_{a}}&=&0
%\nonumber\\
%\Pi_{\lambda_{a}}&=&0
%\end{eqnarray}
 
 From the definition of the phase space variables, we  get the following primary constraints
\begin{eqnarray}
\nonumber\\
\Phi_{1}&=& \Pi_{t} + \lambda_{t} \approx 0
\nonumber\\
\Phi_{2}&=& \Pi_{a} + \lambda_{a} \approx 0
\nonumber\\
\Phi_{3}&=&\Pi_{T} + \frac{a^{2}TA}{\left(T^{2}- A^{2}\right)^{\frac{3}{2}}} \approx 0
\nonumber\\
\Phi_{4}&=& \Pi_{A} - \frac{a^{2}T^{2}}{\left(T^{2}-A^{2}\right)^{\frac{3}{2}}} \approx 0
\nonumber\\
\Phi_{5}&=& \Pi_{\lambda_{t}}\approx0
\nonumber\\
\Phi_{6}&=& \Pi_{\lambda_{a}} \approx 0.
\end{eqnarray}

%Poisson brackets for the system are defined as
%\begin{equation}
%\left\lbrace{x_{\mu}, \Pi_{x_{\nu}}} \right\rbrace = \delta{_{\mu\nu}} 
%\end{equation} 
The nonzero Poisson brackets between the primary constraints are computed as
\begin{eqnarray}
\nonumber\\
\left\lbrace {\Phi_{1}, \Phi_{5}}\right\rbrace &=& 1
\nonumber\\
\left\lbrace {\Phi_{2}, \Phi_{3}}\right\rbrace &=& -\frac{2aTA}{\left(T^{2}-A^{2}\right)^{\frac{3}{2}}}
\nonumber\\
\left\lbrace {\Phi_{2}, \Phi_{4}}\right\rbrace &=& \frac{2aT^{2}}{\left(T^{2}-A^{2}\right)^{\frac{3}{2}}}
\nonumber\\
\left\lbrace {\Phi_{2}, \Phi_{6}}\right\rbrace &=& 1.
\end{eqnarray}
Taking the constraint combination  
\begin{equation}
\Phi_{3}^{\prime} = T\Phi_{3} + A\Phi_{4}, \approx 0,
\end{equation}
 we find that $\Phi_{3}^{\prime}$ commutes with all the constraints. The nonzero poisson brackets between the newly defined primary set of constraints $\Phi_{1},\Phi_{2},\Phi_{3}^{\prime},\Phi_{4},$ $\Phi_{5},\Phi_{6},$ become
 \begin{eqnarray}
\nonumber\\
\left\lbrace {\Phi_{1}, \Phi_{5}}\right\rbrace &=& 1
\nonumber\\
%\left\lbrace {\Phi_{2}, \Phi_{3}}\right\rbrace &=& -\frac{2aTA}{\left(T^{2}-A^{2}\right)^{\frac{3}{2}}}
%\nonumber\\
\left\lbrace {\Phi_{2}, \Phi_{4}}\right\rbrace &=& \frac{2aT^{2}}{\left(T^{2}-A^{2}\right)^{\frac{3}{2}}}
\nonumber\\
\left\lbrace {\Phi_{2}, \Phi_{6}}\right\rbrace &=& 1.
\end{eqnarray}
We can write down the canonical Hamiltonian as  
\begin{eqnarray}
\nonumber\\
H_{can} &=& \Pi_{q_{\mu}}\dot{q}_{\mu} - L^{\prime} 
\nonumber\\
        &=& -\frac{aT^{2}}{\left(T^{2}-A^{2}\right)^{\frac{1}{2}}} + \left(T^{2}-A^{2}\right)^{\frac{1}{2}} a^{3}H^{2} -\lambda_{a}A - \lambda_{t}T.
\end{eqnarray}

 The total Hamiltonian is given by
 \begin{equation}
 H_{T} = H_{can} + \Lambda_{1}\Phi_{1} + \Lambda_{2}\Phi_{2} +  \Lambda_{3}\Phi_{3}^{\prime} + \Lambda_{4}\Phi_{4}+ \Lambda_{5}\Phi_{5} + \Lambda_{6}\Phi_{6}.
 \label{H_T} 
 \end{equation}
 Here $\Lambda_{1},\Lambda_{2}, \Lambda_{3}, \Lambda_{4}, \Lambda_{5}, \Lambda_{6}$ are  undetermined Lagrange multipliers.
% To get the secondary constraints one essentially  needs their time variation set to zero i.e.  brackets of the primary constraints with $H_{T}$ is should be zero .
  Preserving the primary constraints $\Phi_{1}$, $\Phi_{5}$, $\Phi_{6}$  in time ($ \{\Phi_{i}, H_{T}\} \approx 0$) the following Lagrange multipliers get fixed
 \begin{eqnarray}
 \nonumber\\
 \Lambda_{5}&=& 0 
 \nonumber\\
\Lambda_{1} &=& T
 \nonumber\\
 %\Lambda_{4}&=&
 %\nonumber\\
\Lambda_{2} &=& A. 
 \nonumber
 %\Lambda_{6} &=&
\end{eqnarray}
Whereas, conservation of $\Phi_{2}$ gives the following condition between $\Lambda_{4}$ and $\Lambda_{6}$
\begin{equation}
\frac{T^{2}}{\left(T^{2}-A^{2}\right)^{\frac{1}{2}}} - 3a^{2}H^{2}\left(T^{2} - A^{2}\right)^{\frac{1}{2}} + \Lambda_{6} + \Lambda_{4} \frac{2aT^{2}}{\left(T^{2}-A^{2}\right)^{\frac{3}{2}}}=0.
\label{32}
\end{equation}
 
Time preservation of the constraint  $\Phi_{3}^{\prime}$ gives rise to the following secondary constraint
 \begin{equation}
 \Psi_{1} = \frac{aT^{2}}{\left(T^{2}-A^{2}\right)^{\frac{1}{2}}} - a^{3}H^{2} \left(T^{2}-A^{2}\right)^{\frac{1}{2}} + \lambda_{t}T + \lambda_{a}A \approx 0.
\end{equation} 
Likewise, $\Phi_{4}$ yields the following secondary constraint 
\begin{equation}
\Psi_{2}=\frac{aAT^{2}}{\left(T^{2}-A^{2}\right)^{\frac{3}{2}}} - \frac{a^{3}H^{2}A}{\left(T^{2}-A^{2}\right)^{\frac{1}{2}}} - \lambda_{a} \approx 0.
\end{equation}
Nonzero brackets for $\Psi_{1}$ and $\Psi_{2}$  with the  other constraints are given below,
\begin{eqnarray}
\nonumber\\
\left\lbrace {\Phi_{2}, \Psi_{1}}\right\rbrace  &=& -\frac{T^{2} - 3a^{2}H^{2}\left(T^{2}-A^{2}\right)}{\left(T^{2}-A^{2}\right)^{\frac{1}{2}}}
\nonumber\\
\left\lbrace {\Phi_{4}, \Psi_{1}}\right\rbrace  &=& -\frac{aAT^{2}}{\left(T^{2}-A^{2}\right)^{\frac{3}{2}}} - \frac{a^{3}H^{2}A}{\left(T^{2}-A^{2}\right)^{\frac{1}{2}}} - \lambda_{a}
\nonumber\\
\left\lbrace {\Phi_{3}, \Psi_{1}}\right\rbrace  &=& - \frac{aT(2A^{2} - T^2)}{\left(T^{2}-A^{2}\right)^{\frac{3}{2}}} + \frac{a^{3} H^{2} T}{\left(T^{2}-A^{2}\right)^{\frac{1}{2}}} - \lambda_{t} 
\nonumber\\
\left\lbrace {\Phi_{5}, \Psi_{1}}\right\rbrace  &=& -T
\nonumber\\
\left\lbrace {\Phi_{6}, \Psi_{1}}\right\rbrace  &=& -A
\nonumber\\
\left\lbrace {\Phi_{2}, \Psi_{2}}\right\rbrace  &=& -\frac{AT^{2}}{\left(T^{2}-A^{2}\right)^{\frac{3}{2}}} + \frac{3a^{2}H^{2}A}{\left(T^{2} - A^{2}\right)^{\frac{1}{2}}}
\nonumber\\
\left\lbrace {\Phi_{4}, \Psi_{2}}\right\rbrace  &=& - \frac{aT^{2} \left(T^{2} +2A^{2}\right)}{\left(T^{2}-A^{2}\right)^{\frac{5}{2}}} + \frac{a^{3}H^{2}T^{2}}{\left(T^{2} - A^{2}\right)^{\frac{3}{2}}}
\nonumber\\
\left\lbrace {\Phi_{6}, \Psi_{2}}\right\rbrace  &=& 1.
\end{eqnarray}
Time preservation of  $\Psi_{1}$ trivially gives 0 = 0. A similar analysis involving $\Psi_{2}$ yields, on exploiting (\ref{32}),  

\begin{eqnarray}
\nonumber\\
\Lambda_{4} &=& - \frac{\left(T^{2} - 3a^{2}H^{2}\left(T^{2}-A^{2}\right)\right) \left(T^{2}-A^{2}\right)}{a\left(3T^{2} - a^{2}H^{2}\left(T^{2}-A^{2}\right)\right)}
\nonumber\\
\Lambda_{6} &=& - \frac{\left({ T^{2} - 3a^{2}H^{2}\left(T^{2}-A^{2}\right) }\right)  ( T^{2}-a^{2}H^{2}\left(T^{2}-A^{2}\right)^{\frac{1}{2}} )}{\left({T^{2}-A^{2}}\right)^{\frac{1}{2}} \left({3T^{2} -a^{2}H^{2} \left({T^{2}-A^{2}}\right)}\right)}.
\end{eqnarray}
The iterative procedure is thus closed and no more secondary constraints or other relations are generated.

 The above analysis reveals that of all the Lagrange multipliers $\Lambda_i$, only $\Lambda_1$ remains undetermined in (\ref{H_T}) signifying one independent gauge degree of freedom. It is interesting to note that this consistency is not always obvious in the Ostrogradsky formulation, as we have already mentioned in connection with the massive relativistic particle model  \cite{nesterenko}.

 We have now altogether eight primary and secondary constraints. Computation of the Poisson bracket between these constraints shows that only $\Phi_{3}^{\prime}$ is the first class constraint, whereas other seven  constraints are apparently second class. The odd number of apparently second class constraints signals the existence of additional first class constraints.  Indeed, 
the new constraint combination 
\begin{equation}
\Psi_{1}^{\prime} = \Psi_{1} - \Lambda_{1}\Phi_{1}- \Lambda_{2}\Phi_{2}- \Lambda_{4}\Phi_{4}- \Lambda_{5}\Phi_{5}- \Lambda_{6}\Phi_{6},
\label{new_combination}
\end{equation}
leads to a secondary first class constraint.
So now we have two  first class constraints $\Phi_{3}^{\prime}$, $\Psi_{1}^{\prime}$ and six  second class constraints $\Phi_{1}$, $\Phi_{2}$, $\Phi_{4}$, $\Phi_{5}$, $\Phi_{6}$ and $\Psi_{2}$. The total number of phase space variables is twelve. The number of independent phase space variables is therefore $12 - (2\times 2 +6)$ i.e. $2$. Later on we will explicitly identify these two variables. There is no enhancement of degrees of freedom as is customary for the higher derivative systems. This is consistent with the fact that (\ref{hdlag}) is not a genuine higher derivative system. Also, of the two first class constraints of the system,  $\Phi_{3}^{\prime}$ is the sole primary first class constraint. The number of primary first class constraint matches with the residual number of undetermined multiplier in the total Hamiltonian. This fact will be important in the construction of the gauge generator.

 To study gauge symmetry of the system we need to get rid of the second class constraints. This is done by the introduction of the Dirac brackets which enable us to  set these constraints strongly zero. For simplicity of the calculation we remove them pair by pair. The Dirac bracket between the basic fields after removing $\Phi_{1}, \Phi_{2}, \Phi_{5}, \Phi_{6}$ remains same as their corresponding Poisson brackets.    
Solving $\Phi_{1}, \Phi_{2}, \Phi_{5}, \Phi_{6}$ the new constraint structure becomes 
\begin{eqnarray}
\nonumber\\
F_{1}&=&\Phi_{3}^{\prime}  =T\Phi_{3} + A \Phi_{4} \approx 0
\nonumber\\
 F_{2}&=&\Psi_{1}^{\prime}= \Psi_{1} - \Lambda_{4}\Phi_{4} \approx 0
\nonumber\\
S_{1} &=&  \Phi_{4} \approx 0
\nonumber\\
S_{2}&=& \Psi_{2}= \frac{aAT^{2}}{\left({T^{2}-A^{2}}\right)^{\frac{3}{2}}} - \frac{a^{3}AH^{2}}{\left(T^{2}-A^{2}\right)^{\frac{1}{2}}} + \Pi_{a} \approx 0.
\label{secondclass}
\end{eqnarray}
For simplicity we use new notations \{$F_{1}$, $F_{2}$\} and  \{$S_{1}$, $S_{2}$\} where, the first pair denotes the set of first class constraint and second pair denotes the remaining  set of second class constraints. Some details of this reduction are given below.

To calculate Dirac brackets of the theory we first find out the Poisson brackets between the second class constraints which are written as 
\begin{equation}
\Delta_{ij}=\{S_{i}, S_{j}\}= -\frac {aT^{2} \left({3T^{2}-a^{2}H^{2} \left({T^{2} -A^{2}}\right)}\right)}{\left(T^{2}-A^{2}\right)^{\frac{5}{2}}}  \epsilon_{ij},
\end{equation}
with $\epsilon_{12} = 1$ and $ i, j = 1, 2$. %Dirac brackets are defined by
%\begin{equation}
%\{f,g\}_{D} = \{f,g\} -\{f, S_{i}\}\Delta_{ij}^{-1}\{S_{j},g\}.
%\label{dbdef}
%\end{equation}
 Using (\ref{DB_def}), we calculate the Dirac brackets between the basic fields which are given below (only the nonzero brackets are listed)
\begin{eqnarray}
\nonumber\\
\left\lbrace {a,A}\right\rbrace_{D}&=& -\frac {\left(T^{2}-A^{2}\right)^{\frac{5}{2}}}{aT^{2} \left({3T^{2}-a^{2}H^{2} \left({T^{2} -A^{2}}\right)}\right)}
\nonumber\\
\left\lbrace {a, \Pi_{a}}\right\rbrace_{D}&=& \frac{T^{2}+ 2 A^{2} - a^{2}H^{2} \left(T^{2}-A^{2}\right)}{\left(3T^{2} - a^{2}H^{2}\left(T^{2} - A^{2}\right) \right)}
 \nonumber\\
\left\lbrace {a, \Pi_{A}}\right\rbrace_{D}&=& -\frac{3aA}{3T^{2} - a^{2}H^{2} \left(T^{2}-A^{2}\right)}
 \nonumber\\
\left\lbrace {a, \Pi_{T}}\right\rbrace_{D}&=& \frac{a\left(T^{2}+ 2A^{2}\right)}{T\left(3T^{2}-a^{2}H^{2}(T^{2}-A^{2})\right)}
\nonumber\\
\left\lbrace {t,\Pi_{t}}\right\rbrace_{D}&=&1 \nonumber\\
\left\lbrace {A, \Pi_{a}}\right\rbrace_{D}&=& - \frac{A \left(T^{2}-A^{2}\right) \left(T^{2}-3a^{2}H^{2} \left(T^{2}-A^{2}\right)\right)}{aT^{2}\left(3T^{2} - a^{2}H^{2} \left(T^{2}-A^{2}\right)\right)} 
\nonumber\\
\left\lbrace {A, \Pi_{A}}\right\rbrace_{D}&=& \frac{2 \left(T^{2}-A^{2}\right)}{3T^{2} - a^{2}H^{2} \left(T^{2}-A^{2}\right)}
 \nonumber\\
 \left\lbrace{A, \Pi_{T}}\right\rbrace_{D}&=& \frac{A \left(T^{2} + 2A^{2} - a^{2}H^{2} \left(T^{2}-A^{2}\right) \right)}{T\left(3T^{2} - a^{2}H^{2} \left(T^{2}-A^{2}\right)\right)}
 \nonumber\\
\left\lbrace {T, \Pi_{T}}\right\rbrace_{D}&=& 1 \nonumber\\
\left\lbrace {\Pi_{a}, \Pi_{A}}\right\rbrace_{D}&=& -\frac{a \left(2T^{4}+A^{2}T^{2} + a^{2}H^{2}(T^{2}-A^{2})(9A^{2}-2T^{2})\right)}{ \left(T^{2}-A^{2}\right)^{\frac{3}{2}}(3T^{2} - a^{2}H^{2} \left(T^{2}-A^{2}\right)) }
 \nonumber\\
\left\lbrace {\Pi_{a}, \Pi_{T}}\right\rbrace_{D}&=& \frac{aA \left(T^{4}+2T^{2}A^{2} + a^{2}H^{2}(T^{2}-A^{2})(T^{2}+6A^{2})\right)}{T\left(T^{2}-A^{2}\right)^{\frac{3}{2}}(3T^{2} - a^{2}H^{2} \left(T^{2}-A^{2}\right) )}
 \nonumber\\
\left\lbrace {\Pi_{A}, \Pi_{T}}\right\rbrace_{D}&=& -\frac{a^{2}T \left(T^{2} +2A^{2} - a^{2}H^{2} (T^{2}-A^{2})\right)}{\left(T^{2}-A^{2}\right)^{\frac{3}{2}}(3T^{2} - a^{2}H^{2} \left(T^{2}-A^{2}\right))}.
\label{intdb} 
\end{eqnarray}

 The introduction of the above Dirac brackets allows the second class pair \{$S_{1}$, $S_{2}$\} to be strongly implemented. Note that the secondary first class constraint then becomes equal to the canonical Hamiltonian: 
\begin{equation}
F_{2}=\Psi_{1}= -H_{c}=-T\Pi_t - \frac{T^2}{A}\Pi_a \approx 0.
\end{equation} 
Vanishing of the canonical Hamiltonian is a consequence of the reparametrisation invariance of the theory. 

 %Before going to the construction of the  gauge generator in the next section it is worthwhile to note an interesting characteristics of the algebra of the first class constraints with respect to the Dirac brackets (\ref{intdb}). We notice the Dirac brackets between the first class constraints can be written in a compact form  as 
% \begin{equation}
%\{F_{i}, F_{j}\}_{D} = - \epsilon_{ij}F_{2} \ \ \  ;\ \ \ \ \ i, j=1, 2
%\label{frstconsbrak}
%\end{equation}
%This algebra between the constraints has a certain structure which can clearly be seen if we denote them in the following notation  
%\begin{eqnarray}
 %L_{0} &=&   F_{1}\\
 %L_{1} &=&  F_{2}.
%\end{eqnarray} 
%So, the equation (\ref{frstconsbrak}) becomes
%\begin{equation}
%\{ L_{m}, L_{n}\}_{D} = (m-n) L_{m+n},
%\end{equation} 
 %with $m=0$, $n=1$ this is the form of truncated Virasoro algebra  as proposed in \cite {ho} for HD cases.  \\
%The above analysis essentially completes our preparation for the construction of the gauge generator. We notice that there is only one primary first class constraint and one undetermined Lagrange multiplier in the total Hamiltonian. We can thus expect to have only one gauge symmetry.
 
%In the next section we concentrate to find out the gauge symmetry. 

\subsection{Construction of the gauge generator}

  The gauge generator is defined in (\ref{gauge_gen_def}) can be written as,
\begin{equation}
 G= \epsilon_{1} F_{1} + \epsilon_{2} F_{2}.
 \label{gaugegenerator11}
 \end{equation}
 Here $\epsilon_{1}$ and $\epsilon_{2}$ are the gauge parameters. 
% Now, to find out the independent gauge parameters we use equations  (\ref{master2}).
From equations (, \ref{structure constant_V} \ref{structure constant_C}) we find that $C_{122} = -1 = -C_{212}$ and $V_{12}=1$  are the only nonzero structure functions. Now using equation (\ref{master_eqn_2})the following relation between the gauge parameters is obtained 
 \begin{equation}
 \epsilon_{1}= - \Lambda_{3}\epsilon_{2} - \dot{\epsilon}_{2}. 
 \label{parameterrelation1}
 \end{equation}
 So here  $\epsilon_{2}$ may be chosen as the independent gauge parameter. 
%So we found that there is only one independent gauge transformation which essentially is in conformation with the fact that there is only one independent primary first class constraints.

 At this stage we observe that there is one independent parameter in the gauge generator (\ref{gaugegenerator11}). The conditions (\ref{varsgauge}) following from the higher derivative nature is yet to be implemented. As has been mentioned earlier this may or may not impose additional restriction on the gauge parameters. 
 The gauge transformations of  the fields  are given by 
 \begin{eqnarray}
 \delta{a} &=& \{a, G\}_D = -  \epsilon_{2} A \label{gaugetrans_a}\\
 \delta{t} &=& -  \epsilon_{2} T \label{gaugetrans_t}\\
 \delta{A} &=&   \epsilon_{1} A  - \epsilon_{2} \frac{\left(T^{2} - 3a^{2}H^{2} \left(T^{2}-A^{2}\right)\right)\left(T^{2}-A^{2}\right)}{a\left(3T^{2} - a^{2}H^{2} \left(T^{2}-A^{2}\right)\right)}
  \label{gaugetrans_A}\\
 \delta{T} &=&   \epsilon_{1} T 
 \label{gaugetrans_T}
 \end{eqnarray}
After some calculation we find that
 \begin{eqnarray}
 \nonumber\\
 \frac{d}{d \tau}\delta {a} &=& \delta A \label{hd_gauge_a}\\
   \frac{d}{d \tau}\delta {t} &=& \delta T. \label{hd_gauge_t}
 \end{eqnarray}
So the constraints (\ref{varsgauge}) hold identically for the present model and impose no new condition on the gauge parameters. We find therefore that there is only one independent gauge transformation which essentially is in conformation with the fact that there is only one independent primary first class constraint.
 
 The gauge variations obtained from the Hamiltonian analysis can be exactly mapped to the reparametrization invariance of the model. Consider arbitrary infinitesimal  change in the parameter $\tau \rightarrow \tau^{\prime} = \tau + \sigma$. The action is invariant under this reparametrization. Now the fields transform as
\begin{eqnarray}
\nonumber
\delta a &=&  - \sigma a \\
\delta t &=&  - \sigma t.
\end{eqnarray} 
These are identical with the gauge variations (\ref{gaugetrans_a}) and (\ref{gaugetrans_t}) of $a$ and $t$ if $\sigma$ is identified with $\epsilon_{2}$. The equivalence of gauge invariances with the reparametrization invariance of the model is thus established.

\subsection{Gauge fixing and formal quantization}
 After the reduction of phase space by the  Dirac bracket procedure we are left with only the two first class constraints $F_1$ and $F_2$. These first class constraints reflect the redundancy of the theory which are connected by gauge transformations. In the above analysis our focus was on the abstraction of the gauge degrees of freedom. We now elucidate a formal quantisation prescription. A gauge fixing is done and the appropriate WDW equation is written.

           The choice of gauge is arbitrary subject to the conditions that they must reduce the first class constraints to second class. Also the constraint algebra should be nonsingular. As there are two first class constraints we need two gauge conditions. We take one of these to be the cosmic gauge
%The gauges are
\begin{eqnarray}
\varphi_{1} &=& \sqrt{T^2 - A^2} -1 \approx 0.
\label{gauge1} 
\end{eqnarray}
The name derives from the fact that the resultant metric becomes the usual FLRW metric. As the second gauge condition we take  
\begin{eqnarray}
\varphi_{2} &=& T - \alpha a \approx 0.
\label{gauge2}
\end{eqnarray}
where the constant $\alpha$ is chosen so that $\alpha \ne H$. The following calculations will show that these are appropriate gauge conditions.

           As usual the gauge conditions are treated as additional constraints which make the first class constraints of the theory second class. For convenience, renaming the two first class constraints we write the complete set of constraints as
\begin{eqnarray}
\Omega_{1} = F_{1}\\
\Omega_{2} = F_{2}\\
\Omega_{3} = \varphi_{1}\\ 
\Omega_{4} =  \varphi_{2}.
\end{eqnarray}
Modifying the algebra by the Dirac brackets corresponding to this second class system we will be able to put  all the second class constraints ($\Omega_i, i = 1, 2, 3, 4$) to be strongly equal to zero. These will correspond to operator relations in the corresponding quantum theory.

        Using the algebra (\ref{intdb}) we can straightforwardly compute the algebra of the constraints $\Omega_i$. The results are given in the following table
 \begin{table}[ht]
\caption{Constraint brackets} % title of Table
\centering 
\begin{tabular}{c c c c c} 
\hline\hline
 & $\Omega_{1}$ & $\Omega_{2}$& $\Omega_{3}$ & $\Omega_{4}$ \\ [1.0 ex] % inserts table
%heading
\hline % inserts single horizontal line

$\Omega_{1}$ & 0 & 0 & -1 & $-T$  \\
$\Omega_{2}$ & 0 &0 & $\frac{A (\alpha^{2} - 3 H^2)}{a (3\alpha^{2} - H^2)}$ & $- \alpha A$ \\
$\Omega_{3}$ & 1 & $-\frac{A (\alpha^{2} - 3 H^2)}{a (3\alpha^{2} - H^2)}$ & 0 & $\frac{A}{\alpha a^5 (3 \alpha^2 - H^2)}$\\
$\Omega_{4}$ & $T$ & $ \alpha A$&   $-\frac{A}{\alpha a^5 (3 \alpha^2 - H^2)}$ & 0 \\ [1ex] % [1ex] adds vertical space
\hline %inserts single line
\end{tabular}
\label{table:nonlin} % is used to refer this table in the text
\end{table}\\
From the above table we can read off the matrix
\begin{eqnarray}
\nonumber\\
\Delta_{ij} = \{\Omega_i,\Omega_j\}
\end{eqnarray}
Using the definition (\ref{DB_def}) we can calculate the final Dirac brackets.
Nonzero Dirac brackets between the phase space variables are
\begin{eqnarray}
\nonumber\\
\{ t , a\}^{*} &=& \frac{1}{4 \alpha a^3 (\alpha^2 - H^2)}
\nonumber\\
\{ t , A \}^{*} &=& \frac{\alpha}{4  a^2 A (\alpha^2 - H^2)}
\nonumber\\
\{ t , T \}^{*} &=& \frac{1}{4 a^3 (\alpha^2 - H^2)}
\nonumber\\
\{ t , \Pi_{a}\}^{*} &=& \frac{-4 a^2 \alpha^2 + 3}{4 \alpha a A}
\nonumber\\
\{ t , \Pi_{A} \}^{*} &=& \frac{\alpha}{(\alpha^2 - H^2)}
\nonumber\\
\{ t , \Pi_{t}\}^{*} &=& 1
\nonumber\\
\{ t , \Pi_{T}\}^{*} &=& \frac{-4 a^2 \alpha^2 + 3}{4  a A (\alpha^2 - H^2)}.
\label{final_dirac}
\end{eqnarray}
With the introduction of the final Dirac brackets all the constraints (including the gauge conditions) become second class
and strongly zero. We thus have the following conditions on the phase space variables 
\begin{eqnarray}
\Pi_{T} + \frac{a^{2}TA}{\left(T^{2}- A^{2}\right)^{\frac{3}{2}}} &=& 0
\nonumber\\
 \Pi_{A} - \frac{a^{2}T^{2}}{\left(T^{2}-A^{2}\right)^{\frac{3}{2}}} &=& 0
 \nonumber\\
 -\Pi_{t} - \frac{T}{A}\Pi_{a} &=& 0
 \nonumber\\
 \frac{aAT^{2}}{\left({T^{2}-A^{2}}\right)^{\frac{3}{2}}} - \frac{a^{2}AH^{2}}{\left(T^{2}-A^{2}\right)^{\frac{1}{2}}} + \Pi_{a} &=& 0
 \nonumber\\
 \sqrt{T^2 - A^2} &=& 1
 \nonumber\\
 T - \alpha a &=& 0
\end{eqnarray}
 where use has been made of equations (\ref{secondclass}, \ref{gauge1},   \ref{gauge2}). From the final Dirac brackets (\ref{final_dirac}) it is clear that only the pair $(t, \Pi_t)$ is canonical. We thus identify this pair as the two independent phase space degrees of freedom found earlier by a standard count using the constraints of the system (see below {\ref{new_combination}}). To develop a quantum theory  it is necessary to write down the whole theory with respect to the canonical variables in the reduced phase. All the variables can be expressed in favour of $(t, \Pi_t)$ by appropriately solving the constraints which are now strongly implemented. The result is,
 %Solving the set of equations we can express the phase space variables in terms of $T, \Pi_{T}, A, \Pi_{A}, \Pi_{a}$ in terms of $a$ as
 %This enable us to express the phase space variables in terms of a subset containing number of elements equal to the number of degrees of freedom i.e. two.  
\begin{eqnarray}
\nonumber\\
T &=& \alpha a \nonumber\\
A &=& \sqrt{\alpha^{2} a^2 - 1} \nonumber\\
\Pi_{A} &=& \alpha^{2} a^{4} 
\nonumber\\
\Pi_{T} &=& - \alpha a^{3} \sqrt{\alpha^{2} a^{2} - 1} \nonumber\\
\Pi_{a} &=&  - a^{3}(\alpha^{2} - H^2) \sqrt{\alpha^{2} a^2 - 1}.
\label{gaugephase1} 
\end{eqnarray}
where $a$ is expressed as
\begin{equation}
a = \left( {\frac{\Pi_{t}}{\alpha(\alpha^2 - H^2)}} \right) ^{\frac{1}{4}}
\label{gaugephase2}.
\end{equation}
Thus we find that all the phase space variables except $t$ are given as function of $\Pi_{t}$. 
%So only $t$ and $\Pi_{t}$ remain as independent phase space variables which is consistent with the fact that the number of degrees of freedom in the phase space is two.

The passage from the classical to quantum theory proceeds in the usual way. The phase space variables are lifted to operators in some Hilbert space and the conditions (\ref{gaugephase1}, \ref{gaugephase2}) are now treated as operator relations. The Dirac brackets are promoted to commutators according to the prescription.  
%The theory is now quantised by lifting the phase space variables as operators in some Hilbert space and converting the Dirac brackets to commutators by the usual prescription
\begin{equation}
\{B,C\}^* \to \frac{1}{i\hbar}[B,C].
\end{equation}
The fundamental canonical algebra is thus (with $\hbar = 1$ )
\begin{equation}
[t, t] =[\Pi_{t}, \Pi_{t}] = 0, \ \   [t, \Pi_{t}] = i. 
\label{commutator}
\end{equation}
	
	We next proceed to formulate the WDW equation for the universe governed by the Lagrangian (\ref{orglag}). Before that we  write down the first class constraint $F_2$ which is the canonical Hamiltonian as
\begin{equation}
F_2 = -H_{can} = - \frac{-A^{2}\Pi_{t}^{2}+T^{2}\Pi_{a}^{2}}{A\Pi_{a}} = 0
\label{final_canonical}
\end{equation}
 Considering the sate vector $|\Psi\rangle$ in the appropriate Hilbert space, the WDW equation may be written as,
 \begin{equation}
 H_{can} |\Psi\rangle=0.
 \end{equation}
 Using the Schrodinger representation compatible with (\ref{commutator}), we obtain,
 \begin{equation}
 \Pi_t = -i\frac{\partial}{\partial{t}}
 \label{schrodinger}
 \end{equation}
 Exploiting (\ref{final_canonical}-\ref{schrodinger}) and the expression for $A$ given in (\ref{gaugephase1}) we obtain, after some algebra, the following WDW equation,
 \begin{equation}
-\frac{\partial^{2}}{\partial{t^2}} |\Psi\rangle = \alpha^2 a^8 (\alpha^2 - H^2)^2 |\Psi\rangle.
\label{WDW1}
 \end{equation}
 Making a change of variables $\xi=\frac{\alpha^2}{H^2}$, the WDW equation may be reexpressed as,
 \begin{equation}
 -\frac{\partial^{2}}{\partial{t^2}} |\Psi\rangle =  \xi (\xi - 1)^{2}H^{6}a^{8}  |\Psi\rangle.
\label{WDW2}
\end{equation}  
The above equation exactly reproduces one piece of the bifurcated WDW equation found in the first item of \cite{rt_model} \footnote{Note that the other part of the bifurcated WDW involving the $`a'$ variable is nonexistent in the present analysis. This is because here we have only one (configuration space) independent degree of freedom (i.e. $t$) instead of two variables ($t$ and $a$) as occurs in \cite{rt_model}. This mismatch happens because, contrary to \cite{rt_model}, the present analysis is done in a fully reduced space where all constraints are eliminated} .

	Furthermore, introducing the conserved `energy' $\omega$ by,
	\begin{equation}
	 \xi (\xi - 1)^{2}H^{6}a^{8} = \omega^2
	 \label{conserved_energy}
	\end{equation}
	we may reexpress (\ref{WDW2}) by the standard equation,
	\begin{equation}
 -\frac{\partial^{2}}{\partial{t^2}} |\Psi\rangle =  \omega^2  |\Psi\rangle.
\label{WDW3}
\end{equation} 
The expression for the conserved energy $\omega$ in (\ref{conserved_energy}) matches with the form given in \cite{rt_model}. It is now possible to proceed with the quantisation as elaborated in \cite{rt_model}.

	Before concluding this section it is worthwhile to mention the efficacy of the gauge choice (\ref{gauge2}). While the first gauge condition (\ref{gauge1}) is the standard cosmic gauge, the second one (\ref{gauge2}) has not been considered earlier. We have shown that this simple choice (\ref{gauge2}) is a valid choice that yields the fully reduced space of the model. Also, at the quantum level, the WDW equation subjected  to this gauge fixing reproduces the expression obtained earlier in \cite{rt_model}.  
 %The above equation can be identified to be consistent with the temporal component of the WDW equations found by Davidson et al in \cite{davidson1}. By this we can indentify the significance of the second gauge (\ref{gauge2}) which has a relation to the energy conservation equation given in the same paper.
  %The WDW equation corresponding to the $(a, \Pi_a)$ is irrelevant here, as in this reduced phase space they are no more canonical, which is evident from the brackets(\ref{final_dirac}). While if we neglect the surface term in  the model (\ref{gauge2}) only $a(\tau)$ remains as the canonical variable. The corresponding wave function along with the WDW equations are elaborately discussed in  \cite{davidson1}.
%%%%%%%%%%%%%%%%%%%%%%%%%%%%%%%%%%%%%%%%%%%%%%%

%%%%%%%%%%%%%%%%%%%%%%%%%%%%%%%%%%%%%%%%%%%%%
\section{Discussions}
In this chapter, we have considered the Regge Teitelboim (RT) model in the minisuperspace version. The Einstein Hilbert action in 4D is expressed with respect to the minisuperspace variables. Apparently,  the theory is a higher derivative theory, however, a surface term can be identified which actually contains the higher derivative term. As  surface term plays important role in gravity so they cannot be neglected as in usual theories. Keeping this surface term, we considered the higher derivative version of the RT model and the first order formalism is followed to explore its gauge symmetries and quantise the model formally in the reduced phase space. There appears only one independent primary first class constraint and the corresponding gauge degrees of freedom is identified to be raparametrisation symmetry. To quantise the theory we remove the two first class constraints by considering two gauge conditions. One of the suitable gauge conditions we considered is the well known cosmic gauge while the other one is newly  proposed. Under these conditions the  two first class constraints become second class and  the theory is quantised in the fully reduced phase space after removing the second class constraints. The Wheeler DeWitt equation which is the analogue of Schrodinger equation   in cosmology is constructed. Its compatibility with the existing literature\cite{rt_model} is shown.

\chapter{Conclusions}
Higher derivative theories, due to their importance in various fields of the physics, have become very important and interesting field to be investigated.  The higher derivative theories contain many subtle aspects like the enhancement in the degrees of freedom in comparison with the first order theory. In no way, importance of the symmetry studies of these theories can be minimised. There are certain aspects of symmetry which show startling differences from first order theories. For example, it has been reported in the case of massive relativistic particle model with curvature that the number of independent primary first class constraints (PFC)  apparently do not match with the number of independent gauge degrees of freedom \cite{nesterenko}. This matching is known to exist in conventional theories. Notably, later study follows the Ostrogradski approach of Hamiltonian analysis of the higher derivative theories.

 In this thesis, we have followed a first order formalism \cite{BMP} rather than the Ostrogradski approach. We have developed a systematic algorithm of constructing the independent gauge generator using the technique developed by Banerjee et. al. for the usual first order theory. Also, we have shown the quantisation of a specific higher derivative theory.

	 We followed the first order approach for the Hamiltonian analysis  of  the higher derivative theories. The reason to consider the first order approach is that one can easily perform the canonical analysis of these higher derivative theories like the usual first order theories. Moreover, as far we have seen the proper identification of the gauge degrees of freedom can be done without any nontriviality. We already have seen that there appears a mismatch in the number of independent gauge degrees of freedom and number of independent primary first class constraints which indeed was solved by considering this first order approach \cite{BMP}. According to this formalism we consider all the higher time derivatives of the dynamical variables as independent fields and introduce them separately. Due to redefinition of the fields some interrelation between the fields appear which are treated as constraints. These constraints are added to the Lagrangian via some Lagrange multipliers. The Lagrangian is now  a first order one. The gauge symmetries can now be abstracted  via the Hamiltonian formulation. The gauge generator is constructed by using the first class constraints \cite{dirac2}. To get the independent gauge transformations we employ the equations derived in \cite{BRR}. Other conditions appear strictly due to the higher derivative structure. Specifically, it has been noticed that this can give rise to some new relations which can be treated as constraints.  After removing all the constraints by the Dirac bracket technique, now the theory can be quantised.The algorithm developed in this thesis has been applied to numerous pertinent models from different fields of current interest like the relativistic particle model with curvature, the extended Maxwell-Chern-Simon's model , the quantum charged rigid membrane model and  the Regge-Teitelboim model. Below we give a summary of the results for these various models.
	
	\item The action of the relativistic particle model with curvature  consists of a curvature term along with the usual relativistic term \cite{pisarski, plyuschay1, BMP}. Existence of this curvature term makes the theory higher derivative. For the massive case, we have seen that there are two independent primary first class constraints. Whereas, there is only one independent gauge degree of freedom. This mismatch, which appears in the Ostrogradski formalism \cite{nesterenko}, was solved by considering the first order formalism. In the first order formalism, for the massive case, we got one independent primary first class constraint and one independent gauge symmetry. This   gauge symmetry is identified as the diffeomorphism symmetry. Also, when we considered the mass shell condition $p_1^2 - m^2=0$ separately,  the constraint structure changes. There appears one independent PFC. The Dirac brackets between the basic variables , in this case,also get modified. 
	
	\item From the massive relativistic particle model with curvature,  the massless version of this model is obtained by considering only the curvature term \cite{plyuschay2, BMP}. Canonical analysis in the first order formalism is performed for this model. We have seen that, in this case, the constraint structure is different from the massive case. We have found five first class constraints, with no second class constraints, out of which two are primary. Using these first class constraints, we have constructed the Hamiltonian gauge generator.  The method of Banerjee et. al. showed that in the  gauge generator there are two  independent gauge parameters. Restoring to a geometric approach we have identified these symmetries
with the diffeomorphism and $W$-morphisms and explicitly demonstrated the $W_3$-algebra.
Note that previously the symmetries were demonstrated on-shell \cite{ramos1, ramos2} and a consistent Hamiltonian analysis was lacking. By casting the equations of motion of this model in the Bossinesq form it was earlier demonstrated that the model is endowed with a larger symmetry group. It is interesting that our Hamiltonian method straightforwardly exhibits the full symmetry group and the $W_3$- algebra is established in the Hamiltonian approach without taking recourse of the equations of motion.

	\item The BRST formulation for the higher derivative theories were lacking in the literature. For this reason, the BRST formulation for  the massive and massless relativistic particle models  were considered \cite{banerjee_brst}.  As the massive relativistic particle  consists of the diffeomorphsm symmetries, its BRST analogue is constructed. The massless model, on the other hand, has two gauge symmetries viz. diffeomorphism and $W_3$ symmetry. The nilpotent BRST and anti-BRST symmetries for these models have also been investigated. For the massless version, a remarkable feature for such symmetries comes out which is  the manifestation of $W_3$ algebra.
The BRST symmetries for all variables (except for the antighost variable) corresponding to diffeomorphism and W-morphism satisfy the $W_3$-algebra. Likewise, apart from the ghost variable, the anti-BRST symmetry transformations for all other variables also satisfy the same $W_3$ algebra. Thus the full $W_3$ algebra for all variables is obtained by taking into account both BRST and anti-BRST transformations. The finite coordinate-dependent BRST (FCBRST) symmetry, which is a quantum mechanical analogue of finite field-dependent BRST (FFBRST), has  been analyzed in full generality for higher derivative particle models. It has been
shown that, although such a transformation is a symmetry of the effective action, it breaks the invariance of the generating functional of the path integral. The Jacobian of path integral measure changes nontrivially for FCBRST
symmetry transformation. We have shown that FCBRST transformation with a suitable coordinate-dependent parameter changes the effective action from one gauge to
another within a functional integral. Thus, FCBRST formulation is very useful to connect two different Greens functions for models of relativistic particles.

\item The field theoretic model we considered is the Extended Maxwell-Chern-Simons theory.  The Chern-Simon's piece considered here is a modified one and actually contains the higher derivative\cite{deser2}. The field $A_\mu$ and their time derivatives are considered to be dynamical variables to apply  the first order formalism. We see that there is only one independent PFC and the gauge generator also contains  one independent   gauge parameter. The Lagrangian gauge symmetry which is the $U(1)$ symmetry of the model is shown to match the gauge symmetry obtained by the Hamiltonian formulation in the first order approach.  

\item The next model we have considered is the quantum charged rigid membrane which is the modification of the Dirac's original membrane model of the electron \cite{dirac1}. At the Lagrangian level reparametrisation is the inherent gauge symmetry of this model. The action when expressed with respect to the coordinate variables on the membrane shows higher derivative nature. So, we have adopted the  the first order formalism to explore its gauge symmetries. There appears one independent primary first class constraint and the corresponding  gauge symmetry in the model is mapped to the reparametrisation invariance. Farther, to show the viability of this first order approach the equations of motion are shown to match for the higher derivative Lagrangian and the converted  first order model.

\item Finally, we considered a gravity model   to show the  applicability of  the algorithm  developed in this thesis for analysing the higher derivative theories. The model we have considered here is the Regge Teitelboim model of Cosmology. In the Regge Teitelboim model, a 4D gravity theory is considered to be  induced on a hypersurface which is embedded in a 10D Minkowski spacetime.   Unlike usual gravity theories where the  metric is considered as independent variable, in this case, the embedding functions which are induced on the hypersurface are considered to be dynamical variables. The Lagrangian, however, is not a genuine higher derivative one as it contains a surface term which effectively makes the Lagrangian first order. We kept the surface term as in gravity theories they cannot be neglected.  Hamiltonian formulation in the first order formulation is completed. The model has one independent gauge symmetry which is identified as reparametrisation invariance. As there appeared two independent first class constraints, to remove all the redundant degrees freedom, we considered two gauge conditions. One of the gauge condition is the cosmic gauge condition and the  another one is newly proposed here \cite{banerjee_rt}.  The Wheeler DeWitt equation for this cosmological model is constructed and compatibility with the existing literature is shown.

In this thesis we have developed a systematic algorithm for the analysis of gauge symmetries of a higher derivative theory in the canonical framework. There are different approaches towards the Hamiltonian method of treating the higher derivative theories. We follow the equivalent first order approach where the momenta are defined in the usual way and the signature of the higher derivative nature is contained in the existence of the Lagrangian constraints.

 The algorithm developed in the thesis has been applied to higher derivative models from various fields namely, we have considered the mechanical model, field theoretic model, gravity model etc. We hae also connected our approach with the BRST formalism. In all these examples we have rederived already well established results and resolved many issues which were not adequately discussed in the literature. Besides this, we have shown certain remarkable differences in the abstraction of gauge symmetries compared to usual first order systems. The number of independent gauge transformations is no longer strictly equal to  the number of independent primary first class constraints of the higher derivative theory. The robustness and generality of our approach indicates its utility in the discussion and analysis of higher derivative singular system.
  % 	So from the above discussions we conclude  that in this thesis a vivid picture of the canonical formulation of the higher derivative theories emerges. Since,  the adopted equivalent first order formulation shows the results which already are well establish and also resolves issues which were unsolved  in the previous literature we are sure that this will enable one to have a systematic algorithm for the canonical treatment of the higher derivative theories and specify for the abstraction of the independent gauge degrees of freedom.  

%%%%%%%%%%%%%%%%%%%%%%%%%%%%%%%%%%%%%%%%%%%%%%%%%%%%%%%%%%%%
%%%%%%%%%%%%%%%%%%%%%%%%%%%%%%%%%%%%%%%%%%%%%%%%%%%%%%%%%%%%%%%%%%%%%%%%%%%%%%%
 

\addcontentsline{toc}{chapter}{Bibliography}
\begin{thebibliography}{100}\raggedright \small \setlength{\itemsep}{0.0cm}
%%%%%%%%%%%%%%%%%%%%%%%%%%%%%%%%%%%%%%%%%%%%%%%%%%%%%%%%%%%%%%%%%%%%%%%%%%%%%%%
%Indroduction ref.
%%%%%%%%%%%%%%%%%%%%%%%%%%%%%%%%%%%%%%%%%%%%%%%%%%%%%%%%%%%%%%%%%%%%%%%%%%%%%%

\bibitem{thirring1} W. Thiring, \textit{``Regularization as a consequence of higher order equations"}, Phys. Rev. \textbf{77} (1950) 570.

\bibitem{stelle1} K.S. Stelle,  \textit{``Renormalization of higher derivative quantum gravity"}, Phys. Rev. D  \textbf{16 }(1977) 953.

\bibitem{fradkin1} E.S. Fradkin and A.A. Tseytlin, \textit{``Renormalizable asymptotically free quantum theory of gravity"}, Nucl. Phys. B \textbf{201} (1982) 469.
 
\bibitem{pias1} A. Pais and G.E. Uhlenbeck, \textit{``On field theories with nonlocalized action"}, Phys. Rev. \textbf{79} (1950) 145.



\bibitem{reyes1} C. M. Reyes, \textit{``Testing symmetries in effective models of higher derivative field theories"}, Phys. Rev. D \textbf{80 } (2009) 105008.

\bibitem{pisarski} R.D. Pisarski, \textit{``Field theory of paths with a curvature-dependent term"}, Phys. Rev. D \textbf{34} (1986) 670.

\bibitem{podolsky} B. Podolsky, \textit{``A generalised electrodynamics: part I Non-Quantum"}, Phys. Rev. \textbf{62} (1942) 68;

 \textit{``A generalised electrodynamics: part II Quantum "}, Phys. Rev.\textbf{ 65} (1944) 228.

\bibitem{elizer} D. A. Eliezer and R. P. Woodard, \textit{``The problem of nonlocality in string theory"},  Nucl. Phys.B325, 389
(1989).

\bibitem{bergshoeff}  E. A. Bergshoeff and O. Hohm \textit{``Massive Gravity in Three Dimensions"} Phys. Rev. Lett. \textbf{102} (2009) 201301;

 E. A. Bergshoeff  O. Hohm and P. K. Townsend \textit{``More on massive 3D gravity"}, Phys. Rev. D \textbf{79} (2009)124042. 

\bibitem{gullu} I. Gullu and B. Tekin \textit{``Massive higher derivative gravity inD-dimensional antide Sitter spacetimes"}, Phys. Rev. D \textbf{80} (2009) 064033 ;

 I. Gullu, T. C. Sisman and 
B. Tekin \textit{``Canonical structure of higher derivative gravity in 3D"}, Phys. Rev. D \textbf{81} (2010) 104017; 

\bibitem{neupane} I. P. Neupane, \textit{``Consistency of higher derivative gravity in the brane background"}, JHEP 09 (2000) 040.

\bibitem{hawking} S.W  Hawking and J.  C.  Luttrel  \textit{``Higher derivatives in quantum cosmology"}, Nucl. Phys. B \textbf{247} (1984) 250. 


\bibitem{mazitelli} F. D. Mazitelli {``Higher derivatives and renormalization in quantum cosmology"}, Phys. Rev. D 45 (1992) 2814; .



\bibitem{nojiri1} S. Nojiri, S. D. Odintsov, \textit{``Brane-world cosmology in higher derivative gravity or warped compactification in the next-to-leading order of AdS/CFT correspondence"}, JHEP \textbf{07}(2000)049.

\bibitem{sotiriou} T. P. Sotiriou and  V. Faraoni, \textit{``$f(R)$ theories of gravity"},  Rev. Mod. Phys. \textbf{82}  (2010) 451.

\bibitem{paschalidis} V. Paschalidis,S. M. H. Halataei, S. L. Shapiro and I. Sawicki \textit{``Constraint propagation equations of the 3+1 decomposition of f(R) gravity"}, Class. Quantum Grav. \textbf{28} (2011) 085006.

\bibitem{faraoni} V. Faraoni \textit{``f (R) gravity: successes and challenges"},  arXiv: 0810.2602.

\bibitem{fabri}  L. Fabbri, S. Vignolo \textit{``Dirac fields in f(R)-gravity with torsion"}, 	Class. Quant. Grav. \textbf{28} (2011) 125002. 

\bibitem{ohta}  N. Ohta, \textit{``A complete classification of higher derivative gravity in 3D and criticality in 4D"}, Class. Quantum Grav. \textbf{29} (2012) 015002.

\bibitem{brustein} R. Brustein
 and A. J. M. Medved \textit{``Unitarity constraints on the ratio of shear viscosity to entropy density in higher derivative gravity"}, Phys. Rev. D 83 (2011) 126005.

\bibitem{bunch} T. S. Bunch \textit{``Surface terms in higher derivative gravity"}, J. Phys. A: Math. Gen. \textbf{14} (1981) L139.

\bibitem{deser3} S. Deser and P. van Nieuwenhuizen  \textit{``One-loop divergences of quantized Einstein-Maxwell fields"}, Phys. Rev. D \textbf{10} (1974) 401.

\bibitem{dyer} E. Dyer and K. Hinterbichler, \textit{``Boundary terms, variational principles, and higher derivative modified gravity"}, Phys. Rev. D \textbf{79} (2009) 024028 .

\bibitem{nojiri3} S. Nojiri, S. D. Odintsov, \textit{``Strong coupling limit of N=2 SCFT free energy and higher derivative AdS/CFT correspondence"},  Phys. Lett. B  \textbf{471}(1999)155.

\bibitem{fukuma} M. Fukuma, S. Matsuura, T. Sakai, \textit{``Higher-Derivative Gravity and the AdS/CFT Correspondence"}, Prog. Theor. Phys. \textbf{105} (2001) 1017 [hep-th/0103187].

\bibitem{ostro}  M. Ostrogradsky, \textit{Mem. Ac. St. Petersbourg} {\bf V 14} (1850) 385.

\bibitem{nesterenko} V.V. Nesterenko, \textit{``The singular lagrangians with higher derivatives"}, J. Phys. A \textbf{22} (1989) 1673.

\bibitem{mulish} S. I. Muslih, H. A. El-Zalan \textit{`` Hamiltonian Formulation of Systems with Higher Order
Derivatives"}, Int J Theor Phys  \textbf{46} (2007) 3150.

\bibitem{BMP} R. Banerjee, P. Mukherjee, B. Paul, \textit{``Gauge symmetry and W-algebra in higher derivative systems" }, JHEP \textbf{08} (2011) 085. 

\bibitem{plyuschay1}M.S. Plyushchay, \textit{``Canonical quantisation and mass spectrum of relativistic particle analogue of relativistic string with rigidity"}, Mod. Phys. Lett. A \textbf{3} (1988) 1299.

\bibitem{plyuschay2} M.S. Plyushchay, \textit{``Massless point particle with rigidity"}, Mod. Phys. Lett. A \textbf{4 }(1989) 837.

\bibitem{BRR} R. Banerjee, H. J. Rothe, and K. D. Rothe, \textit{``Hamiltonian approach to lagrangian gauge symmetries"},  Phys. Lett. B \textbf{463}  (1999) 248;

 \textit{``Master equation for lagrangian gauge symmetries"}, Phys. Lett B.  \textbf{479} (2000) 429 .


\bibitem{dirac2} P. A. M. Dirac, \textit{``The Fundamental Equations of Quantum Mechanics"}, Poc. R. Soc. Lond. A 1925, 109. ``Lectures on Quantum Mechanics", New York : Belfer Graduate School of Science, Yeshiva University, 1964.


\bibitem{henneaux1} M. Henneaux and C. Teitelboim, \textit{``Quantization of Gauge  Systems"},  (Princeton University Press, Princeton, New Jersey, 1994).

\bibitem{gitman} D.M. Gitman and I.V. Tyutin, \textit{``Quantization of Fields with constraints"}, Springer, U.S.A.
(1990).
\bibitem{hanson1} A. Hanson, T. Regge, and C. Tietelboim, \textit{``Constrained
Hamiltonian System"}, (Accademia Nazionale Dei Lincei, Roma, 1976).

\bibitem{rothe1} H. J. Rothe and K. D. Rothe, \textit{``Classical and Quantum Dynamics of Constrained Hamiltonian Systems"}, Lecture Notes in Physics Vol. 81 (World Scientific, New York, 2010).

\bibitem{sundermeyer1} K. Sundermeyer, Lecture Notes in Physics 169,
\textit{``Constrained Dynamics"}, (Springer-Verlag, Berlin, 1982).

\bibitem{prokhorov} L. V. Prokhorov and S. V. Shabanov, \textit{``Hamiltonian Mechanics of Gauge Systems"}, Cambridge University Press,  Cambridge CB2 8RU, UK, 2011.



\bibitem{dirac1} P. A. M. Dirac, \textit{``An Extensible Model of the Electron"}, Proceedings of the Royal Society of London. Series A, Mathematical and Physical Sciences \textbf{268} (1962) 57.

\bibitem{regge1} T. Regge and C. Teitelboim, in \textit{``Proceedings of the First Marcel Grossman Meeting"}, Trieste, Italy, (1975), ed. R. Ruffini (NorthHolland, Amsterdam) 77 (1977).

\bibitem{polyakov} A. Polyakov, \textit{``Fine structures of strings"}, Nucl.  Phys.  B \textbf{268} (1986)  406.



\bibitem{peliti}  L. Peliti and S. Leibler, \textit{``Effects of Thermal Fluctuations on Systems with Small Surface Tension"}, Phys. Rev. Lett. 54, 1690. 

\bibitem{plyuschay3} M. S. Plyuschay, \textit{``Massless particle with rigidity as a model for the description of bosons and fermions"}, Phys. Lett. B  \textbf{243} (1990) 383.


\bibitem{ramos1} E. Ramos and J. Roca, \textit{``W-symmetry  and  the  rigid particle"}, Nucl. Phys. B \textbf{436} (1995)529; 

\textit{``Extended gauge invariance in geometrical particle models and the geometry of $W$-symmetry"},  Nucl. Phys. B \textbf{452} (1995) 705.
 

\bibitem{ramos2} E. Ramos and J. Roca,\textit{``On W3-morphisms and the geometry of plane curves"},  Phys. Lett. B \textbf{ 366}  (1996) 113.

\bibitem{capovilla1} R. Capovilla, J. Guven and E. Rojas, \textit{``Hamiltonian FrenetSerret dynamics"},  Class. Quantum Grav. \textbf{19}  (2002) 2277.

\bibitem{kunzetsov} Yu. A. Kuznetsov and M.S. Plyushchay, \textit{``The model of the relativistic particle with curvature and torsion"}, Nucl. Phys. B \textbf{389} (1993) 181.

\bibitem{bakas}  I. Bakas, \textit{``The Structure of the $W_\infty$ Algebra"}, Comm. Math. Phys. \textbf{134} (1990) 487.

\bibitem{hull} C. M. Hull, \textit{``The geometry of \textit{W}-gravity"}, Phys. Lett. B  \textbf{269 } ( 1991 ) 257.

\bibitem{gervais}  J. L. Gervais and Y. Matsuo, `\textit{`\textit{W}-geometries"}, Phys. Lett. B \textbf{274} (1992) 309.

 \bibitem{becchi} C. Becchi, A. Roue, and R. Stora,  \textit{``Renormalization of gauge theories"}, Annl. Phys. \textbf{98} (1976) 287.

\bibitem{brst} S. D. Joglekar and  B. P. Mandal,  \textit{``Finite field-dependent BRS transformations"}, Phys. Rev. D \textbf{51} (1995) 1919; 

 \textit{``Application of finite field dependent BRS transformations to the problems of the Coulomb gauge"}, Int. J. Mod. Phys. A \textbf{17} (2002) 1279;
 
  
 S. Upadhyay and B. P. Mandal, \textit{``Finite BRST transformation and constrained systems"}, Ann. Phys. (N.Y.) \textbf{327} (2012) 2885 ; 
 
    
     R. Banerjee and B. P. Mandal,\textit{``Quantum gauge symmetry from finite field dependent BRST transformations"},  Phys. Lett. B \textbf{488} (2000) 27.
 
 \bibitem{banerjee_brst} R. Banerjee, B. Paul, S. Upadhyay,  \textit{``BRST symmetry and W-algebra in higher derivative models" }, Phys. Rev. D \textbf{88} (2013) 065019. 

%\bibitem{sdj} S. D. Joglekar,  B. P. Mandal, \textit{``Finite field-dependent BRS transformations"}, Phys. Rev. D {\bf 51}  (1995)  1919. 

 
\bibitem{deser2} S. Deser and R. Jackiw, \textit{``Higher derivative ChernSimons extensions"}, Phys. Lett. B \textbf{451}  (1999) 73.

\bibitem{sarmishtha} S. Kumar \textit{``Lagrangian and Hamiltonian formulations for Higher derivative chern simons theory"}, Int. J. Mod. Phys. A, 18, 1613 (2003) [hep-th/0112121]. 
 
 \bibitem{mukherjee_exmcs} P. Mukherjee, B. Paul,  \textit{``Gauge invariances of higher derivative Maxwell-Chern-Simons field theory -- a new Hamiltonian approach"}, Phys. Rev. D \textbf{85 }(2012) 045028.

\bibitem{dirac_membrane} P. A. M. Dirac,\textit{ ``An Extensible Model of the Electron"}, Proc. R. Soc. Lond. A \textbf{268} (1962)  57.

\bibitem{onder}  M. Onder, R.W. Tucker, \textit{``Membrane interactions and total mean curvature"}, Phys. Lett B \textbf{202} (1988) 501.


\bibitem{napoli} G. Napoli and L. Vergori \textit{``Extrinsic Curvature Effects on Nematic Shells"}, Phys. Rev. Lett \textbf{108 }(2012) 207803.

\bibitem{cordero_qcrm}R. Cordero, M. Cruz, A. Molgado, and E. Rojas \textit{``Quantum modified Regge-Teitelboim cosmology"}, 	Gen. Relativ. Grav. \textbf{46} (2014) 1761.




 \bibitem{christensen} M. Christensen, V. P. Frolov and A. L. Larsen, \textit{``Soap bubbles in outer space: Interaction of a domain wall with a black hole"}, Phys. Rev. D   \textbf{58} (1998) 085008.
 
  \bibitem{hope} J. Hoppe, \textit{``Relativistic membranes"}, J. Phys. A  Math. Theor. \textbf{46} (2013) 023001.
  
   \bibitem{trzetrzelewski}  M. Trzetrzelewski, \textit{``Spiky membranes"}, Phys. Lett. B \textbf{684 } (2010) 256.
 
 \bibitem{gnadig} P. Gnadig and Z. Kunszt, \textit{``Diracs  Extended  Electron  Model"}, Annl. Phys.  \textbf{116}  (1978)  380.
 
 \bibitem{capovilla}  R. Capovilla and J. Guven, \textit{``Geometry of deformations of relativistic membranes"}, Phys. Rev. D \textbf{51}(1995) 6736.
 
  \bibitem{lopez} C. A. Lopez, \textit{``Extended model of the electron in general relativity"}, Phys. Rev. D \textbf{30} (1984) 313;
 
     \textit{``Dynamics of charged bubbles in general relativity and models of particles"}, Phys. Rev. D \textbf{38} (1988) 3662;
  
    \textit{``Stability of an extended model of the electron"}, Phys. Rev. D \textbf{33} (1986) 2489. 
  
\bibitem{paul_membrane} B. Paul, \textit{``Gauge symmetry and Virasoro algebra in quantum charged rigid membrane: A first order formalism" }, Phys. Rev. D \textbf{87} (2013) 045003.  
  
  \bibitem{rt_model} A. Davidson, D. Karasik, and Y. Lederer, \textit{`` Wavefunction of a brane-like universe"}, Class.
Quantum Grav. \textbf{16} (1999) 1349.

 \bibitem{davidson2} A. Davidson, D. Karasik, and Y. Lederer, \textit{``Geodesic evolution and nucleation of a de Sitter brane"}, Phys. Rev. D \textbf{72} (2005) 064011.
 
  \bibitem{davidson3} D. Karasik and A. Davidson \textit{``Geodetic brane gravity"}, Phys. Rev. D \textbf{67} (2003) 064012.

\bibitem{cordero_rt} R. Cordero, A. Molgado, and E. Rojas, \textit{``Ostrogradski approach for the Regge-Teitelboim type cosmology"}, Phys. Rev. D \textbf{79} (2009) 024024

\bibitem{banerjee_rt} R. Banerjee, P. Mukherjee, B. Paul, \textit{``New Hamiltonian analysis of Regge Teitelboim minisuperspace cosmology"},  Phys. Rev. D \textbf{89} (2014) 043508 [arXiv:1307.4920].


\bibitem{WDW} B. S. DeWitt \textit{``Quantum Theory of Gravity. I. The Canonical Theory"}, Phys. Rev. \textbf{160} (1967) 113;

 \textit{``Quantum Theory of Gravity. II. The Manifestly Covariant Theory"}, Phys Rev. \textbf{162} (1967) 1195;
 
  \textit{``Quantum Theory of Gravity. III. Applications of the Covariant Theory"}, Phys. Rev.\textbf{ 162} (1967).



\bibitem{deser1} S. Deser and P. van Nieuwenhuizen,  \textit{``One-loop divergences of quantized Einstein-Maxwell fields"}, Phys. Rev. D \textbf{10 } (1974) 401;

\textit{``Nonrenormalizability of the quantized Dirac-Einstein system"}, Phys. Rev. D \textbf{10 } (1974) 411.

\bibitem{ho} P. M. Ho, \textit{``Virasoro algebra for particles with higher derivative interactions"}, Phys. Lett. B \textbf{ 558 }(2003) 238.

\bibitem{cordero} R. Cordero, A. Molgado, E. Rojas,  \textit{``Quantum charged rigid membrane"}, Class. Quantum Grav. \textbf{28} (2011) 065010.
 
 \bibitem{barut}  A.O. Barut, M. Pavic \textit{``Dirac's shell model of the electron and the general theory of moving relativistic charged membrabes"}, Phys. Lett. B \textbf{306} (1993) 49.
 %\begin{center}

 %\end{center}
%%%%%%%%%%%%%%%%%%%%%%%%%%%%%%%%%%%%%%%%%%%%%%%%%%%%%%%%%%%%%%%%%%%%%%%%%%%%%%%%%%%%%%%%%%%%%%%%%%%%%
\end{thebibliography}
\end{document}